\documentclass{aa}

\usepackage{graphicx}
\usepackage{subcaption}
\usepackage{mwe}

\usepackage{txfonts}
\usepackage{hyperref}
\usepackage{url}
\usepackage{xcolor}
\hypersetup{
colorlinks,
linkcolor={red!80!black},
citecolor={blue!80!black},
urlcolor={blue!80!black}
}

\newenvironment{myfont}{\fontfamily{ppl}\selectfont}{\par}
\DeclareTextFontCommand{\textmyfont}{\myfont}

\newcommand{\code}[1]{\texttt{#1}}

\def\nifs{\iso{56}Ni}
\def\cofs{\iso{56}Co}

\def\cm3{cm$^{-3}$}
\def\kms{\mbox{km~s$^{-1}$}}
\def\msunyr{$M_{\odot}$\,yr$^{-1}$}

\def\msun{$M_{\odot}$}

\def\one{\ts {\,\sc i}}
\def\two{\ts {\,\sc ii}}
\def\three{\ts {\,\sc iii}}
\def\four{\ts {\,\sc iv}}

\def\beq{\begin{equation}}
\def\eeq{\end{equation}}

\def\lesssim{\mathrel{\hbox{\rlap{\hbox{\lower4pt\hbox{$\sim$}}}\hbox{$<$}}}}
\def\gtrsim{\mathrel{\hbox{\rlap{\hbox{\lower4pt\hbox{$\sim$}}}\hbox{$>$}}}}

\def\one{{\,\sc i}}
\def\two{{\,\sc ii}}
\def\three{{\,\sc iii}}
\def\four{{\,\sc iv}}

\def\mic{\,$\mu$m}

\def\v1d{{\code{V1D}}}

\def\cmfgen{{\code{CMFGEN}}}

\def\photb{{\code{P-HOTB}}}

\def\ergs{erg\,s$^{-1}$}

\def\ciiuv{[C\two]\,$\lambda$\,$2325.4$}

\def\niidoub{[N\two]\,$\lambda\lambda$\,$6548.0,\,6583.5$}

\def\oidoub{[O\one]\,$\lambda\lambda$\,$6300.3,\,6363.8$}

\def\oiidoub{[O\two]\,$\lambda\lambda$\,$7319.5,\,7330.2$}
\def\oiiuv{[O\two]\,$\lambda$\,$3726.0,\,3728.8$}

\def\oiiidoub{[O\three]\,$\lambda\lambda$\,$4958.9,\,5006.8$}

\def\nad{Na\one\,D}

\def\mgifs{Mg\one]\,$\lambda$\,$4571.1$}
\def\mgiiuv{Mg\two\,$\lambda\lambda$\,$2795,\,2802$}

\def\siiinz{[S\three]\,$\lambda$\,$9068.6$}
\def\siiinf{[S\three]\,$\lambda$\,$9530.6$}

\def\caiidoub{[Ca\two]\,$\lambda\lambda$\,$7291.5,\,7323.9$}

\def\neiifs{[Ne\two]\,12.81\,$\mu$m}

\def\nkiifs{[Ni\two]\,$\lambda$\,7377.8}

\def\nkiimir{[Ni\two]\,6.634\,$\mu$m}
\def\ariimir{[Ar\two]\,6.983\,$\mu$m}
\def\neiiifs{[Ne\three]\,15.55\,$\mu$m}

\newcommand{\iso}[2]{\ensuremath{^{#1}\rm{#2}}}

\begin{document}

   \title{The long-term influence of a magnetar power in stripped-envelope supernovae}   
   \subtitle{Radiative-transfer modeling of He-star explosions from 1 to 10 years}

   \author{
   Luc Dessart\inst{1}
}

   \institute{
 Institut d'Astrophysique de Paris, CNRS-Sorbonne Universit\'e, 98 bis boulevard Arago, F-75014 Paris, France\\
 \email{dessart@iap.fr}
             }

   \date{}

  \abstract{Much interest surrounds the nature of the compact remnant formed in core collapse supernovae (SNe). One means to constrain its nature is to search for signatures of power injection from the remnant in the SN observables years after explosion. In this work, we conduct a large grid of 1D nonlocal thermodynamic equilibrium radiative transfer calculations of He-star explosions under the influence of magnetar-power injection from post-explosion age of about one to ten years. Our results for SN observables vary with He-star mass, SN age, injected power, or ejecta clumping. At high mass (model he12p00), the ejecta coolants are primarily O and Ne, with \oidoub, \oiidoub, and \oiiidoub\ dominating in the optical, and with strong \neiifs\ in the infrared -- this line may carry more than half the total SN luminosity. For lower He-star masses (models he6p00 and he3p30), a greater diversity of coolants appear, in particular Fe, S, Ar, or Ni from the Si- and Fe-rich regions. All models tend to rise in ionization in time, with twice-ionized species (i.e., O\three, Ne\three, S\three, or Fe\three) dominating at $\sim$\,10\,yr, although this ionization is significantly reduced if clumping is introduced. Our treatment of magnetar power in the form of high-energy electrons or X-ray irradiation yields similar results -- no X-rays emerge from our ejecta even at ten years because of high-optical depth in the keV range. An uncertainty of our work concerns the power deposition profile, which is not known from first principles, although this profile could be constrained from observations. Our magnetar-powered model he8p00 with moderate clumping yields a good match to the optical and near-infrared observations of Type Ib SN\,2012au at both 289--335\,d (power of $1-2\times10^{41}$\,\ergs) and 2269\,d (power of 10$^{40}$\,\ergs). Unless overly ionized (i.e., if the optical spectrum shows only strong \oiiidoub), we find that all massive magnetar-powered ejecta should be infrared luminous at 5--10\,yr through strong \neiifs\ line emission.
}


   \maketitle


\section{Introduction}
\label{sect_intro}

Massive star explosions captured by transient surveys as Type II, Ib, and Ic supernovae (SNe) should in general be forming a neutron star (see, e.g., \citealt{burrows_lattimer_86}, \citealt{sukhbold_ccsn_16}, \citealt{ertl_ibc_20}, \citealt{burrows_rev_21}). This compact object generally rests as a dark, dormant body at the center of the SN ejecta, with no observable impact on the SN properties for years, decades, or centuries after the explosion of the progenitor star \citep{fransson_87A_24}.  At least, photospheric -and nebular-phase observations of standard core-collapse SNe can at all epochs up to a few years be explained by the release of a combination of shock-deposited energy (only relevant during the photospheric phase) and radioactive energy from the decay of \nifs\ (generally dominant during the nebular phase). A power contribution from the compact remnant cannot be excluded, but in general there is no need to invoke this extra source. For example, the late-time brightness of Type II SNe is compatible with of order 0.01\,\msun\ of \nifs\ which core collapse SNe routinely produce through explosive nucleosynthesis (see, e.g., \citealt{sukhbold_ccsn_16}).

However, a fraction of core-collapse SNe exhibit properties that are in tension with current expectations for neutrino-driven explosions of massive stars. One critical diagnostic is the excess peak brightness (i.e., relative to some mean value built from a large sample) of some transients during the photospheric phase, which may occur in Type II SNe (e.g., \citealt{terreran_slsn2_17}, although this seems quite rare), or in Type Ib or Ic SNe (see for example \citealt{drout_11_ibc}; \citealt{prentice_ibc_16}; \citealt{lyman_ibc_16}; \citealt{anderson_nifs_19}; \citealt{meza_anderson_ni56_20}), leaving aside here the transients that show obvious signatures of ejecta interaction with circumstellar material (CSM). Some extreme outliers are SNe Ic associated with $\gamma$-ray bursts (e.g., \citealt{woosley_98bw_99}, \citealt{patat_98bw_01}, \citealt{foley_02ap_03}) or superluminous SNe Ic \citep{pasto_10gx_10,quimby_slsnic_11, nicholl_slsn_14}, but some objects are also found in the intermediate domain between superluminous SNe and the large population of ``standard" SNe (e.g., SN\,2005bf, \citealt{folatelli_05bf_06,maeda_05bf_07}; or SN2012au, \citealt{milisavljevic_12au_13,pandey_12au_21}). Explaining the peak of the light curve with radioactive decay requires uncomfortably large \nifs\ masses of several 0.1\,\msun, well outside the reasonable range expected for garden-variety core-collapse SNe (see, e.g., \citealt{sukhbold_ccsn_16}). Another source of tension is the persisting brightness, or even the rebrightening of some transients at late times, in contrast to the expectation of a luminosity following the exponential decline of the radioactively-decaying power source. Examples of such phenomena for Type II SNe are SN\,2013by \citep{black_13by_17} and SN\,2017eaw \citep{weil_17eaw_20}, and for stripped-envelope SNe we can name SN\,2014C \citep{milisavljevic_14C_15,margutti_14C_16}, SN\,2012au \cite{milisavljevic_12au_18}, SN\,2017dio \citep{kuncarayakti_17dio_18}, SN\,2021ocs \citep{kuncarayakti_21ocs_22}, or SN2022xxf \citep{kuncarayakti_22xxf_23}.

In some of these cases, the additional source of power at late times is thought to arise from interaction with CSM (e.g., SN\,2017eaw, \citealt{weil_17eaw_20}; SN\,2023ixf, \citealt{bostroem_23ixf_24}; SN\,1993J, \citealt{matheson_93j_00a}; SN\,2014C, \citealt{margutti_14C_16}; or SN\,2019yvr, \citealt{ferrari_19yvr_24}), and radiative-transfer simulations support this interpretation \citep{chevalier_fransson_94,dessart_csm_22,dessart_late_23}. This seems particularly well suited for Type II SNe because of their slow, dense winds, but less so in stripped-envelope SNe because of their more tenuous, faster expanding winds. Stripped-envelope SNe with signatures of interaction should then arise from interaction with material lost in previous mass transfer events (see for example the recent simulations of \citealt{ercolino_bin_24}), but in general this material would have been lost long before core collapse and may be too distant from the exploding star to produce any detectable signature.

An alternative is for the power to arise from the compact remnant, not in quantities sufficient to produce a superluminous SN but strong enough to alter modestly the SN radiation properties during the photospheric phase (to cause the marginally overluminous early peak, say produce an excess luminosity by a factor of two or three) and long-lived to generate a late-time power that far exceeds any contribution from the radioactive decay of unstable isotopes produced through explosive nucleosynthesis. Indeed, the spin-down of a magnetized neutron star with an initial field of order 10$^{14}$\,G can deliver a power of order 10$^{41}$\,\ergs\ for decades after explosion, which is orders of magnitude larger than any reasonable contribution from \nifs, \iso{57}Ni, or \iso{44}Ti at such times. Finding such transients is however a challenge (for an X-ray survey, see \citealt{margutti_slsn_xray_18}), although it is not clear how the injected power should channel into different regions of the electromagnetic spectrum, from X-rays to the infrared.

Previous work on radiative-transfer modeling of SNe influenced through power injection from a compact remnant are few. Some previous work focused on the spectral properties during the photospheric phase, for both H-rich \citep{dessart_audit_18,d18_iptf14hls} and H-poor configurations \citep{d12_magnetar,mazzali_slsn_16,d19_slsn_ic}. Other investigations considered the nebular phase but up to about one or two years after explosion   \citep{jerkstrand_slsnic_17,d18_iptf14hls,d19_slsn_ic}. More recently, \citet{omand_pm_23} performed radiative-transfer modeling of O-rich SN ejecta with masses in the range 1 to 10\,\msun\ and subject to power injection from the compact remnant or associated wind nebula. They treat this power in the form of a photoionizing flux and study the resulting properties for the gas and radiation at late times of 1 to 6\,yr post explosion. Focusing on the spectral signatures in the optical, they document how the strength of the \oidoub, \oiidoub\ and \oiiidoub\ vary with the model parameters such as the photoionization spectrum or ejecta mass. They obtain a good match to the observed optical spectra of SN\,2012au at 1 and 6\,yr although not for the same ejecta mass.

Here, we extend previous work on radiative-transfer modeling of magnetar-powered SNe by performing full-ejecta simulations from 250--500\,d until about 4000\,d after explosion. We employ the He-star explosion models of \citet{woosley_he_19} and \citet{ertl_ibc_20}, mixed macroscopically but not microscopically using a shuffled-shell technique \citep{DH20_shuffle}. Radiative-transfer simulations for such ejecta at nebular epochs and under the influence of radioactive decay only were presented in \citet{dessart_snibc_21,dessart_snibc_23}. Unlike \cite{omand_pm_23}, our simulations cover a range of He-star masses and retain the full complexity of the ejecta composition in each, in particular the marked differences between preSN shells. We also document the observables from the ultraviolet to infrared so direct comparison to multiwavelength observations of stripped-envelope SNe at late times is possible.

In the next section, we present the numerical setup for our calculations, including the selection of explosion models and the assumptions used for the radiative transfer modeling. In Section~\ref{sect_rad_to_pwn}, we present the photometric evolution of various models subject to a constant magnetar-power of 10$^{39}$\,\ergs\ and discuss the transition from radioactively powered to magnetar powered at about 700\,d. We review the cooling processes and key emission lines of our H-free ejecta at nebular times in Section~\ref{sect_cool_proc}. We then present our results for the gas and radiation properties for different He-star models, first with the he12p00 model (Section~\ref{sect_he12p00}), the he6p00 model (Section~\ref{sect_he6p00}), and the he3p30 model (Section~\ref{sect_he3p30}). We then discuss the dependency of our results on a number of parameters, including the level of ejecta clumping (Section~\ref{sect_dep_fvol}), the nature of the power injected (i.e., high-energy electrons or X-ray irradiation; Section~\ref{sect_pwr_vs_xray}), the adopted deposition profile for the power injection (Section~\ref{sect_dep_edep}), the magnitude of the power injected (Section~\ref{sect_pwr}). In Section~\ref{sect_comp_obs}, we select the best suited models from our grid for a comparison to the magnetar-powered  candidate SN\,2012au. Section~\ref{sect_pm_vs_csm} confronts the magnetar and CSM interaction scenarios for explaining the late-time observations of SNe like 2012au. We present our conclusions in Section~\ref{sect_conc}.\footnote{All simulations in this work will be uploaded at \url{https://zenodo.org/communities/snrt}.}

\begin{table*}
    \caption{Ejecta model properties.
\label{tab_prog}
}
    \vspace{-0.5cm}
    \begin{center}
\begin{tabular}{|l@{\hspace{2mm}}|c@{\hspace{2mm}}c@{\hspace{2mm}}c@{\hspace{2mm}}c@{\hspace{2mm}}c@{\hspace{2mm}}c@{\hspace{2mm}}c@{\hspace{2mm}}c@{\hspace{2mm}}c@{\hspace{2mm}}c@{\hspace{2mm}}c@{\hspace{2mm}}c@{\hspace{2mm}}c@{\hspace{2mm}}|}
\hline
Model  &  $M_{\rm preSN}$  & $M_{\rm ej}$  &     $E_{\rm kin}$   & $V_m$   &  \iso{4}He & \iso{12}C   & \iso{14}N  & \iso{16}O &  \iso{24}Mg &  \iso{28}Si &  \iso{40}Ca &  \iso{56}Ni$_{t=0}$ & $V_{{\rm max},\nifs}$ \\
       &     [\msun] & [\msun]   &        [foe]    & [\kms]    &    [\msun] & [\msun] & [\msun] & [\msun] & [\msun] & [\msun] & [\msun] & [\msun] & [\kms] \\
\hline
   he3p30      & 2.67 &   1.20  &     0.55   &      6777   &      0.84    &      0.06  &   6.21(-3)   &   1.51(-1)   &   1.75(-2)   &   2.76(-2)   &   1.00(-3)   &   4.00(-2)  &  3712   \\
   he6p00      & 4.44 &   2.82  &     1.10   &      6269   &      0.95    &      0.25  &   6.20(-3)   &   9.74(-1)   &   1.01(-1)   &   5.88(-2)   &   2.12(-3)   &   7.04(-2)  &  4990   \\
   he8p00      & 5.63 &   3.95  &     0.71   &      4251   &      0.84    &      0.49  &   5.17(-3)   &   1.71       &   1.10(-1)   &   4.89(-2)   &   2.00(-3)   &   5.46(-2)  &  3435   \\
   he12p00     & 7.24 &   5.32  &     0.81   &      3911   &      0.23    &      1.00  &   1.42(-4)   &   3.03       &   8.73(-2)   &   7.41(-2)   &   3.42(-3)   &   7.90(-2)  &  2531   \\
\hline
\end{tabular}
\end{center}
    {\bf Notes:} The table columns correspond to the preSN mass, the ejecta mass, the ejecta kinetic energy (1 foe $\equiv 10^{51}$\,erg), the mean
    expansion rate $V_{\rm m} \equiv \sqrt{2E_{\rm kin}/M_{\rm ej}}$, the cumulative yields of \iso{4}He, \iso{12}C, \iso{14}N, \iso{16}O, \iso{24}Mg,
    \iso{28}Si, \iso{40}Ca, and \nifs\ prior to decay, as well as the ejecta velocity that bounds 99\% of the total \nifs\ mass in the corresponding
    model. Numbers in parenthesis correspond to powers of ten.
\end{table*}

\begin{figure*}
   \centering
    \begin{subfigure}[b]{0.49\textwidth}
       \centering
       \includegraphics[width=\textwidth]{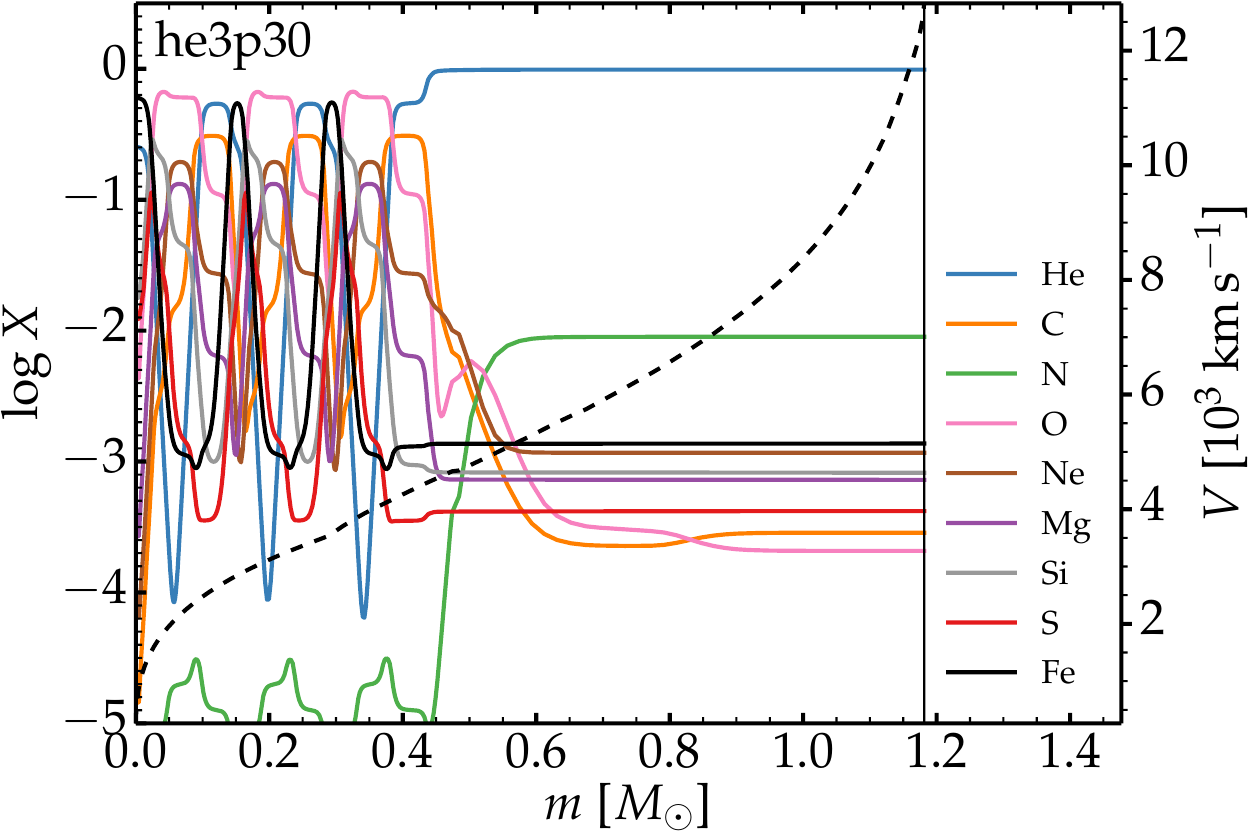}
    \end{subfigure}
    \hfill
    \begin{subfigure}[b]{0.49\textwidth}
       \centering
       \includegraphics[width=\textwidth]{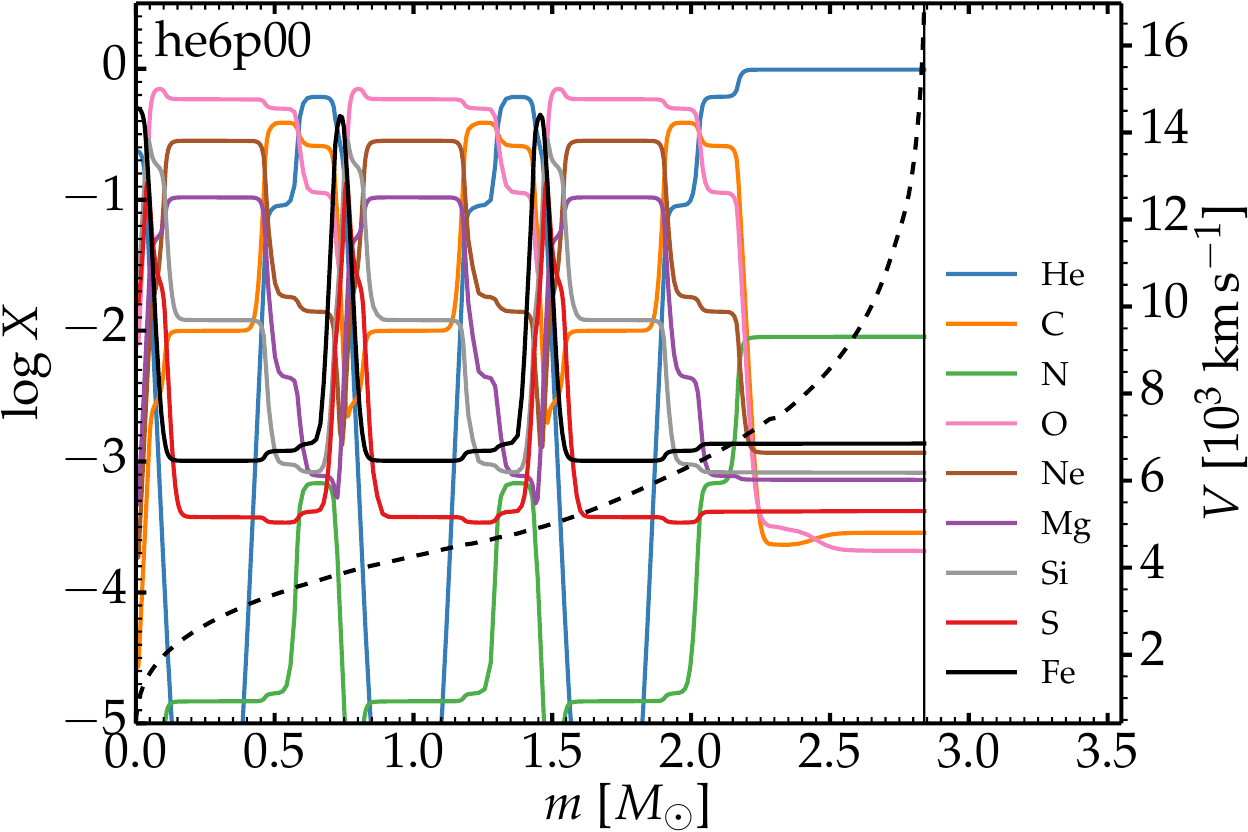}
    \end{subfigure}
     \hfill
    \begin{subfigure}[b]{0.49\textwidth}
       \centering
       \includegraphics[width=\textwidth]{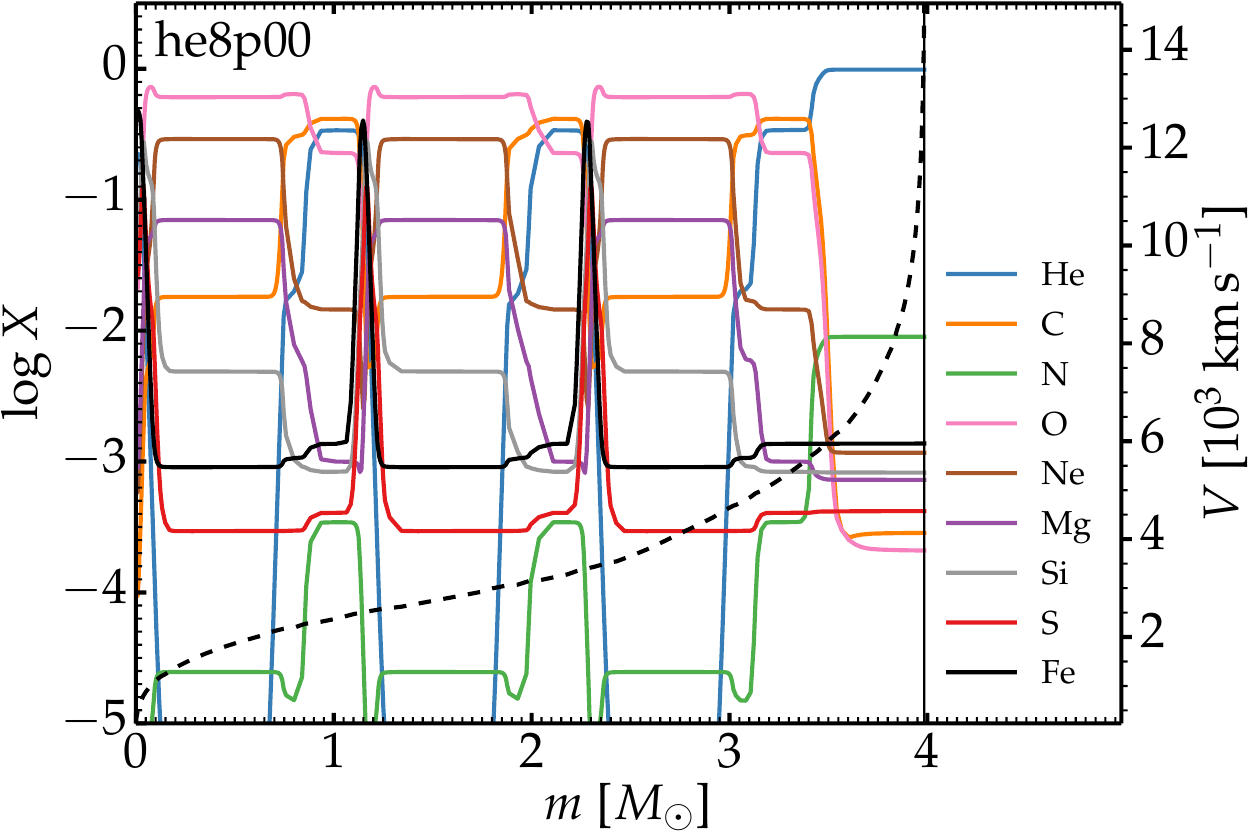}
    \end{subfigure}
     \hfill
     \begin{subfigure}[b]{0.49\textwidth}
       \centering
        \includegraphics[width=\textwidth]{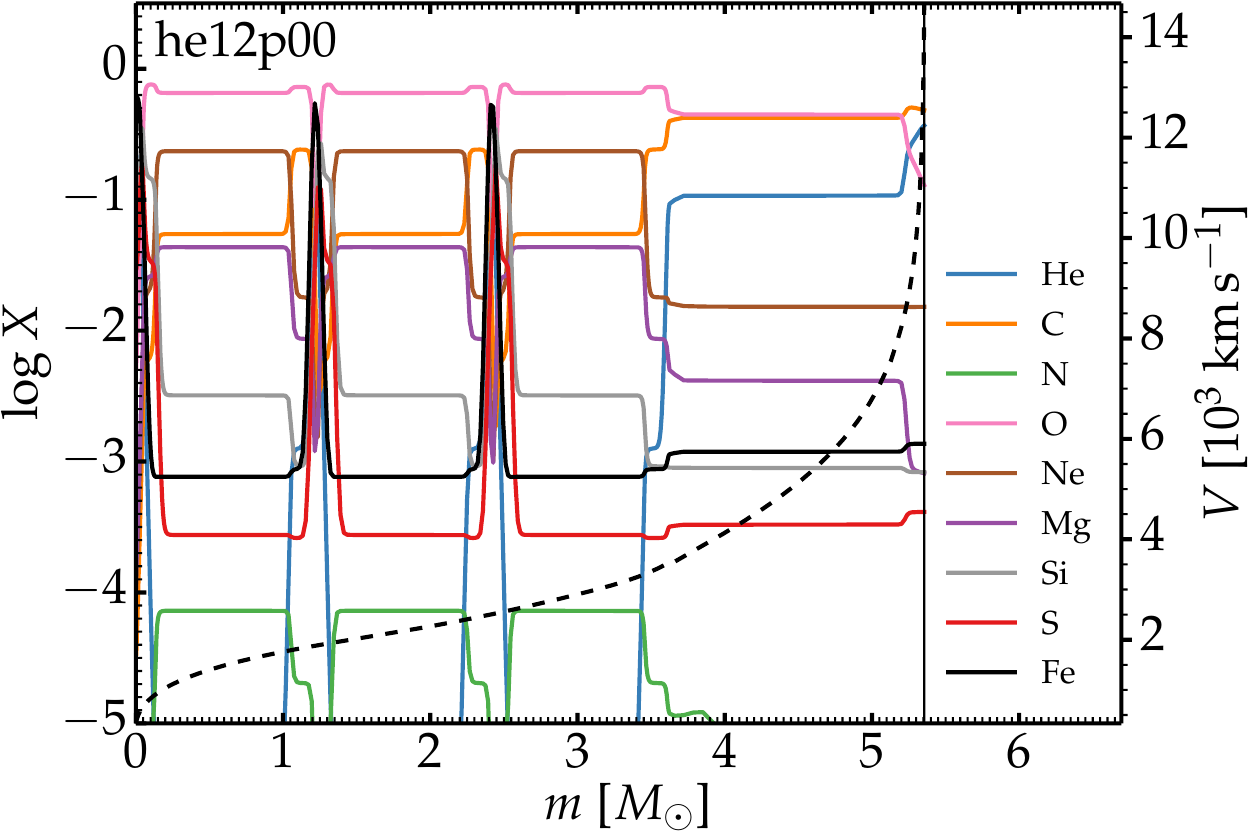}
    \end{subfigure}
\caption{Initial ejecta properties. From top-left to bottom-right panels, we show the composition versus mass for the dominant species, including He, C, N, O, Ne, Mg, Si, S, and Fe, for models he3p30, he6p00, he8p00, and he12p00. We also show the ejecta velocity (dashed line, with labelling on the y-axis at right).
\label{fig_init}
}
\end{figure*}

\section{Numerical setup}
\label{sect_setup}

\subsection{PreSN and explosion models}

For this study, we focus on stripped-envelope SNe and use the same explosion models as those discussed in \citet{dessart_snibc_21,dessart_snibc_23}. They correspond to stars that were evolved at solar metallicity and without rotation from the He zero-age main sequence \citep{woosley_he_19} and subsequently exploded following core collapse \citep{ertl_ibc_20}. We select their He-star models with initial masses between 2.6 and 12\,\msun, corresponding to H zero-age main sequence masses between about 14 and 36\,\msun. Here, we limit our set to four He-star masses, which are representative of ejecta compositions and masses for stripped-envelope SNe, namely with he3p30 (composition dominated by helium), he6p00, he8p00, and he12p00 (composition dominated by oxygen) in order of increasing preSN mass, ejecta mass, or oxygen mass (see, e.g., Fig.~3 in \citealt{dessart_snibc_23}).  The 1D explosion models were produced  with \photb\ using a calibrated explosion engine (for details, see \citealt{ertl_ibc_20}) so that the explosion energy and explosive yields bear a greater physical realism than if they were essentially prescribed arbitrarily in a piston-driven explosion. For the present conceptual exploration, this is not critical either way. A summary of model properties (yields, masses, energetics) is provided in Table~\ref{tab_prog}.

To keep in line with the previous studies of  \citet{dessart_snibc_21,dessart_snibc_23}, we  use the same shuffled-shell ejecta composition (see discussion in \citealt{DH20_shuffle}). With this approach, individual shells (e.g., the Si/S or the O/Ne/Mg shell) in the unmixed ejecta of the 1D explosion model are redistributed or shuffled in mass (or equivalently in velocity) space but without any alteration to the composition mixture of the original shells (i.e., the relative abundances of all elements is the same in the original and shuffled-shell model).\footnote{The outermost He/C or He/N shell does not take part in the shuffling.} This introduces macroscopic mixing but no microscopic mixing, and thus captures in a 1D treatment the multidimensional mixing that takes place first during a neutrino-driven explosion (see, e.g., \citealt{wongwathanarat_15_3d}, \citealt{gabler_3dsn_21}) and exacerbated by the prolonged power injection from a compact remnant \citep{suzuki_pm_2d_17,chen_3d_pm_20}. We show the composition stratification of our four shuffled-shell ejecta models in Fig.~\ref{fig_init}. A full description of these models is given in \citet{dessart_snibc_21} and is thus not repeated here.

The presence of a magnetized, rotating compact object may not be justified in these four models but our goal here is primarily to document how a long-lived power source would impact the radiation emerging from ejecta of distinct mass and composition. Ultimately, observations should be sought to qualify or disqualify a given progenitor mass or composition. Progenitors evolved with fast rotation and magnetic fields, which evolve chemically homogeneously (see, e.g., \citealt{yoon_grb_06}, \citealt{WH06}, \citealt{georgy_rot_12}, \citealt{aguilera_evol_ic_18}) were used for the magnetar-powered models presented in \citet{dessart_98bw_17} and \citet{d19_slsn_ic} -- such models were He-deficient and thus more suitable to explain broad-lined, superluminous, or GRB/SNe of Type Ic. One advantage of the present models (and also the motivation for their use in \citealt{dessart_snibc_21,dessart_snibc_23}) is the detailed and physically-consistent nucleosynthesis (both secular and explosive; \citealt{woosley_he_19}; \citealt{ertl_ibc_20}), suppressing the shortcomings of older simulations performed with a 13-, 19-, or 21-isotope network. Both the he6p00 and he8p00 models are He-rich and may be compatible with the Type Ib classification of SN\,2012au (see, e.g., the He-star explosion models studied in their photospheric phase in \citealt{dessart_snibc_20}). 

\begin{table*}
   \caption{Summary of main parameters covered by our magnetar-powered simulations. [See Section~\ref{sect_setup} for details.]
\label{tab_sum}
}
\vspace{-0.5cm}
\begin{center}
  \begin{tabular}{|l@{\hspace{8mm}}|c@{\hspace{8mm}}c@{\hspace{8mm}}c@{\hspace{8mm}}c@{\hspace{8mm}}|}
\hline
Model          &  Power                               &         $dV$       & $f_{\rm vol}$              &  Epochs  \\
               &  [\ergs]                             &        [\,\kms]    &                           &    [d]   \\
\hline
   he3p30      &  1e39                                &         2000       &  0.01, 0.1, 0.2, 0.5, 1.0 &    250--4300 \\
   he6p00      &  1e39, 1e40, 1e41                    &         2000--2800 &  0.01, 0.1, 0.2           &    300--4300 \\
      he8p00    &  [1,2,4,8]e39, 1e40, [1,2]e41  &         2000--2800 &  0.1, 0.2, 0.5, 1.0       &     300--4300 \\
   he12p00     &  1e39, 1e40                          &         2000--2800 &  0.01, 0.1, 0.2, 0.5, 1.0 &     500--4300 \\
\hline
\end{tabular}
\end{center}
{\bf Notes:} The columns list for each model the power employed, the parameter $dV$ caracterizing the associated deposition profile, the volume filling factor
    $f_{\rm vol}$, and the epochs for the calculation. Not all parameter permutations are covered.  Not included in the table are models in which the
    power is injected in the form of X-rays, which was explored with model
    he6p00 at 3100\,d with an injection power of 10$^{39}$\,\ergs\ and adopting a variety of injection profiles (see Section~\ref{sect_pwr_vs_xray} for discussion).
\end{table*}

\subsection{Radiative transfer with \cmfgen}

The simulations presented in this work were carried out with the 1D nonlocal thermodynamic equilibrium (nonLTE) time-dependent radiative transfer code \cmfgen\ \citep{hm98,HD12}. In contrast to standard SN simulations, we use a steady-state mode (each model is calculated without time dependence and thus ignores the previous history of the radiation field or level populations). In the study of \citet{dessart_snibc_23}, some simulations were run in a time-dependent model, others assumed steady state, and all produced essentially the same results.\footnote{We leave to future work a proper testing of this assumption at these very late times out to 10\,yr since \citet{dessart_snibc_23} covered out to 1.5\,yr at most. At sufficiently late times, the ever decreasing density might inhibit recombination and cooling, which could affect the luminosity \citep{fransson_87A_93} or the ejecta ionization.} However, with steady state comes a greater flexility for simulating ejecta influenced by different levels of magnetar power or ejecta clumping. Furthermore, although all simulations include the contribution of radioactive decay, we focus on the influence of a prolonged injection of power from the compact remnant on the SN ejecta and radiation properties. Because \cmfgen\ is a radiative-transfer code, we ignore any dynamical effects related to this power injection.

The radiation from magnetars and more generally from magnetized fast rotating neutron stars, may be a complex mix of high-energy particles and radiation (see, e.g., \citealt{kouveliotou_pm_98}, \citealt{abdo_fermi_13}, \citealt{guillot_pulsar_19}). For the most part of this work, we inject the magnetar power in the form of high-energy electrons. These are similar in nature to the electrons that Compton-scatter with the $\gamma$ rays released in radioactive decays. We can thus use the standard nonthermal solver in \cmfgen\ generally employed for the treatment of radioactive decay (the same ansatz is used for the treatment of shock power from ejecta interaction with circumstellar material (CSM) in \citealt{dessart_csm_22}). In Section~\ref{sect_pwr_vs_xray}, we show the results obtained when the power injection is instead treated as an X-ray emitting source (see Dessart \& Hillier, in prep., for details) but we find that it essentially yields the same results under the current assumptions (e.g., spherical symmetry) and epochs considered. For the sake of simplicity and flexibility, we inject this power at a prescribed rate without any consideration of how it translates into a specific magnetic field and rotation rate for the protomagnetar at birth.\footnote{For that approach, see simulations for iPTF14hls presented in \citet{d18_iptf14hls} and for superluminous SNe Ic presented in \citet{d19_slsn_ic}.} 

The power from the compact remnant is given a prescribed deposition profile using the approach of  \citet{dessart_csm_22}.  Versus velocity $V$, it goes as $\exp \left(- [(V-V_0)/dV]^2 \right)$ where $V_0$ is the innermost ejecta velocity and $dV$ is some prescribed velocity scale that depends on the model and epoch. A normalization is then applied to yield the desired instantaneous injection power at that time. The choice of value for $dV$ has a strong impact on the results and is thus discussed specifically in Section~\ref{sect_dep_edep}. In most simulations, $dV$ is chosen to be a few 1000\,\kms\ in order to allow the magnetar power to influence a large volume of the ejecta. Assuming that all the power were absorbed in the innermost ejecta layers would lead to the formation of a temperature spike compromising the convergence of the code (it is also most likely unphysical -- see below). Unless otherwise stated, $dV$ is set to 2000\,\kms. Some models use an increased value of 2800\,\kms, in particular for later times when one may expect the magnetar radiation to penetrate further into the outer ejecta.  The original set of simulations performed for this study used a power of 10$^{39}$\,\ergs\ since this was of the order of the total luminosity recorded in the optical in SN\,2012au at $\sim$\,6\,yr \citep{milisavljevic_12au_18}. We later found out from our simulations that a large fraction of the flux comes out in the infrared after about 1--2\,yr, so many simulations were later performed with an increased power up to 10$^{40}$\ergs. Additional simulations with even larger power up to several 10$^{41}$\ergs\ were also performed for comparison to SN\,2012au at times prior to 1\,yr (these models are presented in Dessart \& Hillier, in prep.).  

To better design our simulations, the magnitude of the power injected should be informed from observations, ideally covering from the ultraviolet to infrared. Unfortunately, such observations are lacking -- the infrared flux may also arise in part from newly formed or pre-existing dust (see, e.g., \citealt{shahbandeh_jwst_23}) so spectra rather than photometry would be required to distinguish continuum-like dust emission from line emission by the ejecta. The other shortcoming in the design of our simulations is the adopted rather than derived deposition profile for the power from the compact remnant. There is evidence that overluminous or superluminous stripped-envelope SNe, likely powered by a magnetar, are asymmetric explosions on large scales \citep{mazzali_97ef_00,maeda_2d_98bw_02,maeda_98bw_03,dessart_98bw_17,barnes_grb_snicbl_18,suzuki_grb_22}. Numerical simulations for ejecta influenced by such power injection also suggest considerable structure on small scales, along with turbulence, chemical mixing, and clumping \citep{chen_pm_2d_16,chen_slsnic_2d_17,suzuki_pm_2d_17,chen_3d_pm_20,suzuki_pm_2d_21}. These works suggest that the power from the remnant is thus not deposited in a narrow, spherical shell at the inner edge of the ejecta (this might at best hold at the earliest times) but more likely over a range of depths encompassing a broad range of velocities in the inner parts of the ejecta (and likely with some angular dependence, which we cannot account for in our 1D approach). This is the motivation for using a relatively large value for $dV$.

All simulations with \cmfgen\ performed in this study assume steady state. Calculations at one epoch, or for one injection power or for some different level of clumping, are used as initial conditions for other simulations with similar parameters. This speeds up the convergence of these CPU-intensive calculations which typically employ 200\,000 frequency points and 2000 atomic levels whose populations are solved for at each of the 350 radial grid points. We treat the following atoms and ions: He\one--\two, C\one--\three, N\one--\three, O\one--\three, Ne\one--\three, Na\one, Mg\one--\three, Al\two--\three, Si\one--\three, S\one--\three, Ar\one--\three, K\one, Ca\one--\two, Sc\one--\three, Ti\two--\three, Cr\two--\three, Fe\one--\four, Co\two--\three, and Ni\one--\three. This model atom departs from those normally used in core-collapse SNe at late times with the additional treatment here of twice ionized C, N, O, Ne, or Mg, which are normally neglected. 

Although the original goal of this work was to investigate the properties at the origin of the observed spectrum of SN\,2012au at $\sim$\,6\,yr, we have extended the parameter space not just by including four different He-star models, a range of magnetar powers, but also a much broader range of epochs spanning from 250--500\,d out to 4300\,d. We also introduced clumping in numerous models since it is known to mitigate the ejecta ionization \citep{d18_fcl} and was even found to be essential for the reproduction of nebular-phase spectra of superluminous, likely magnetar powered, Type Ic SNe \citep{jerkstrand_slsnic_17,d19_slsn_ic}. Clumping is also a natural consequence of power injection in the inner ejecta material \citep{suzuki_pm_2d_17,chen_3d_pm_20}. For simplicity, when introduced in our calculations, this clumping is uniform at all depths and corresponds to a volume filling factor of the gas between 1\,\% (highest clumping used) and 100\,\% (i.e., a smooth ejecta). This broad grid of models allows us to test the sensitivity of these ejecta to power injection and may be of use to study objects other than SN\,2012au. Furthermore, because magnetar-power injection may not be fundamentally so different in nature from shock power injection in a young SN remnant or interaction with CSM (i.e., in all cases power is injected in the form of high-energy particles and photons), these simulations may have a wider range of relevance than just magnetar-powered SNe. Table~\ref{tab_sum} presents a summary of all the simulations that we carried out, including their key distinguishing parameters.

\begin{figure*}[h]
\centering
\includegraphics[width=0.4\hsize]{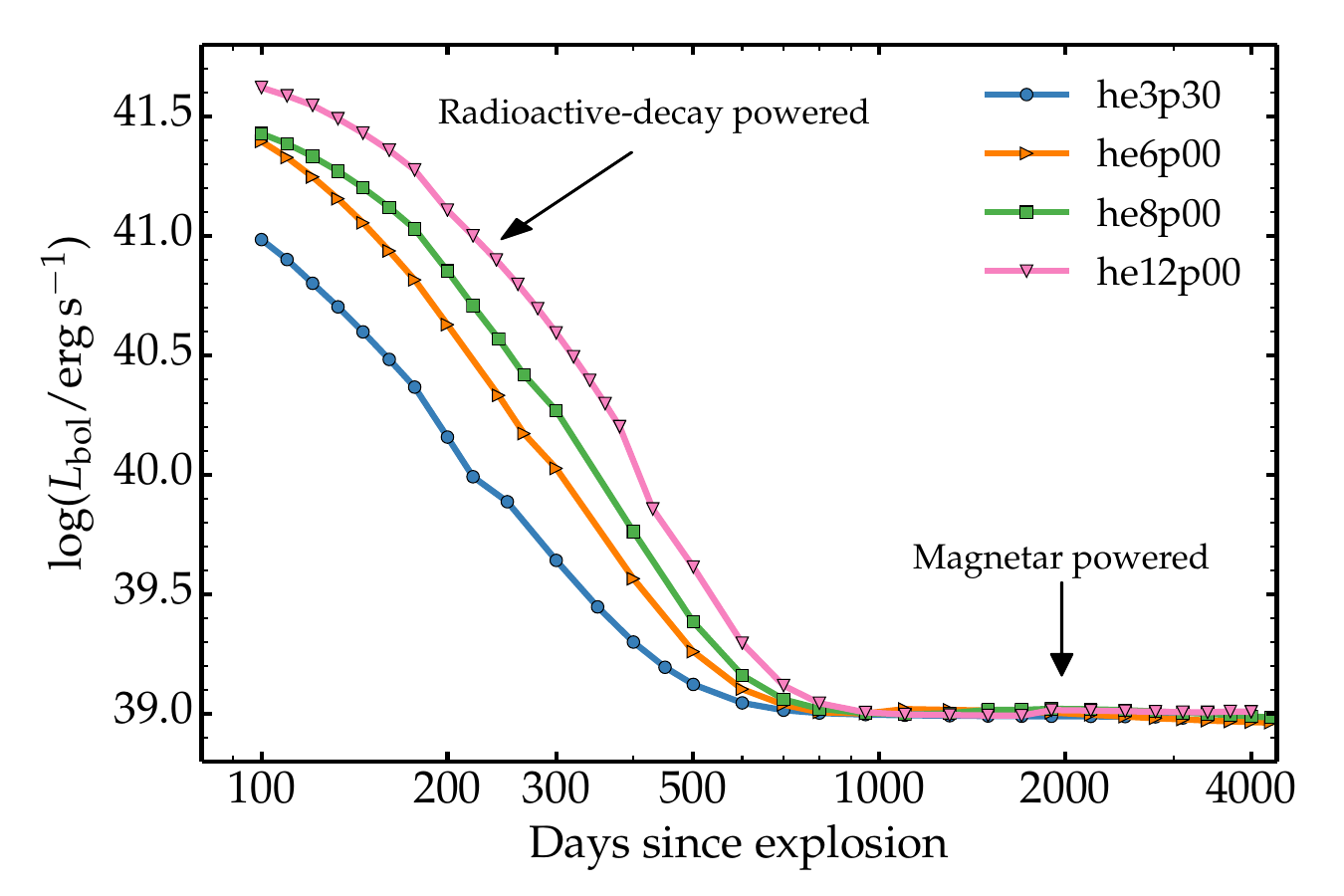}
\includegraphics[width=0.4\hsize]{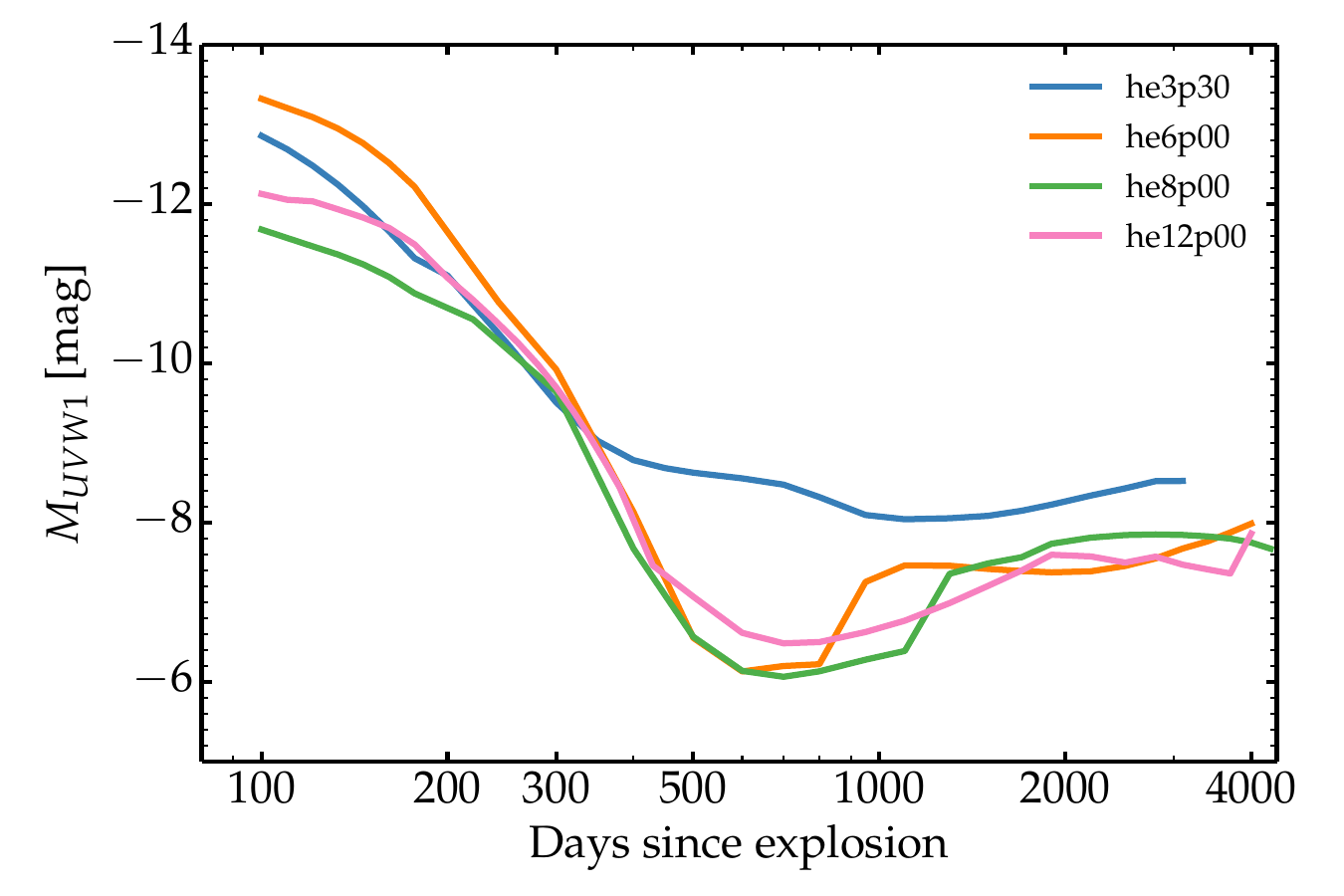}
\includegraphics[width=0.4\hsize]{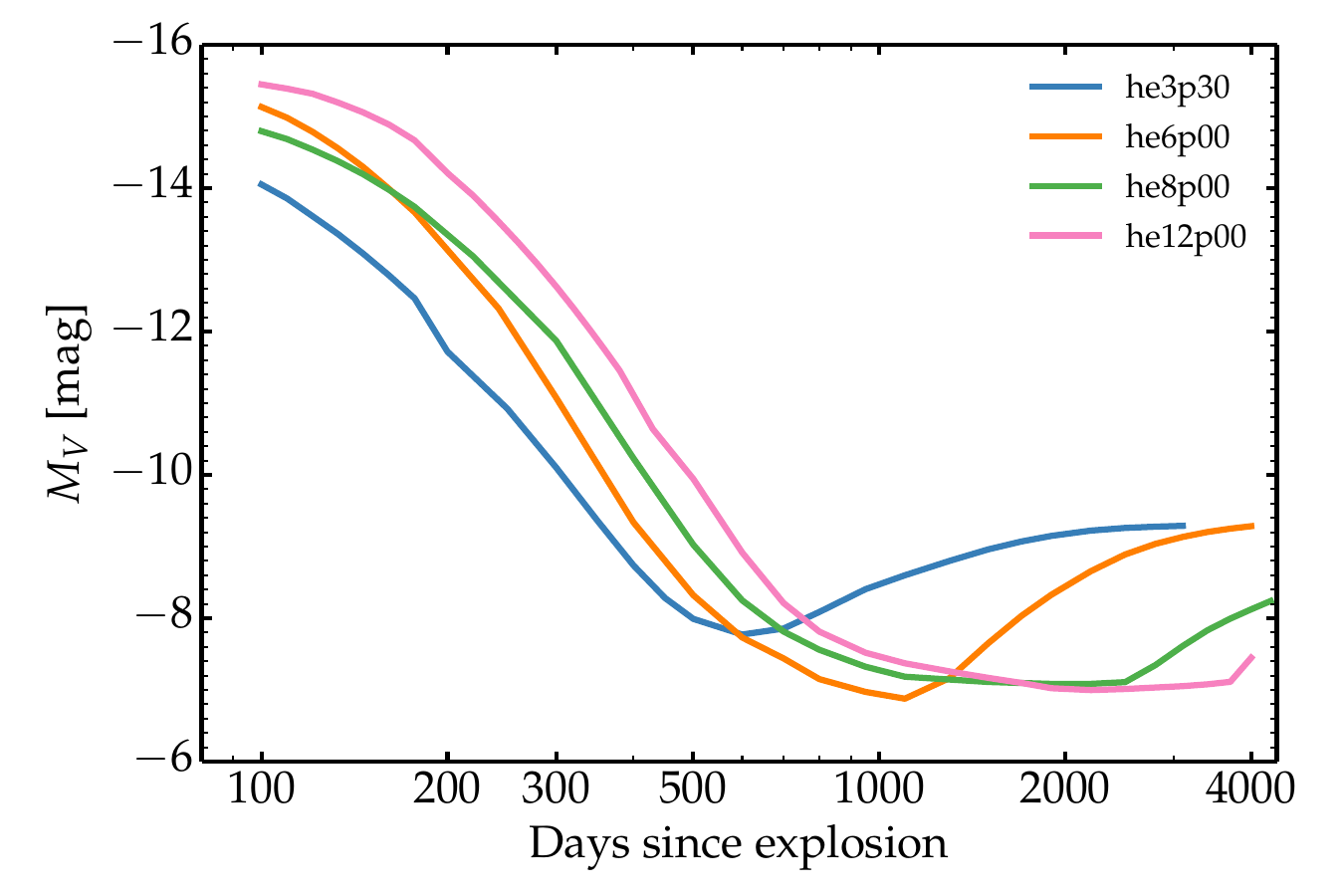}
\includegraphics[width=0.4\hsize]{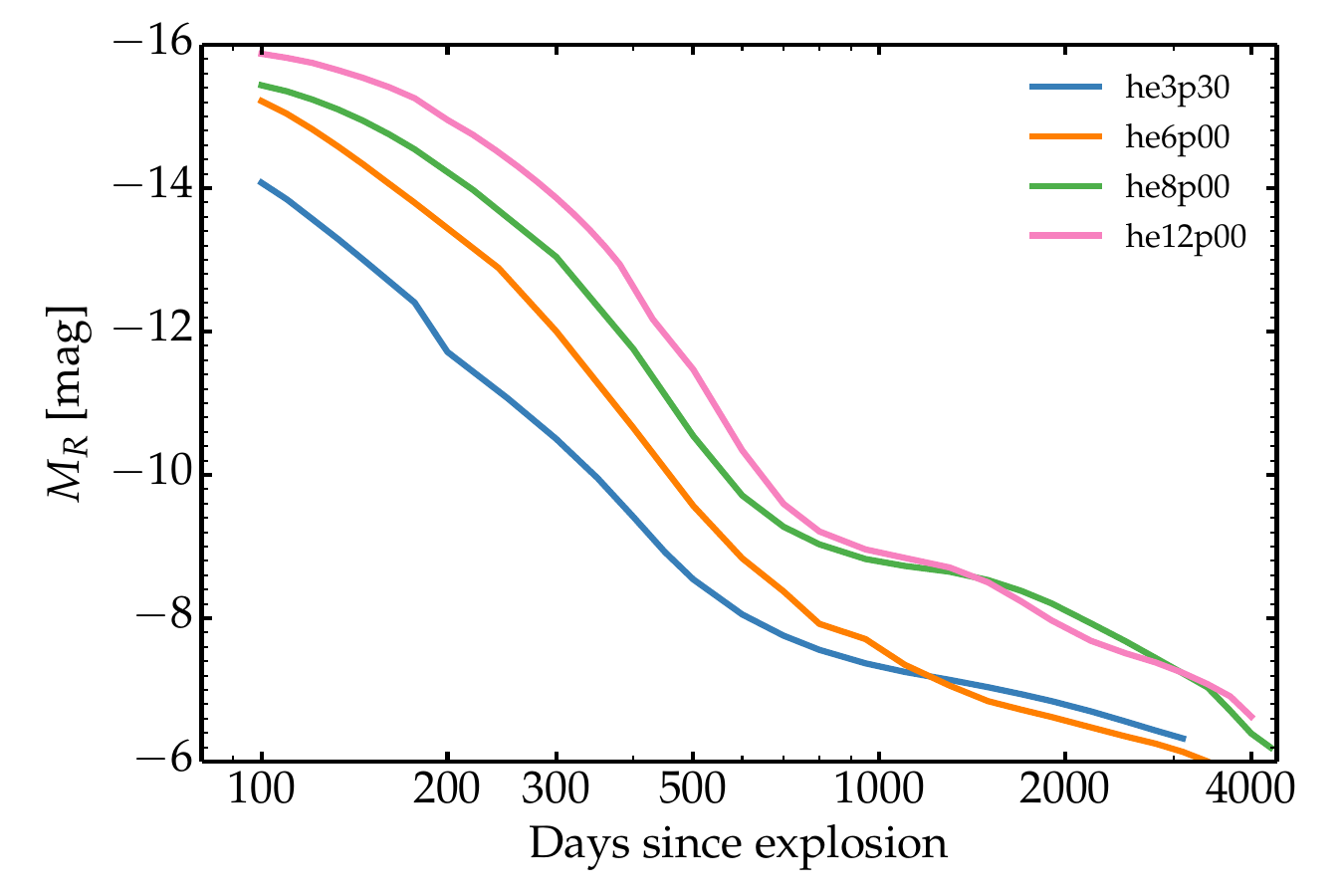}
\caption{Bolometric and multiband light curves for He-star explosion models he3p30, he6p00, he8p00, and he12p00 covering from 100 to $\sim$\,4000\,d after explosion. A constant magnetar power of 10$^{39}$\,\ergs\ is introduced after 250--500\,d, depending on the model, and becomes the dominant power source beyond about 700\,d. Light curves at early times, prior to magnetar power injection, are taken from \citet{dessart_snibc_23}. [See Section~\ref{sect_rad_to_pwn} for discussion.]}
\label{fig_lbol}
\end{figure*}

\section{The transition from radioactively-powered to magnetar-powered at late times}
\label{sect_rad_to_pwn}

The radioactive decay of \nifs\ and \cofs\ is included at all times but the nominal magnetar power, set to a constant value of 10$^{39}$\,\ergs, is introduced only after 250\,d in model he3p30, after 300\,d in models he6p00 and he8p00, and after 500\,d in model he12p00. At those times, the main power source for the ejecta is radioactivity with a total SN luminosity of about 10$^{40}$\,\ergs. In these models, the ejecta switch from being primarily radioactively powered to being primarily magnetar powered at about 700\,d. This is visible in the bolometric light curve (top-left panel of Fig.~\ref{fig_lbol}), with a clear inflection from an exponentially declining to a flat curve. With a greater or lower power injection, this inflection in the light curve would have occurred earlier or later. Similar inflections may occur in Type II SNe at 1--3\,yr after explosion, as observed in SNe 2013by \citep{black_13by_17} or 2017eaw \citep{weil_17eaw_20}, and obtained in 
simulations of the interaction of the SN ejecta with a progenitor wind even with a modest mass loss rate of 10$^{-6}$\,\msunyr\ \citep{dessart_late_23}.

While the influence of the power injection on the bolometric light curve is obvious, the multi-band light curves show a more complicated structure (top-right panel and bottom row of Fig.~\ref{fig_lbol}). Magnetar power eventually leads to a rebrightening in $UVW1$ and $V$, here at about 700\,d, but the $R$ band exhibits instead a persistent brightness decline. The light curves are therefore strongly filter dependent, which arises from the fact that the spectral energy distribution at these nebular times is dominated by line emission, rather than continuum emission. Hence, the multi-band light curves reflect which lines (whose wavelength, and thus associated filter, is dictated by atomic physics) dominate the cooling and evolve in strength (see next section and Table~\ref{tab_coolants_processes_lines}). 

\begin{table}
\caption{Summary of important coolants in our late-time magnetar-powered ejecta models. We list the ions and processes of relevance, together with the dominant associated lines. The list is ordered in increasing atomic number, and then in increasing ionization state, for all species. [See Section~\ref{sect_cool_proc} for discussion.]
\label{tab_coolants_processes_lines}
}
\vspace{-0.5cm}
\begin{center}
\begin{tabular}{|l|l|c|}
\hline
    Ion      & Proc. & Main lines \\
\hline
He\,{\sc i}  & NT   &       3888.6\,\AA; 1.083, 2.058\mic\ \\
He\,{\sc ii} & NT   &       4685.7\,\AA\ \\
\hline
C\,{\sc ii}  & COL  &       2325.4\,\AA\ \\
C\,{\sc ii}  & NT   &       1335.7, 6578.0--6582.9, 7231.3--7236.4\,\AA\   \\
C\,{\sc iii} & COL  &       1906.7--1908.7\,\AA\  \\
\hline
N\,{\sc ii}  & COL  &       6548.0, 6583.4\,\AA \\
\hline
O\,{\sc i}   & COL  &       6300.3, 6363.8\,\AA  \\
O\,{\sc i}   & NT   &       8446.4\,\AA\ \\
O\,{\sc ii}  & COL  &       3726.0--3728.8\,\AA\  \\
O\,{\sc ii}  & NT   &       7318.9, 7320.0, 7329.7, 7330.7\,\AA\ \\
O\,{\sc iii} & COL  &       4958.9, 5006.8\,\AA\ \\
O\,{\sc iii} & NT   &       4363.2\,\AA\   \\
\hline
Ne\,{\sc ii} & COL  &       12.81\mic\ \\
Ne\,{\sc iii}& COL  &       15.55\mic\ \\
\hline
Si\,{\sc ii} & COL  &       1.691\mic\ \\
Si\,{\sc ii} & NT   &       3856.0, 5056.0, 6347.1\,\AA\ \\
Si\,{\sc iii}& COL  &       1882.7, 1892.0\,\AA\ \\
\hline
S\,{\sc i}   & COL  &       25.242\mic\ \\
S\,{\sc ii}  & COL  &       6716.4\,\AA\ \\
S\,{\sc iii} & COL  &       9068.6, 9530.6\,\AA; 18.708\mic\ \\
\hline
Ar\,{\sc ii} & COL  &       6.983\mic\ \\
Ar\,{\sc iii}& COL  &       7135.8\,\AA; 8.989, 21.826\mic\ \\
\hline
Fe\,{\sc i}  & COL  &       24.036\mic\   \\
Fe\,{\sc ii} & COL  &       17.931, 24.512, 25.981\mic\  \\
Fe\,{\sc iii}& COL  &       4658.0, 5011.2\,\AA; 2.218, 7.788, 22.919\mic\ \\
Fe\,{\sc iv} & COL  &       2829.4, 2835.7\,\AA, 2.836\mic \\
\hline
Ni\,{\sc i}  & COL  &       7.505\mic\ \\
Ni\,{\sc ii} & COL  &       6.634, 10.679, 12.725\mic\ \\
Ni\,{\sc iii}& COL  &       7.347, 10.999\mic\ \\
\hline
\end{tabular}
\end{center}
{\bf Notes:} The term ``COL'' stands for collisional excitation and ``NT'' for nonthermal excitation.
\end{table}

\section{Cooling processes at nebular times}
\label{sect_cool_proc}

We explore the late-time evolution of He-star explosion models over a timespan of $\sim$\,1 to $\sim$\,10\,yr post explosion. Our ejecta have representative velocities $V_{\rm m}$ of about 4000--6000\,\kms\ (see Table~\ref{tab_sum}), which translates into a spatial extent at late times in the range 10$^{16}$--10$^{17}$\,cm. The representative electron density is 10$^{4}$--10$^{6}$\,cm$^{-3}$ and the temperature of order 1000\,K, with a strong variation for shells of distinct composition across the ejecta. The ejecta optical depth is typically between 10$^{-2}$ and 10$^{-4}$ so any power absorbed by the ejecta at any given time is reradiated away instantaneously. The low densities of our ejecta models are typically below the critical density of numerous strong forbidden transitions so the power injected by the magnetar (and for up to about 700\,d by radioactive decay) is radiated away by means of collisional excitation and radiative de-excitation of the strongest forbidden lines, which vary in nature depending on the composition and ionization at each ejecta location. This is the dominant cooling process of the ejecta. Another process giving rise to ejecta cooling is nonthermal excitation caused by the injected high-energy electrons and positrons from the magnetar (i.e., this is the adopted treatment in our setup). This process is important to populate levels with a greater excitation energy above the ground state. This is for example what distinguishes \oiiuv, which forms through collision excitation, and \oiidoub, which benefits from nonthermal excitation in our simulations. 

Table~\ref{tab_coolants_processes_lines} lists the main processes and associated atomic transitions contributing to the cooling in our magnetar-powered He-star explosion models at late times. Collisional excitation is denoted by ``COL'' and nonthermal excitation by ``NT''. We list all important transitions identified in our spectra from the ultraviolet out in the infrared. Not all lines are reported and our model atom may not be complete enough to include all weak transitions, although these would contain a negligible flux -- observations in the far-infrared would help with such identifications.  

The lines listed in Table~\ref{tab_coolants_processes_lines} vary considerably in strengths in our simulations, reflecting differences in ejecta mass, kinetic energy (or expansion rate), and composition. Overall, the strongest lines tend to be \oidoub, \oiiuv\, \oiiidoub, \siiinf, and \neiifs. In the next sections, we document how these various coolants strengthen and weaken from model to model, and as function of the power injected and its profile, or as a function of the adopted clumping.


\begin{figure*}
\centering
\includegraphics[width=\hsize]{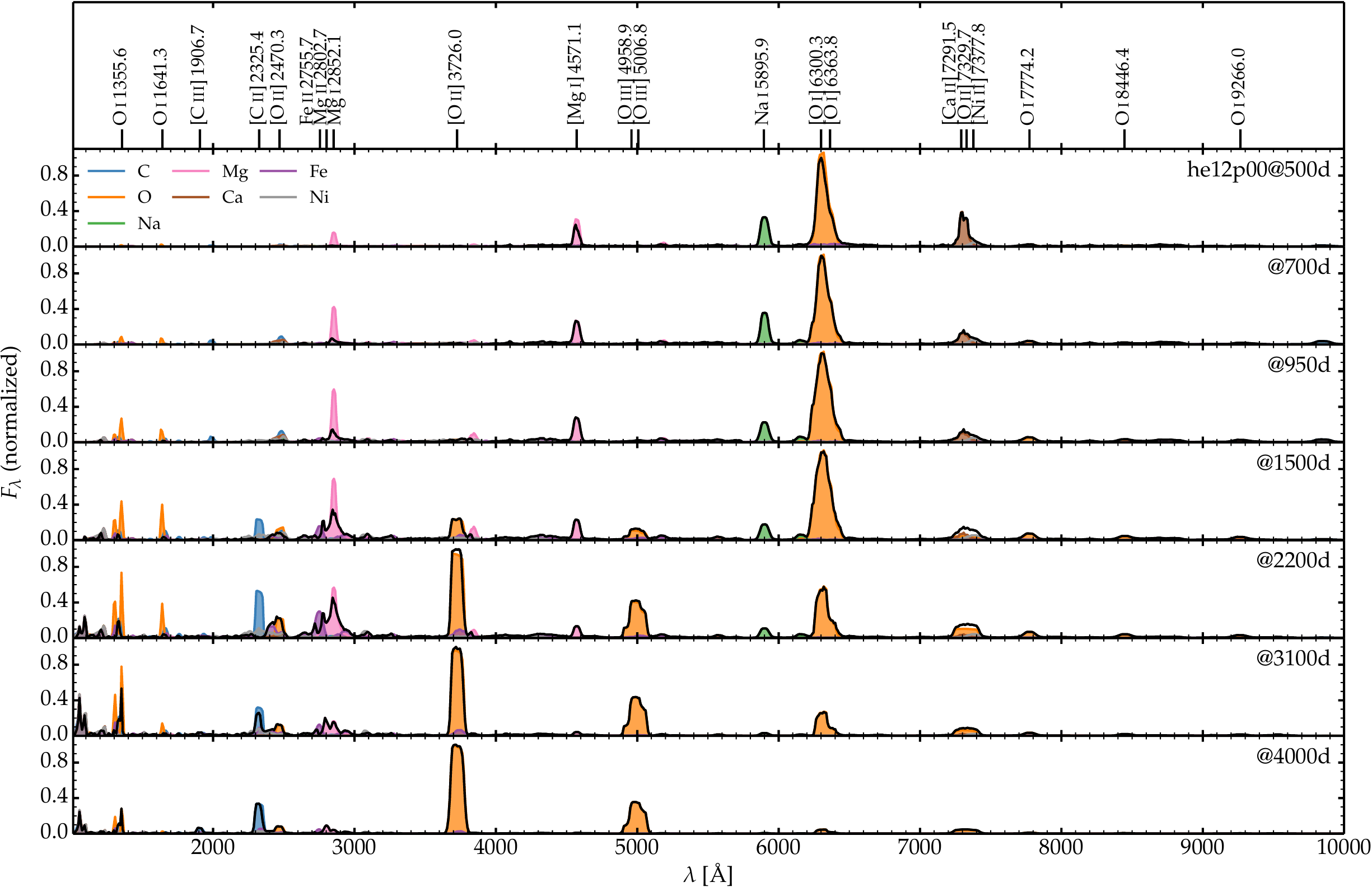}
\caption{Spectral evolution between 1000\,\AA\ and 1\,$\mu$m and from 500 to 4000\,d for model he12p00 under the influence of a magnetar power of 10$^{39}$\,\ergs\ and $dV$ of 2000\,\kms\ at all times. All spectra are normalized to a maximum flux of unity at each epoch. Contributions by species (e.g., those associated with oxygen include contributions from O\one, O\two, and O\three) are indicated with a colored shading. Identification of the main lines is provided in the top panel (in case of multiple contributors, the strongest one is given). [See Section~\ref{sect_he12p00} for discussion.]
\label{fig_spec_he12p00}
}
\end{figure*}

\section{Results for different progenitor masses: models he12p00, he6p00, and he3p30}
\label{sect_var_mass}

In this section, we discuss the results for three He-star models that differ in preSN mass, and consequently in preSN composition, explosion energy, ejecta mass (and thus in expansion rate; Table~\ref{tab_prog}). This translates into different ejecta conditions and  coolants at late times. In the next section, we describe the results for model he12p00, followed by model he6p00 (Section~\ref{sect_he6p00}) and he3p30 (Section~\ref{sect_he3p30}) -- model he8p00 is not discussed in this section because its properties are intermediate between those of he12p00 and he6p00. The same magnetar power of 10$^{39}$\,\ergs\ (with a $dV$ of 2000\,\kms) is employed in all three models at all times and the ejecta are considered unclumped (i.e., the volume filling factor of the material is 100\,\%). 

\begin{figure*}
\centering
\includegraphics[width=\hsize]{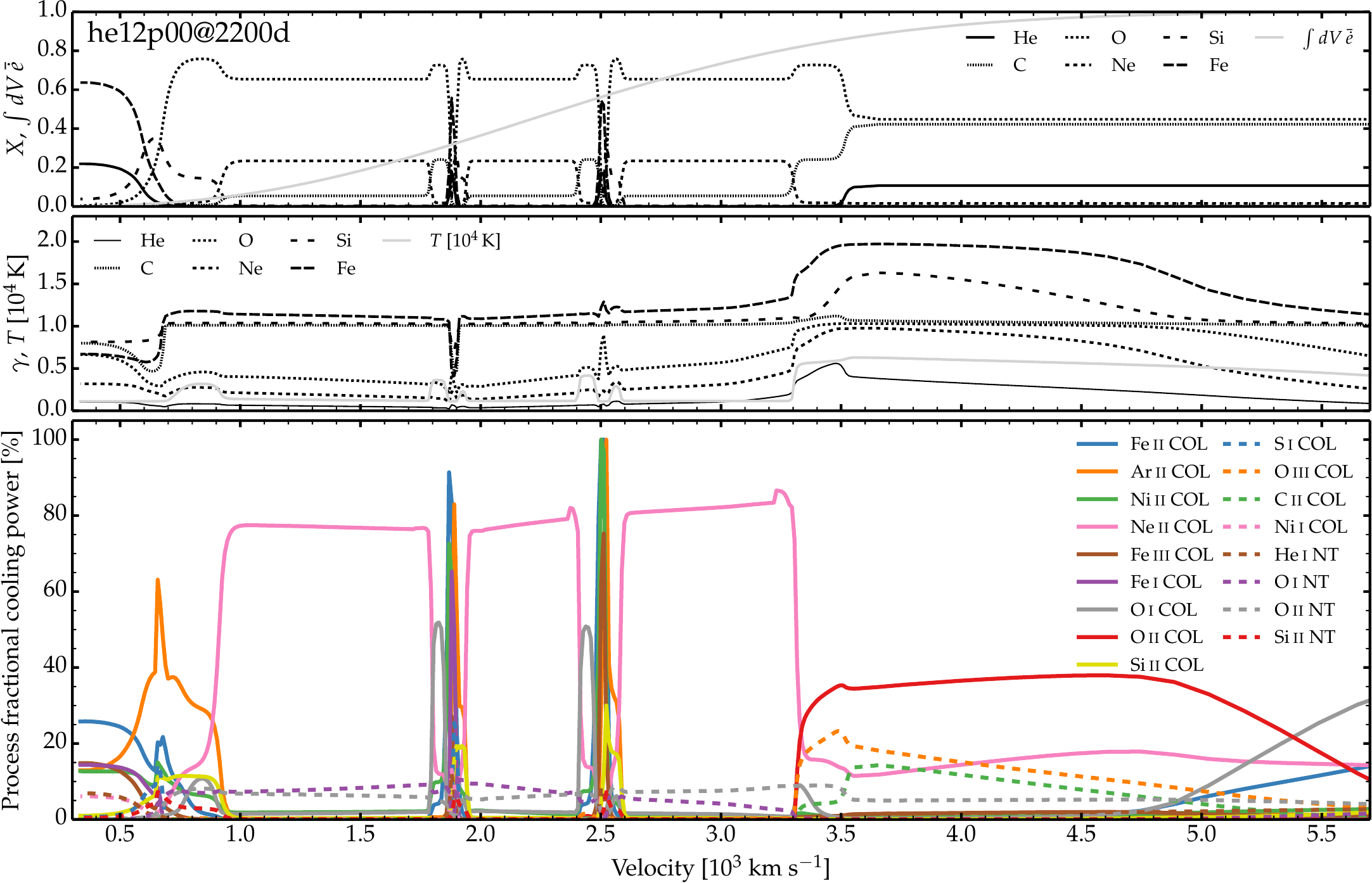}
\caption{Illustration of ejecta and radiative properties for model he12p00 at 2200\,d and influenced by a magnetar power of 10$^{39}$\,\ergs. The top row shows the mass fraction of He, C, O, Ne, Si, and Fe as a function of velocity, which reveals the location and extend of the main shells (e.g., the O/Ne/Mg shell etc). Also shown is the normalized cumulative absorbed power from the magnetar, integrated from the inner ejecta (light grey line). The middle row shows the ionization state for the same species (i.e., zero corresponds to the neutral state, one to once ionized etc). Also shown is the gas or free-electron temperature (light grey curve). The bottom row shows the variation of the local cooling power, as a percentage of the local total heating power, for the 17 top coolants ejecta-wide (these total cooling powers are spelled out in Table~\ref{tab_cooling_heating_he12p00opt_fvol1p0_pwr1e39_2200d_dv2e8}). The term ``COL'' refers to collisional excitation and ``NT'' to nonthermal excitation. [See Section~\ref{sect_he12p00} for discussion.]
\label{fig_cool_he12p00_2200d}
}
\end{figure*}

\subsection{Model he12p00}
\label{sect_he12p00}

The ejecta model he12p00 was evolved from 500 to 4000\,d under the influence of a constant power of 10$^{39}$\,\ergs\ and $dV$ of 2000\,\kms\ at all times. To complement the photometric light curves discussed above, Fig.~\ref{fig_spec_he12p00} shows the spectral evolution from 1000\,\AA\ to 1\,$\mu$m over that time span (not all epochs computed are shown for better visibility). At 500\,d, the spectrum is very similar to what it was over the previous year \citep{dessart_snibc_23} with a dominance of emission lines from neutral or once-ionized species, namely with \mgiiuv, \mgifs, \nad, \oidoub, \caiidoub, and \nkiifs. Over the next 1000\,d, the spectrum changes little, showing only a modest strengthening of emission lines in the ultraviolet (e.g., \mgiiuv) and the burgeoning of some lines like \oiiuv, \oiidoub, and \oiiidoub\ (there is also a transitory appearance of the O\one\,7774\,\AA\ multiplet). Beyond 1500\,d, the model spectrum undergoes a significant change towards higher ionization and exhibits a nearly pure oxygen spectrum in the optical. The strong \oidoub\ at 1500\,d progressively weakens and eventually disappears by 4000\,d whereas \oiiuv\ strengthens considerably to become the strongest oxygen line between 1000\,\AA\ and 1\,$\mu$m -- \oiidoub\ is comparatively very weak (the \oiiuv\ line is much stronger than \oiidoub\ because it is tied to the ground state). The \oiiidoub\ also strengthens. In Fig.~\ref{fig_spec_he12p00}, the individual species contributions (indicated by the color shading) are computed by performing a formal solution of the radiative transfer equation and accounting for bound-bound transitions of all ionization states of a given species treated in the calculation (e.g., O\one, O\two, and O\three\ for oxygen). At nearly all wavelengths, the total flux (shown as a black line) is greater than the sum of species' contributions, which suggests optically-thin line emission. One exception is \ciiuv\ at 1000--2200\,d, which clearly suffers absorption by metal-line blanketing (most likely by Fe\two\ and Fe\three\ lines), and it is only at times past 2200\,d that it becomes detectable. This indicates that metal-line blanketing continues to impact photon escape in the ultraviolet at late times. Nearly all lines shown here correspond to forbidden transitions with only few exceptions (e.g., \mgiiuv\ and \nad). The overall evolution indicates a rise in ionization at around 1500\,d. Although that specific timing arises from a complex combination of processes, it naturally follows from our injection of a constant magnetar power in an ejecta of progressively decreasing density (i.e., the density declines everywhere as $1/t^3$, where $t$ is the time since explosion).

To gain some insight into the underlying processes, Fig.~\ref{fig_cool_he12p00_2200d} shows some gas and radiation properties for the he12p00 model at 2200\,d after explosion. Versus velocity, we show the composition and ionization for important species in the ejecta (top two rows) as well as the cooling rate (shown as a fraction of the total, local heating rate) for the top 17 coolants ejecta wide. The volume-integrated power from these processes, which are either associated with collisional excitation or nonthermal excitation, is given in Table~\ref{tab_cooling_heating_he12p00opt_fvol1p0_pwr1e39_2200d_dv2e8}. Figure~\ref{fig_cool_he12p00_2200d} also shows the profile for the normalized cumulative magnetar power absorbed in the ejecta as well as the temperature (light-grey lines in the top two rows). Shown in this manner, it becomes apparent that the strong chemical stratification of the ejecta translates into strong modulations in ionization, temperature, and coolants versus velocity whereas the actual power is smoothly distributed throughout the ejecta, with about 80\,\% of this power falling between 1000 and 3500\,\kms.
  
\begin{table}
\caption{Summary of the strongest, volume-integrated cooling powers compared to the injected magnetar power for model he12p00 at 2200\,d. The list is ordered from the stongest to the weakest power out of the 17 top cooling powers ejecta-wide.
\label{tab_cooling_heating_he12p00opt_fvol1p0_pwr1e39_2200d_dv2e8}
}
\vspace{-0.5cm}
\begin{center}
  \begin{tabular}{|lc|c|c|}
\hline
       \multicolumn{2}{|c|}{Process}       &   Power [\ergs]&     [\%]  \\
\hline
            \multicolumn{2}{|c|}{Magnetar} &     1.00(39) &      96.91  \\
            \multicolumn{2}{|c|}{Art. HT} &     3.19(37) &       3.09  \\
\hline
    Ne\,{\sc ii}  & COL &     5.71(38) &      55.30 \\
      O\,{\sc ii} & NT &     6.42(37) &       6.22 \\
     O\,{\sc ii}  & COL &     5.60(37) &       5.42 \\
       O\,{\sc i} & NT &     5.48(37) &       5.31 \\
    Fe\,{\sc ii}  & COL &     4.98(37) &       4.82 \\
    Ar\,{\sc ii}  & COL &     4.85(37) &       4.70 \\
    Ni\,{\sc ii}  & COL &     4.53(37) &       4.39 \\
      O\,{\sc i}  & COL &     4.12(37) &       3.99 \\
    O\,{\sc iii}  & COL &     3.01(37) &       2.91 \\
    Si\,{\sc ii}  & COL &     1.76(37) &       1.71 \\
     C\,{\sc ii}  & COL &     1.67(37) &       1.62 \\
   Fe\,{\sc iii}  & COL &     1.61(37) &       1.56 \\
     Fe\,{\sc i}  & COL &     8.73(36) &       0.85 \\
      He\,{\sc i} & NT &     5.65(36) &       0.55 \\
     Si\,{\sc ii} & NT &     4.27(36) &       0.41 \\
      S\,{\sc i}  & COL &     3.52(36) &       0.34 \\
     Ni\,{\sc i}  & COL &     2.33(36) &       0.23 \\
\hline
\end{tabular}
\end{center}
{\bf Notes:} The first two columns lists all cooling processes (``COL'' for collisional excitation and ``NT'' for nonthermal excitation) balancing the
    heating from the magnetar power injected. The term ``Art. HT'' corresponds to an artificial heating component introduced to prevent the
    temperature from dropping below a floor temperature of 1000\,K (this occurs at sporadic locations). The third column gives the corresponding power in \ergs. The last column gives that quantity as the percentage fraction of the total cooling power. Numbers in parenthesis correspond to powers of ten.
\end{table}

The ionization tends to rise with velocity, with a marked jump beyond the inner edge of the He/C/O shell at about 3500\,\kms, except for He, which is essentially neutral throughout (He is only abundant beyond 3500\,\kms). Below that threshold velocity, O and Ne are nearly neutral, and Si and Fe are once ionized. Above that threshold velocity, species are partially ionized or once ionized, with Fe nearly twice ionized. This trend in ionization is opposite that of the electron-density profile, which peaks in the inner ejecta regions where the mass density is greatest -- ionization is instead greater in the outer, low density ejecta. The temperature shows a similar pattern, being marginally above 1000\,K in the inner ejecta and rising to about 5000\,K beyond 3500\,\kms\ in the He/C/O shell. In our models, the temperature can drop to low values in some regions, which causes numerical difficulties. Thus, to prevent the temperature from dropping below 1000\,K, we introduce an artificial heating term (denoted ``Art. HT'' in Table~\ref{tab_cooling_heating_he12p00opt_fvol1p0_pwr1e39_2200d_dv2e8}). This artificial heating contributes a negligible 3.1\,\% of the total power injected.

The magnetar power absorbed in the ejecta exhibits a smooth curve. The deposition profile and the magnitude of the magnetar power are prescribed rather than informed from observations -- they are parameters of the simulation. So, the profile would likely deviate from that pertaining in a magnetar-powered SN in nature but the general offset with the decay power is a key feature of such models at late times (i.e., several years after explosion) because only 3.5\,\% of $\gamma$ rays from \cofs\ decay are absorbed (87\,\% of the decay power comes from positron absorption). The \iso{44}Ti decay chain, which is ignored here, would dominate over that of \nifs\ at such late times \citep{jerkstrand_87a_11,dessart_late_23} but the associated power released would be orders of magnitude below the adopted magnetar power here. With our adopted magnetar-power deposition profile in model he12p00 at 2200\,d, we find that 14.7\,\% of the total power is absorbed in the He/C/O shell, 10.3\,\% in the O/C shell, 64.5\,\% in the O/Ne/Mg shell, 5.9\,\% in the O/Si shell, 2.8\,\% in the Si/S shell and 1.5\,\% in the Fe/He shell. So, it is the massive O/Ne/Mg shell that absorbs the bulk of the available power. This is as expected given the relatively low mass of the O/C, O/Si, Si/S, and Fe/He shells in model he12p00 (and in general in massive preSN stars, with the exception here of model he3p30). A deposition profile less confined to the inner ejecta regions would give greater power to the outer He/C/O shell. Instead, a more confined deposition profile would not alter much the identity of shells that absorb the power because of the shuffled-shell structure (the main effect would then be to deposit power in slower and denser regions).

The main coolants of each of these shells in the he12p00 model at 2200\,d are shown in the bottom-row panel of  Fig.~\ref{fig_cool_he12p00_2200d} and listed in Table~\ref{tab_cooling_heating_he12p00opt_fvol1p0_pwr1e39_2200d_dv2e8}. Collisional excitation is the main cooling process in all shells, with a small contribution from nonthermal excitation. The species/ions involved in the cooling differs between shells. The He/C/O shell cools primarily by collisional excitation of O\two\ (i.e., \oiiuv), O\three\ (i.e., \oiiidoub), and Ne\two\ (i.e., \neiifs). The O/C shell cools through O\one\ (i.e., \oidoub) and Ne\two\ (i.e., \neiifs). However, we find that \neiifs\ is the main coolant of the O/Ne/Mg shell, and because this shell is the most massive in the he12p00 ejecta model, it is responsible for 55.3\,\% of the total ejecta cooling rate. \neiifs\ is the dominant coolant of the material originally in the O/Ne/Mg shell because O is partially ionized there and because this Ne\two\ transition is stronger than either \oidoub\ or \oiiuv. A similar importance of the cooling by \neiifs\ was also found by \citet{omand_pm_23}. In the O/Si and Si/S shells, Ar\two, Si\two, and Fe\two\ are the primarily coolants, whereas Ni\two\ and Fe\three\ also contribute in the Fe/He shell. This distribution of coolants suggests that the resulting spectrum should be dominated by one strong line in the infrared (i.e., \neiifs) and a collection of weak and near equal strength lines from secondary coolants spread across the ultraviolet, optical, and infrared (see Table~\ref{tab_coolants_processes_lines}).

\begin{figure}
\centering
\includegraphics[width=\hsize]{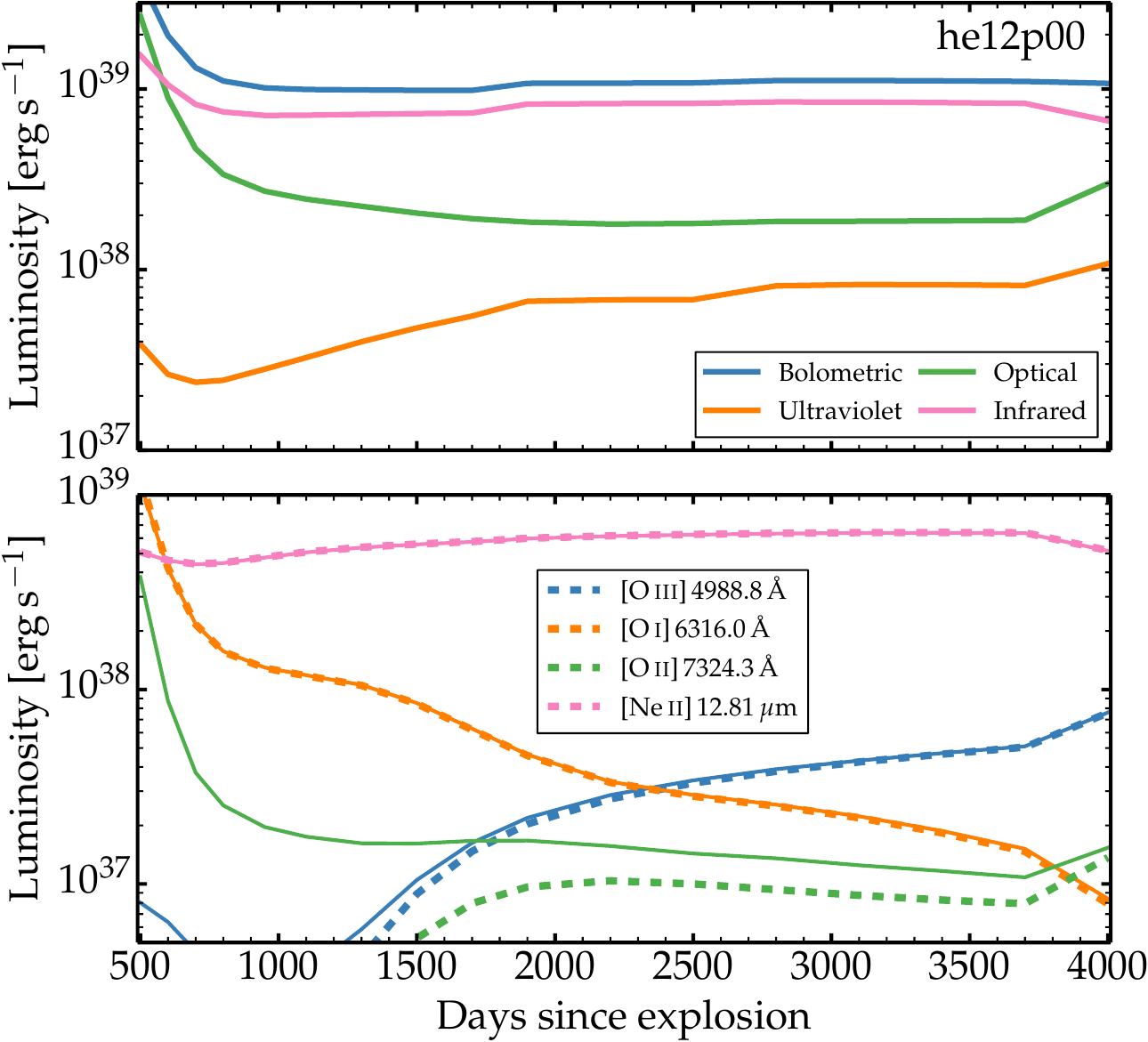}
\caption{Evolution of various luminosities in model he12p00  under the influence of a constant magnetar power of 10$^{39}$\,\ergs. The top panel shows the bolometric, ultraviolet, optical, and infrared luminosities from 500 to 4000\,d. The bottom panel shows the luminosity in \oidoub, \oiidoub, \oiiidoub, and \neiifs\ (dashed line) together with the total flux (i.e., including all contributing lines) for the associated emission feature at 6300,  7300, 5000\,\AA\, and 12.81\,$\mu$m (solid line). In most cases, overlapping lines are subdominant (i.e., the dashed and solid lines overlap).
\label{fig_fluxes_he12p00}
}
\end{figure}

Figure~\ref{fig_fluxes_he12p00} shows the evolution of the bolometric, ultraviolet, optical, and infrared luminosities from 500 to 4000\,d for the he12p00 model (top panel), together with various line luminosities (bottom panel). The optical luminosity dominates until 600\,d, beyond which the infrared luminosity dominates and remains so until the end of the sequence. The ultraviolet never represents more than 10\,\% of the total flux, even at the latest times. At all times after 700\,d, \neiifs\ dominates the cooling in the infrared. In the optical, \oidoub\ dominates up to about 2200\,d whereas \oiiidoub\ dominates at later times. \oiidoub\ tends to be a weak coolant at most times (even under optimal ionization conditions) because these O\two\ transitions have a relatively high excitation energy \citep{omand_pm_23}. \oiiuv, which sits at the blue edge of the optical, is stronger than \oiiidoub\ and is the main O\two\ coolant. 

\begin{figure*}
\centering
\includegraphics[width=\hsize]{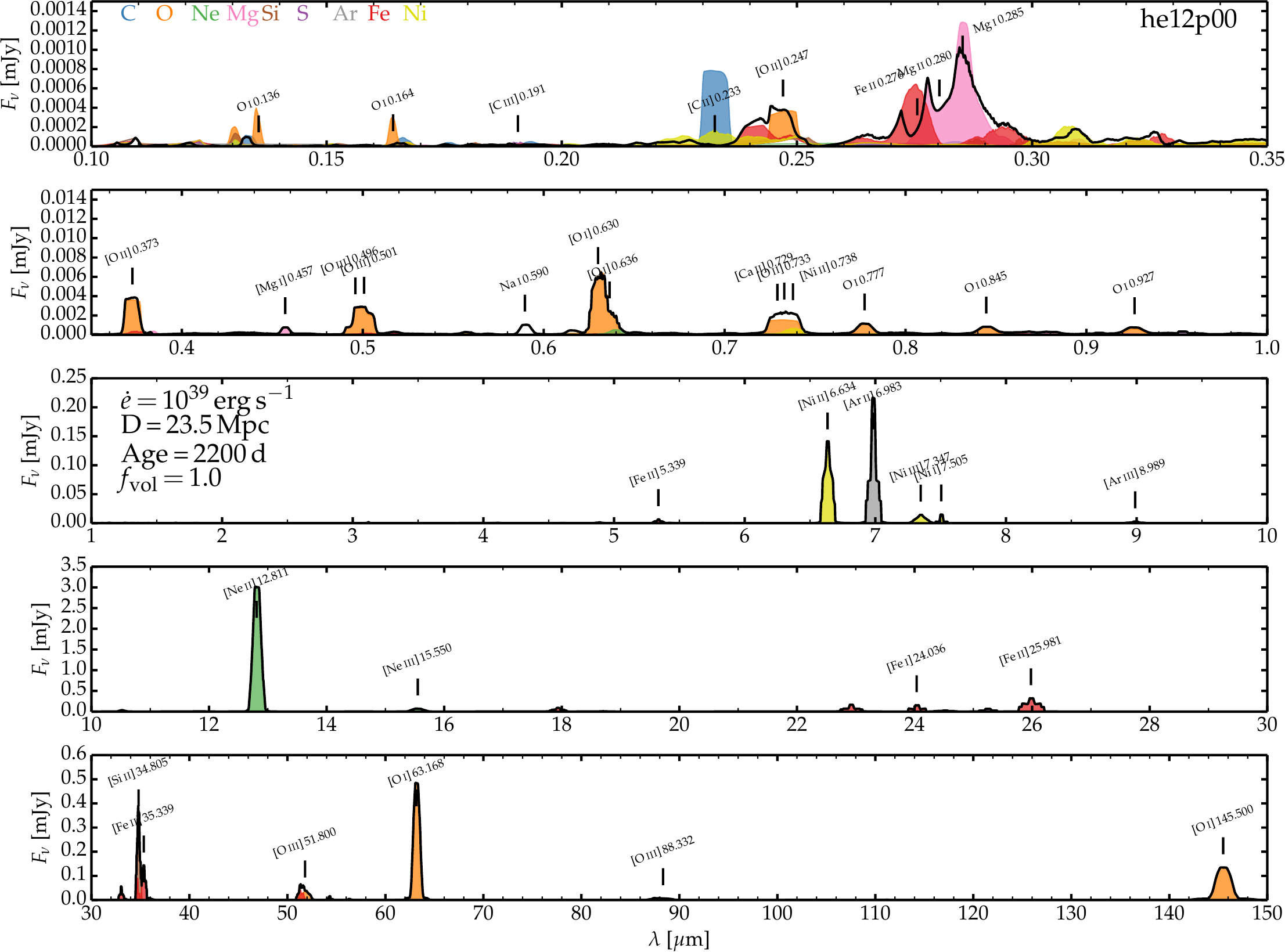}
\caption{Ultraviolet to infrared spectrum of model he12p00 at 2200\,d and influenced by a magnetar power of 10$^{39}$\,\ergs\ (see Section~\ref{sect_he12p00}). We show the flux  $F_\nu$ in units of mJy and for the inferred distance of 23.5\,Mpc to SN\,2012au (this flux $F_\nu$ contrasts with the flux $F_\lambda$  shown in Fig.~\ref{fig_spec_he12p00}). Line identifications are provided, together with a species-dependent shading.
\label{fig_spec_uv_to_fir_he12p00_2200d}
}
\end{figure*}

Figure~\ref{fig_spec_uv_to_fir_he12p00_2200d} illustrates the full ultraviolet to far-infrared properties of the he12p00 model at 2200\,d and influenced by a magnetar power of 10$^{39}$\,\ergs. Here, the flux is scaled to the distance of SN\,2012au to put it into the observational context of a magnetar-powered SN candidate \citep{milisavljevic_12au_18}. This illustrates how oxygen lines dominate the optical whereas carbon and magnesium lines are present, albeit weak in the ultraviolet. Similarly, the mid-infrared shows emission lines due to \nkiimir, \ariimir, [Ni\three]\,7.347\mic\ (weak), \neiifs\ (and a weak \neiiifs), [Fe\,\two]\,25.981\,$\mu$m, [Si\,\two]\,34.805\,$\mu$m, [Fe\,\two]\,35.339\,$\mu$m, [O\,\three]\,51.8\,$\mu$m,
[O\,\one]\,63.168\,$\mu$m, [O\,\three]\,88.332\,$\mu$m, and [O\,\one]\,145.5\,$\mu$m. The lines from iron-group elements are the dominant coolant of the Si/S and Fe/He shells, but given the small power absorbed in those low-mass shells, these lines contain a small fraction of the total model flux. The lines predicted here at $\sim$\,6\,yr in the far-infrared are observed in some SN remnants like the 480\,yr old Kes 75 \citep{temim_kes75_19}, despite the large age difference.

\begin{figure*}
\centering
\includegraphics[width=\hsize]{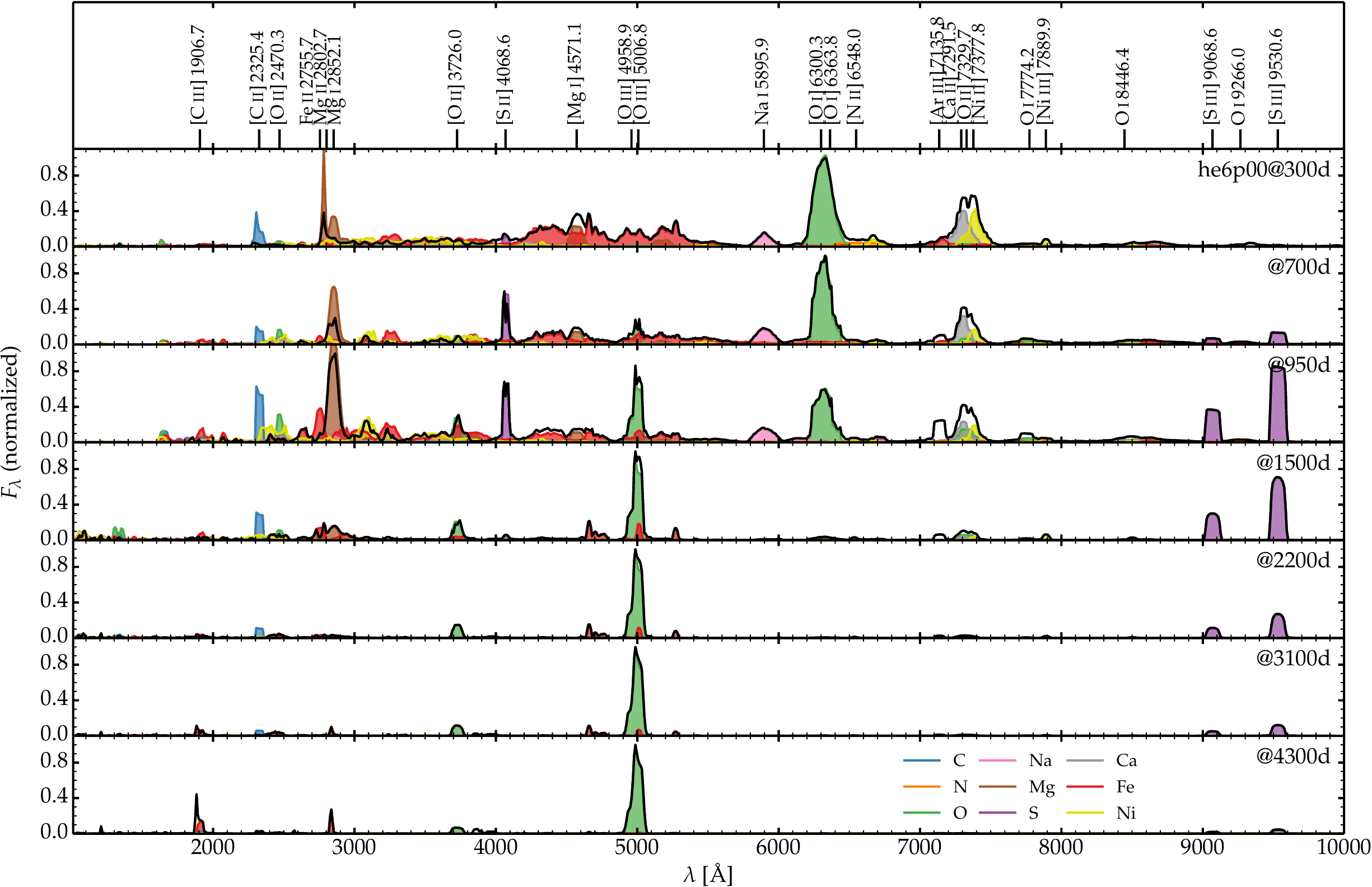}
\caption{Same as Fig.~\ref{fig_spec_he12p00}, but now for model he6p00 and for times covering from 300 to 4300\,d. A constant magnetar power of 10$^{39}$\,\ergs\ is adopted. 
\label{fig_spec_he6p00}
}
\end{figure*}

\subsection{Model he6p00 model}
\label{sect_he6p00}

We now turn to model he6p00 influenced by a magnetar power of 10$^{39}$\,\ergs\ and the same value $dV$ of 2000\,\kms\ characterizing the power deposition profile at all times. Because this model is less massive and has a greater kinetic energy, the density is typically lower and the velocity span greater than for model he12p00 (see properties in Table~\ref{tab_prog} and Fig.~\ref{fig_init}). This also translates into a more confined deposition of the power within the ejecta since the deposition profile is the same. Below, we describe this he6p00 model but not in as much detail as for model he12p00 since the qualitative aspects and concepts are the same. 

\begin{figure}
\centering
\includegraphics[width=\hsize]{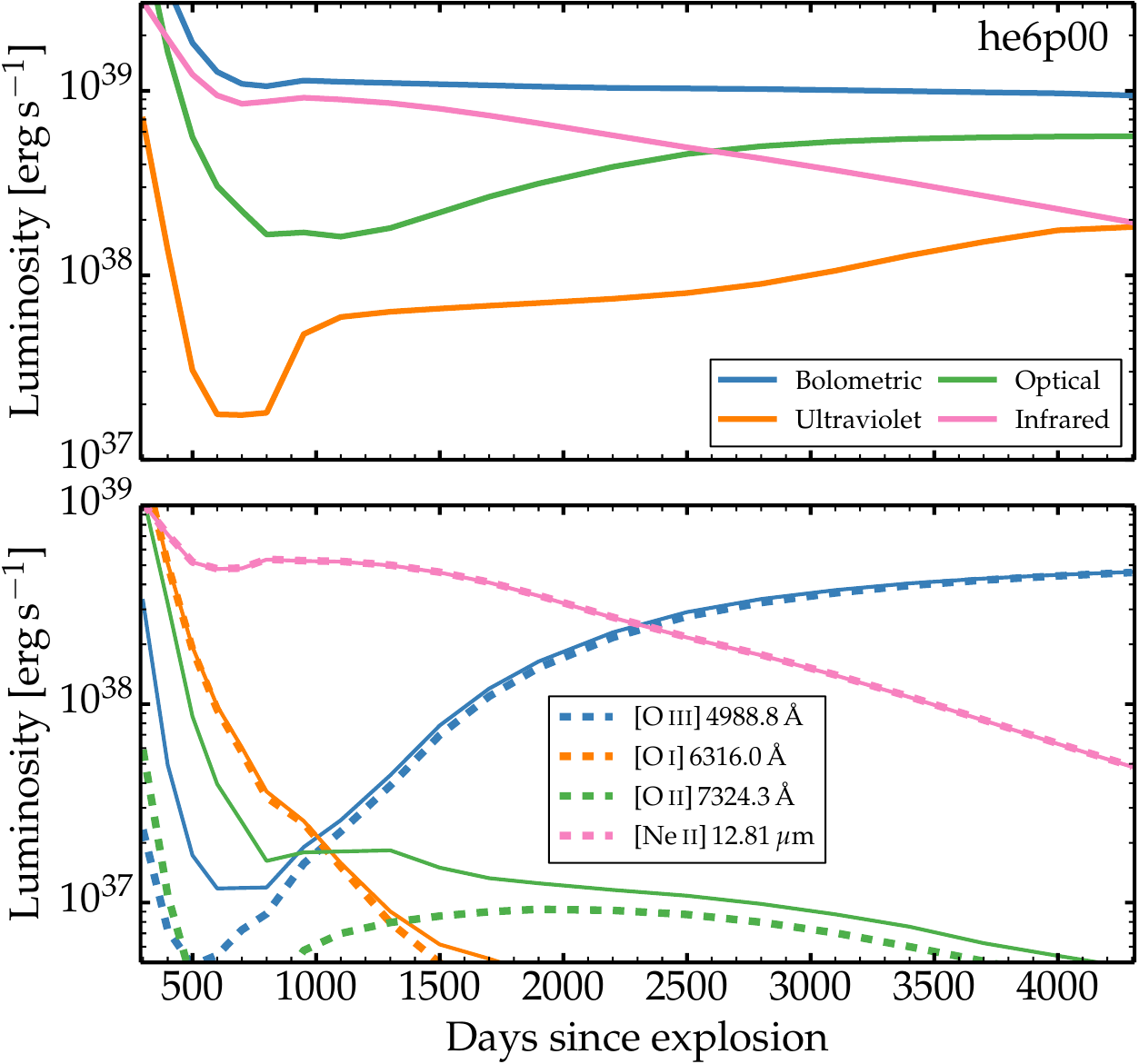}
\caption{Same as Fig.~\ref{fig_fluxes_he12p00} but now for the he6p00 model.
\label{fig_fluxes_he6p00}
}
\end{figure}

Figure~\ref{fig_spec_he6p00} shows the spectral evolution from 1000\,\AA\ to 1\mic\ for model he6p00 over a time spanning 300 to 4300\,d after explosion. At 300\,d, the magnetar power is only 10\,\% of the total luminosity so the decay power dominates. The spectrum appears essentially unchanged from that without magnetar power and as obtained in \citet{dessart_snibc_23}. The spectrum exhibits strong \oidoub, a weak \niidoub, \caiidoub, \nkiifs, and a forest of Fe\two\ lines between 4000\,\AA\ and 5500\,\AA. At 300\,d, 65\,\% of the total flux falls in the optical, 6.5\,\% in the ultraviolet and 28.6\,\% in the infrared. As time proceeds, the spectrum shows a weakening of \oidoub, and the appearance and strengthening of \oiiidoub, \siiinz, \siiinf\ -- this reflects a shift to a higher ionization. By 1500\,d after explosion, the optical spectrum is essentially composed of \oiiidoub, \siiinz, \siiinf\ and little else. With further evolution, the sulfur lines weaken and the entire optical flux is channeled into \oiiidoub, and that line alone radiates 60\,\% of the total flux -- the remaining 40\,\% go in equal shares to the ultraviolet and the infrared (Fig.~\ref{fig_fluxes_he6p00}).

\begin{figure*}
\centering
\includegraphics[width=\hsize]{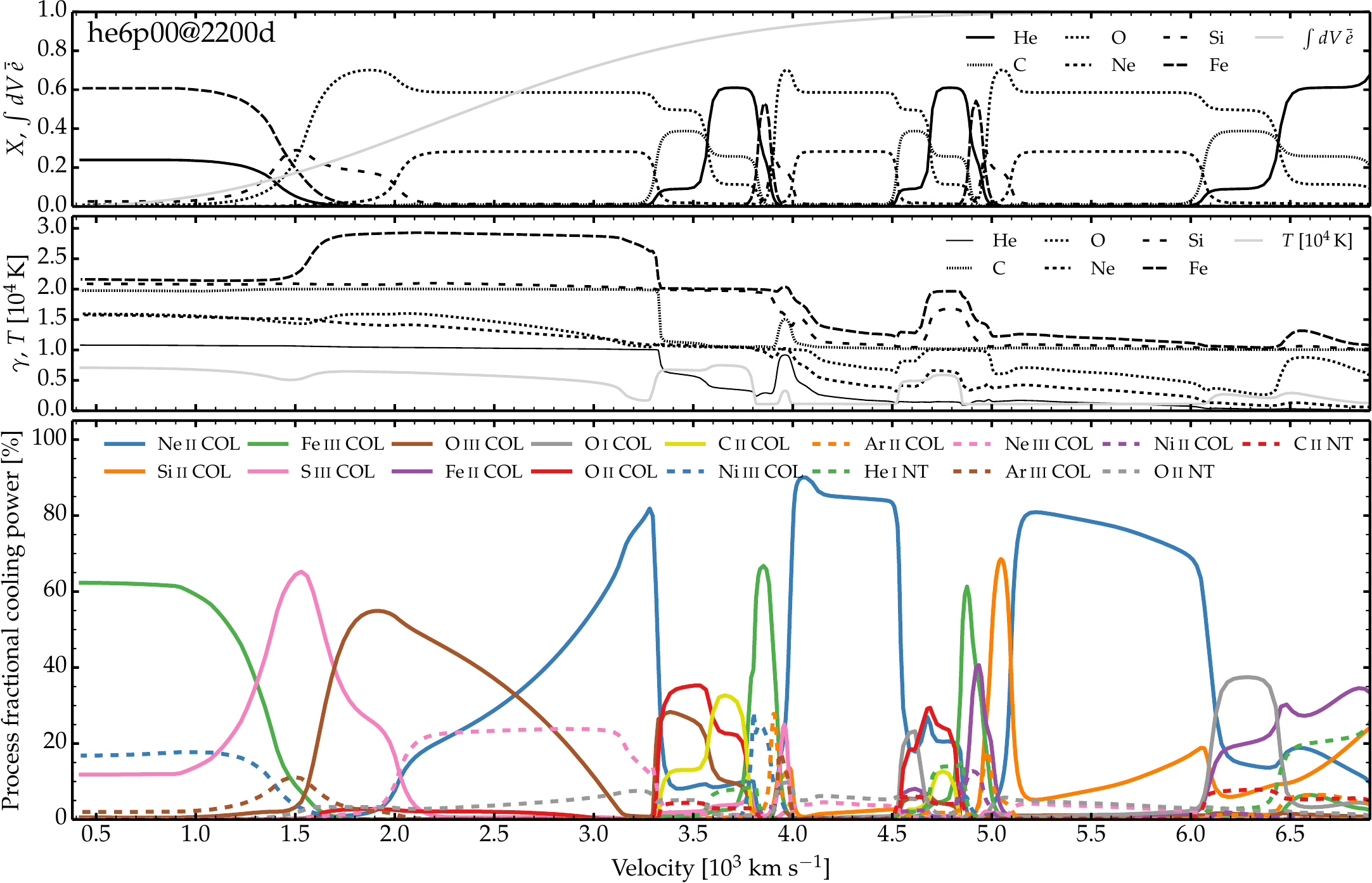}
\caption{Same as Fig.~\ref{fig_cool_he12p00_2200d} but now for model he6p00 at 2200\,d, influenced by a power of 10$^{39}$\,\ergs\ ($dV$ is 2000\,\kms). Volume-integrated cooling powers are given for this model in Table~\ref{tab_cooling_heating_he6p00opt_fvol1p0_pwr1e39_2200d_dv2e8}. [See Section~\ref{sect_he6p00} for discussion.]
\label{fig_cool_he6p00_2200d}
}
\end{figure*}

Figure~\ref{fig_cool_he6p00_2200d} illustrates the variation of the ionization of various species and the dominant cooling rates (given as a percentage of the total heating rate) versus velocity for the model he6p00 at 2200\,d (the mass fractions of key species in the dominant shells are given in the top-row panel). Because the O/Ne/Mg shell is not as massive as in model he12p00, the various shells are better identified and one can see how much the various coolants differ between shells, in part controlled by the dominant ionization state of the various species. We find that all species are once or twice ionized (even triply ionized for Fe in the interval 1500--3200\,\kms) below about 4000\,\kms\ and the ionization continuously drops beyond that velocity. In this model, nearly 100\,\% of the magnetar power is absorbed below 5000\,\kms, and about 50\,\% below 2500\,\kms\ (light-grey curve in top-row panel), and the distribution of the absorbed power between shells at 2200\,d is as follows: 6.1\,\% is absorbed in the He/C/O shell, 5.6\,\% in the O/C shell, 50.2\,\% in the O/Ne/Mg shell, 16.1\,\% in the O/Si shell, 9.5\% in the Si/S shell, and 12.6\,\% in the Fe/He shell. Only 3.3\,\% of the decay power is absorbed in the ejecta, of which 93.7\,\% comes from the local deposition of positrons.

Table~\ref{tab_cooling_heating_he6p00opt_fvol1p0_pwr1e39_2200d_dv2e8} lists the top ejecta-wide coolants for this he6p00 model at 2200\,d. \neiifs\ radiates 24.25\,\% of the total power, \oiiidoub\ 21.45\,\%, \siiinz\ and \siiinf\ a total of 11.87\,\%, [Fe\three]\,4658.0 and 5011.2\,\AA, and \neiiifs\ carry 10.12\,\%. The bottom row panel of Fig.~\ref{fig_cool_he6p00_2200d} shows how these coolants vary in relative strength at different location in the ejecta. O\two, O\three, and C\two\ dominate the cooling of the O/C shell. C\two\, O\two, and O\three\  dominate the cooling of the He/C/O shell (i.e., same ions but different order of importance relative to the O/C shell).  Ne\two\ and to a lesser extent Ne\three\ dominate the cooling of the O/Ne/Mg shell. In the O/Si shell, O\three\ and S\three\ dominate the cooling. S\three\ is the main coolant for the Si/S shell. Finally, in the Fe/He shell, Fe\three\ dominate the cooling, with Ni\three\ and S\three\ taking a secondary role. A myriad of other, but secondary ions contribute to the cooling, each at the 1\,\% level, and give rise to weak lines spread over the ultraviolet, optical, and infrared ranges. Although some coolants are strong in the outer ejecta (e.g., collisional excitation of O\one, which corresponds to \oidoub), they give rise to a negligible flux in the emergent spectrum because they arise from regions where essentially no power is absorbed (see the deposition profile shown in the top panel of Fig.~\ref{tab_cooling_heating_he6p00opt_fvol1p0_pwr1e39_2200d_dv2e8}).

\begin{table}
  \caption{Same as Table~\ref{tab_cooling_heating_he12p00opt_fvol1p0_pwr1e39_2200d_dv2e8} but now for model he6p00 at 2200\,d.
\label{tab_cooling_heating_he6p00opt_fvol1p0_pwr1e39_2200d_dv2e8}
}
\vspace{-0.5cm}
\begin{center}
    \begin{tabular}{|lc|c|c|}
\hline
 \multicolumn{2}{|c|}{Process}        &   Power [\ergs]&     [\%]     \\
\hline
 \multicolumn{2}{|c|}{Magnetar} &     1.00(39) &      93.23 \\
 \multicolumn{2}{|c|}{Art. HT} &     7.26(37) &       6.77 \\
\hline
    Ne\,{\sc ii}  & COL &     2.60(38) &      24.25 \\
    O\,{\sc iii}  & COL &     2.30(38) &      21.45 \\
    S\,{\sc iii}  & COL &     1.27(38) &      11.87 \\
   Fe\,{\sc iii}  & COL &     1.16(38) &      10.78 \\
   Ne\,{\sc iii}  & COL &     1.09(38) &      10.12 \\
     O\,{\sc ii}  & COL &     3.86(37) &       3.60 \\
      O\,{\sc ii} & NT &     3.58(37) &       3.34 \\
   Ni\,{\sc iii}  & COL &     3.55(37) &       3.31 \\
   Ar\,{\sc iii}  & COL &     1.86(37) &       1.74 \\
     C\,{\sc ii}  & COL &     1.77(37) &       1.65 \\
    Fe\,{\sc ii}  & COL &     9.33(36) &       0.87 \\
    Si\,{\sc ii}  & COL &     8.62(36) &       0.80 \\
    Ar\,{\sc ii}  & COL &     6.41(36) &       0.60 \\
      He\,{\sc i} & NT &     5.82(36) &       0.54 \\
      C\,{\sc ii} & NT &     4.48(36) &       0.42 \\
    Ni\,{\sc ii}  & COL &     3.97(36) &       0.37 \\
      O\,{\sc i}  & COL &     1.93(36) &       0.18 \\
\hline
\end{tabular}
\end{center}
\end{table}

\begin{figure*}
\centering
\includegraphics[width=\hsize]{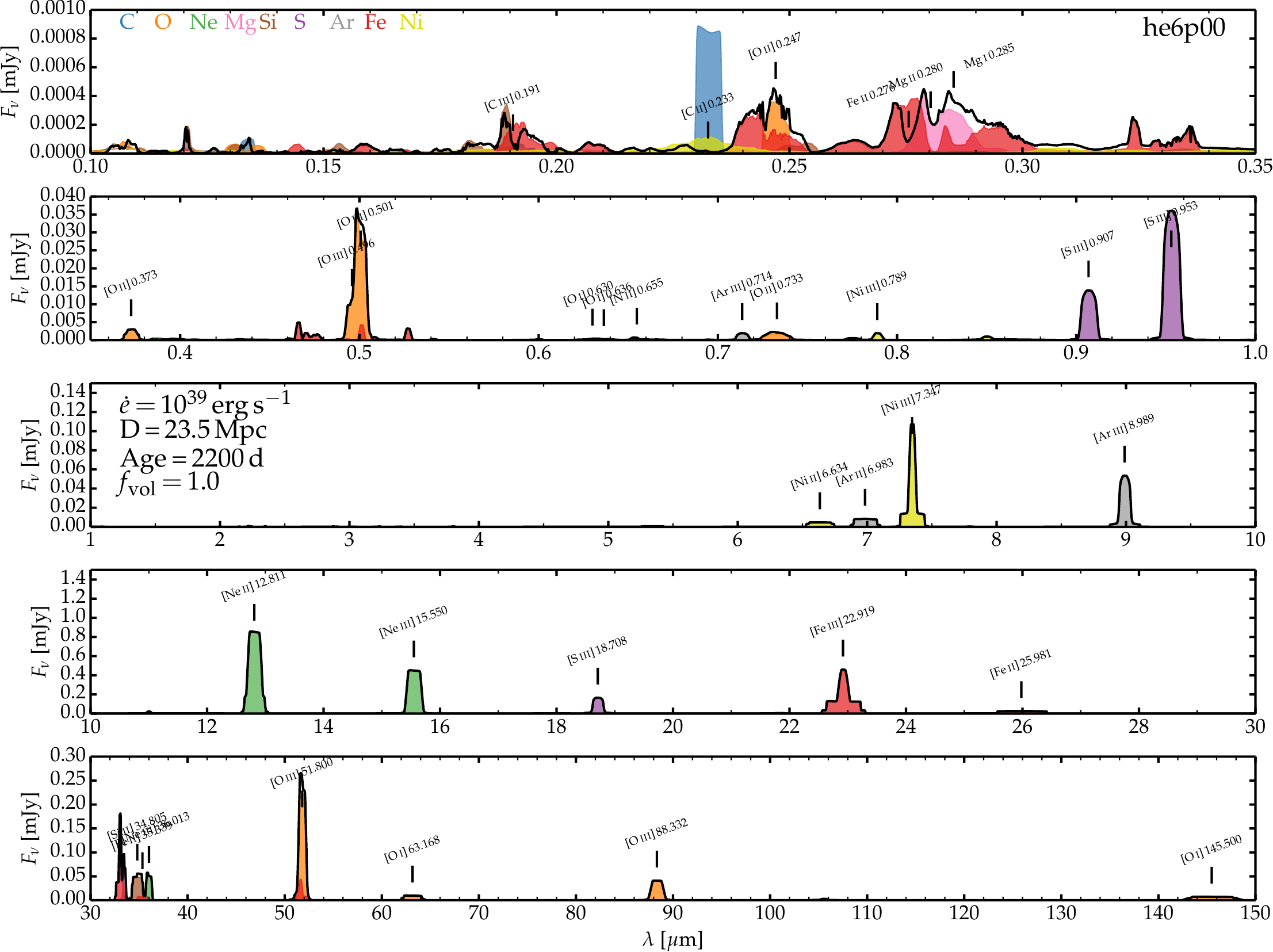}
\caption{Same as Fig.~\ref{fig_spec_uv_to_fir_he12p00_2200d} but now for model he6p00 at 2200\,d.
\label{fig_spec_uv_to_fir_he6p00_2200d}
}
\end{figure*}

Figure~\ref{fig_spec_uv_to_fir_he6p00_2200d} illustrates the full ultraviolet to far-infrared properties of the he6p00 model at 2200\,d and influenced by a magnetar power of 10$^{39}$\,\ergs. As for Fig.~\ref{fig_spec_uv_to_fir_he12p00_2200d}, the flux (i.e., $F_\nu$) is scaled to the distance of SN\,2012au. 
The main differences with model he12p00 is the overall higher ionization of the ejecta leading to stronger lines of O\three, Ne\three, S\three,  Ni\three, and correspondingly a weakening of the lines from neutral and once-ionized ions.

\subsection{Model he3p30}
\label{sect_he3p30}

\begin{figure*}
\centering
\includegraphics[width=\hsize]{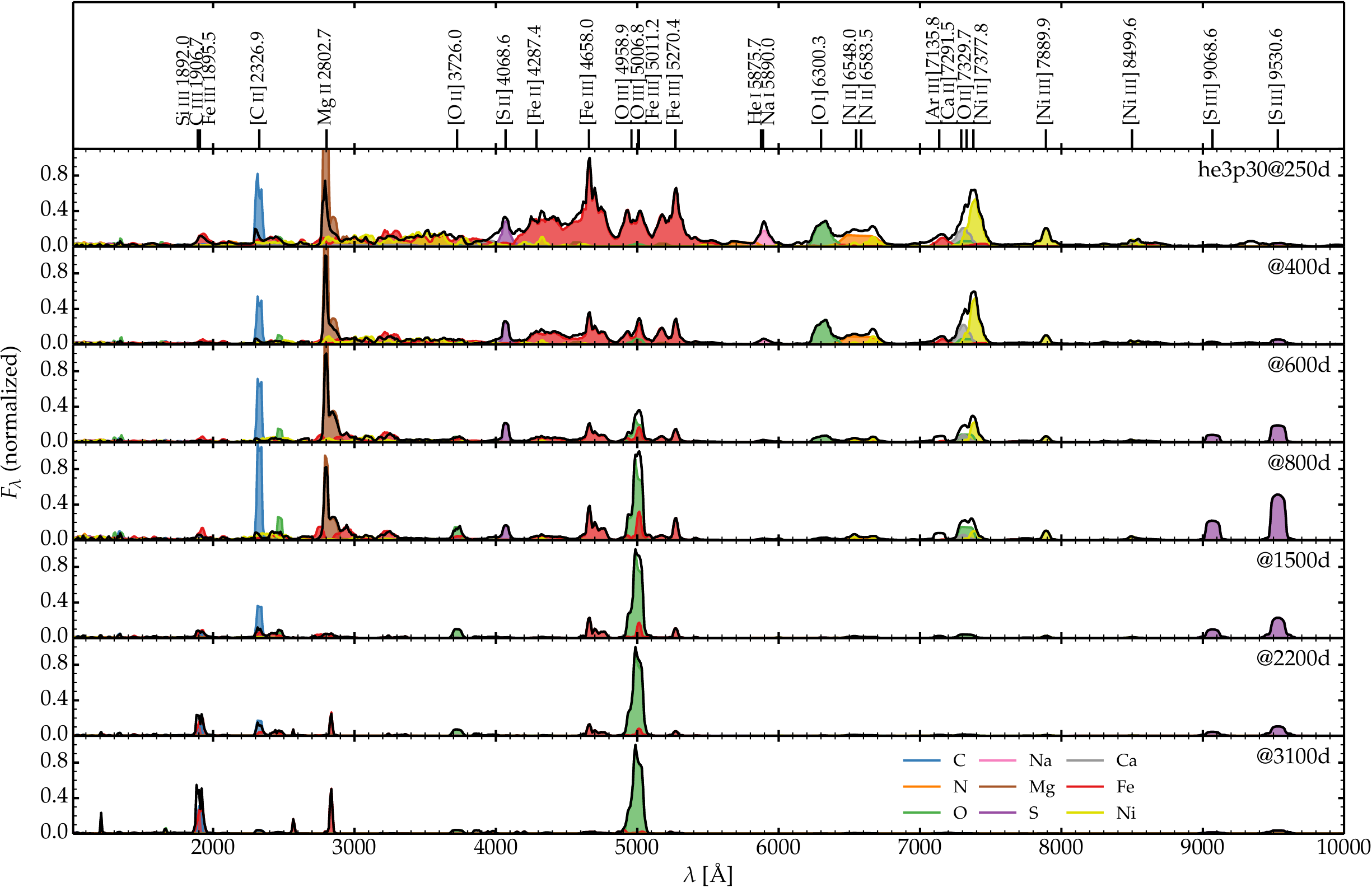}
\caption{Spectral evolution between 1000\,\AA\ and 1\,$\mu$m and from 250 to 3100\,d for model he3p30 under the influence of a constant power of 10$^{39}$\,\ergs. All spectra are normalized to a maximum flux of unity at each epoch. Contributions by species (e.g., those associated with oxygen include contributions from O\one, O\two, and O\three) are indicated with a colored shading.
\label{fig_spec_he3p30}
}
\end{figure*}

\begin{figure}
\centering
\includegraphics[width=\hsize]{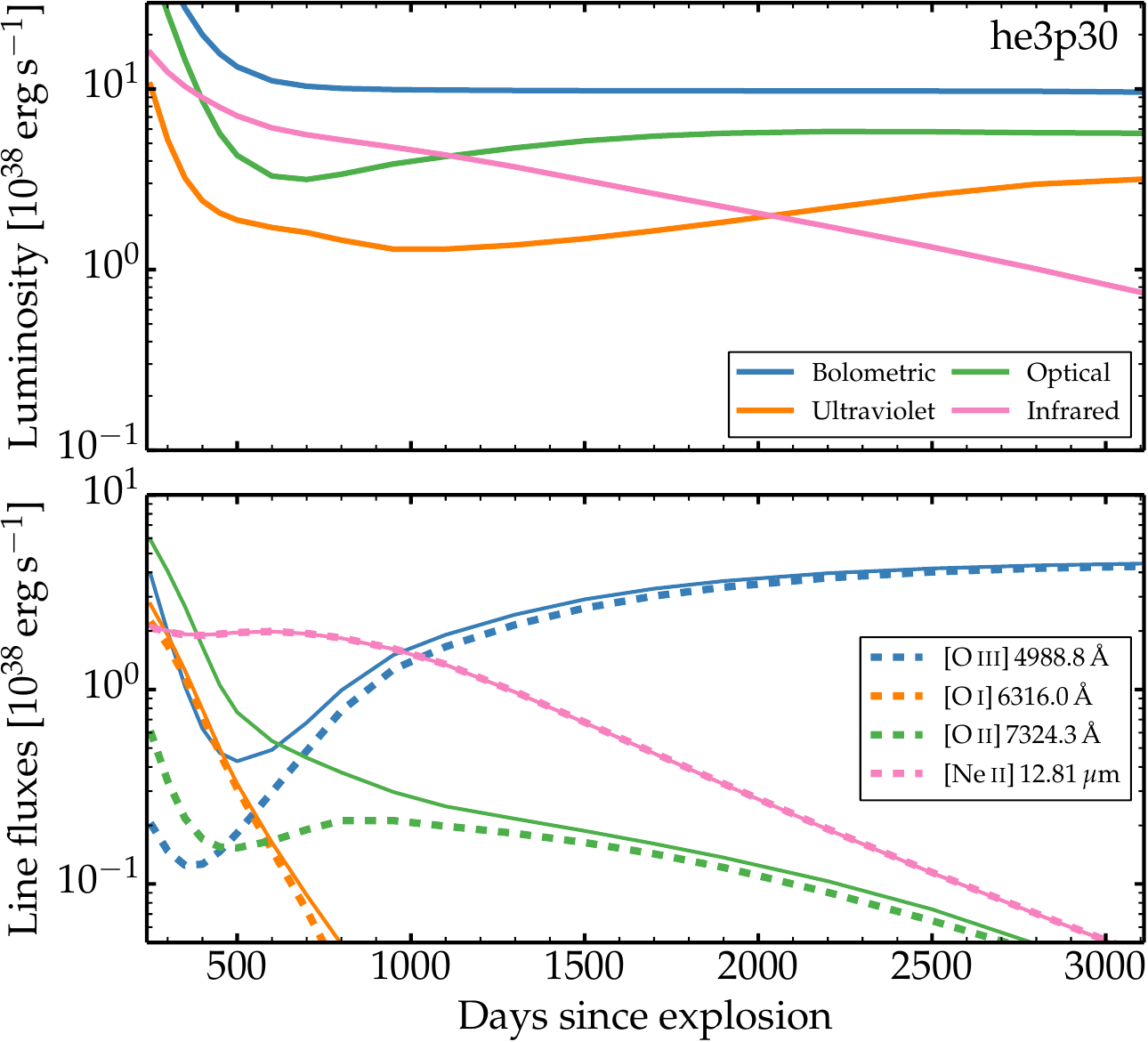}
\caption{Same as Fig.~\ref{fig_fluxes_he12p00} but now for the he3p30 model.
\label{fig_fluxes_he3p30}
}
\end{figure}

\begin{table}
\caption{Same as Table~\ref{tab_cooling_heating_he12p00opt_fvol1p0_pwr1e39_2200d_dv2e8} but now for model he3p30 at 2200\,d.
\label{tab_cooling_heating_he3p30opt_fvol1p0_pwr1e39_2200d_dv2e8}
}
\vspace{-0.5cm}
\begin{center}
    \begin{tabular}{|lc|c|c|}
\hline
       \multicolumn{2}{|c|}{Process}       &   Power [\ergs]&     [\%]  \\
\hline
            \multicolumn{2}{|c|}{Magnetar} &     1.00(39) &    100.00 \\
\hline
    O\,{\sc iii}   &  COL &   3.87(38) &      38.71 \\
   Fe\,{\sc iii}   &  COL &   1.42(38) &      14.21 \\
    S\,{\sc iii}   &  COL &   1.13(38) &      11.28 \\
   Si\,{\sc iii}   &  COL &   5.15(37) &       5.15 \\
   Ne\,{\sc iii}   &  COL &   4.33(37) &       4.33 \\
     C\,{\sc ii}   &  COL &   4.29(37) &       4.29 \\
     O\,{\sc ii}   &  COL &   3.87(37) &       3.87 \\
    Fe\,{\sc iv}   &  COL &   3.71(37) &       3.71 \\
   Ni\,{\sc iii}   &  COL &   2.43(37) &       2.43 \\
    C\,{\sc iii}   &  COL &   2.03(37) &       2.03 \\
     O\,{\sc iii}  &   NT &   1.94(37) &       1.94 \\
    Ne\,{\sc ii}   &  COL &   1.89(37) &       1.89 \\
   Ar\,{\sc iii}   &  COL &   1.19(37) &       1.19 \\
     He\,{\sc ii}  &   NT &   1.10(37) &       1.10 \\
     N\,{\sc ii}   &  COL &   8.23(36) &       0.82 \\
      O\,{\sc ii}  &   NT &   7.83(36) &       0.78 \\
      He\,{\sc i}  &   NT &   3.89(36) &       0.39 \\
    Ar\,{\sc ii}   &  COL &   5.31(35) &       0.05 \\
\hline
\end{tabular}
\end{center}
\end{table}

We now explore the influence of a magnetar power of 10$^{39}$\,\ergs\ ($dV$ is set to 2000\,\kms\ at all times) in the much lighter he3p30 model. Its total ejecta mass is 1.2\,\msun, of which 70.0\,\% is helium and only 12.6\,\% is oxygen. Despite its modest kinetic energy of 0.55\,foe, its mean expansion rate is large and slightly greater than that of model he6p00 (Table\ref{tab_prog}). 

Figure~\ref{fig_spec_he3p30} shows the spectral evolution from 1000\,\AA\ to 1\mic\ for model he3p30 over a time spanning 250 to 3100\,d after explosion (later times are not computed because the spectra were no longer changing). At 250\,d, decay power dominates over magnetar power (see Fig.~\ref{fig_lbol}) and the model spectrum is essentially the same as obtained with power only from radioactive decay \citep{dessart_snibc_23}. The presence of multiple small and near equal-mass shells of distinct composition gives rise to a myriad of lines from different elements. Unlike model he12p00 and he6p00, the spectrum exhibits a relatively weak \oidoub, with numerous lines of equal strengths from \nad, \niidoub, \caiidoub, \nkiifs, [Ni\three]\,7889.9\,\AA, and prominent Fe\two\ and Fe\three\ lines in the blue part of the optical (e.g., [Fe\three]\,4658.0 and 5270.4\,\AA). \mgiiuv\ is the only strong line in the ultraviolet (\ciiuv\ is predicted in the model when only carbon lines are included in the spectral calculation, but fully blocked by metal-line blanketing when all lines are included). At 250\,d, 65.8\,\% of the total flux falls in the optical, 13.6\,\% in the ultraviolet and 20.5\,\% in the infrared. As time proceeds, the spectrum shows a weakening of all the lines present at 250\,d and the spectrum shows a similar evolution as model he6p00 with the appearance and strengthening of \oiiidoub, \oiidoub, \siiinz, \siiinf\ after about 600\,d. By the end of the sequence at 3100\,d, \oiiidoub\ radiates the entire optical luminosity, which is 59.2\,\% of the total, whereas 32.9\,\% goes to the ultraviolet and 7.8\,\% to the infrared (Fig.~\ref{fig_fluxes_he3p30}).

\begin{figure*}
\centering
\includegraphics[width=\hsize]{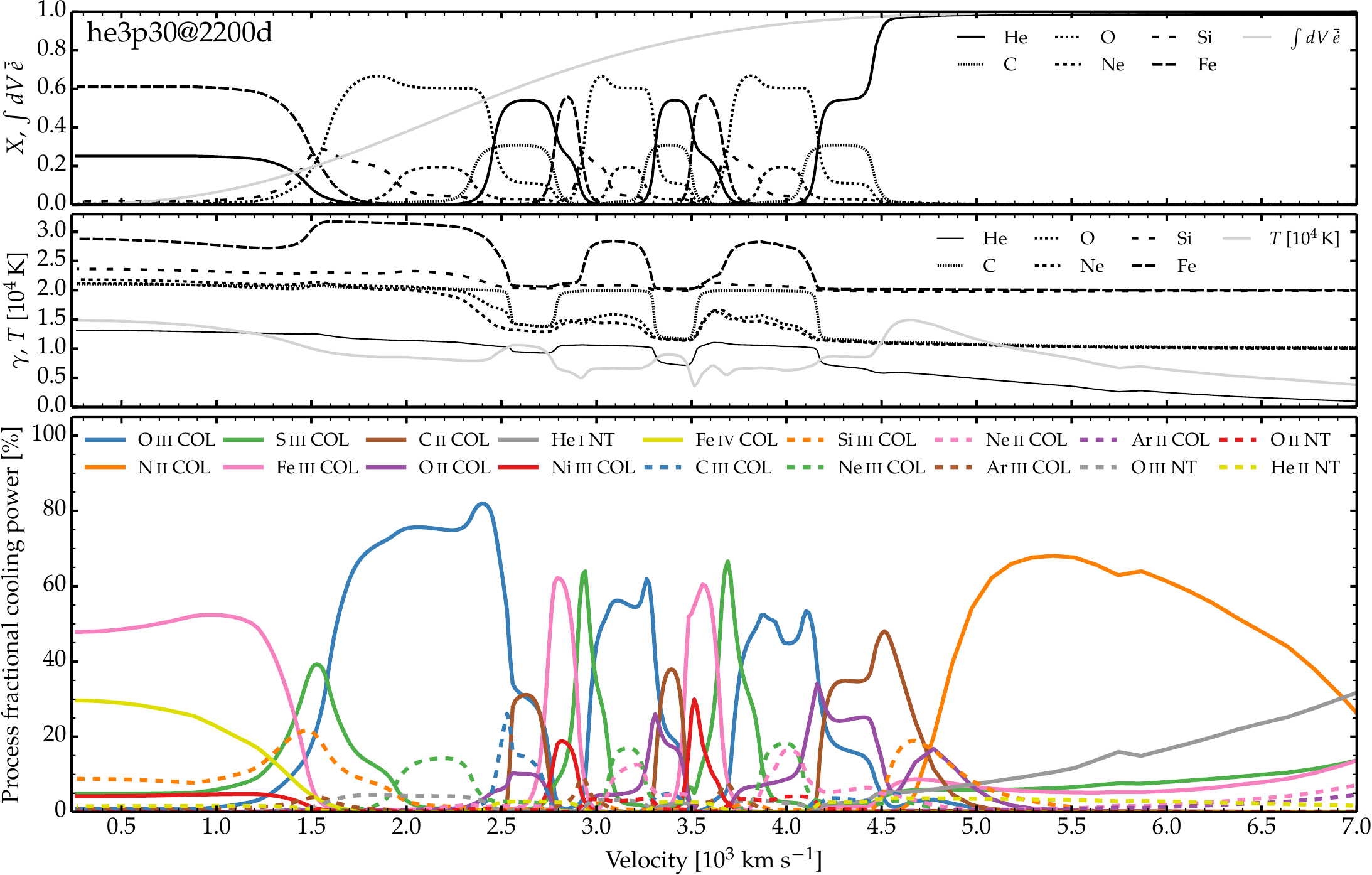}
\caption{Same as Fig.~\ref{fig_cool_he12p00_2200d} but now for model he3p30 at 2200\,d, influenced by a power of 10$^{39}$\,\ergs. Volume-integrated cooling powers are given for this model in Table~\ref{tab_cooling_heating_he3p30opt_fvol1p0_pwr1e39_2200d_dv2e8}. [See Section~\ref{sect_he3p30} for discussion.]
\label{fig_cool_he3p30_2200d}
}
\end{figure*}

Figure~\ref{fig_cool_he3p30_2200d} illustrates the variation of the ionization of various species and the dominant cooling powers (given as a percentage of the total heating power) versus velocity for the model he3p30 at 2200\,d. The ionization for most species is greater than in the previous he6p00 model with most elements twice ionized within the region where the magnetar power is absorbed. He, C, O and Ne are between singly and doubly ionized, whereas Si and Fe are between doubly and triply ionized. The temperature is higher than in models he12p00 and he6p00, covering between about 5000 and 15000\,K. In the outer regions beyond 4000\,\kms, where little power is absorbed, the ionization remains roughly constant. The distribution of the absorbed power between shells at 2200\,d is 26.6\,\% for He/C/O shell (it is hard to distinguish the He/C, O/C, and He/C/O shells so we consider them as grouped in a He/C/O shell), 22.1\,\% for the O/Ne/Mg shell, 17.9\,\% for the O/Si shell, 20.7\,\% for the Si/S shell, 12.0\,\% for the Fe/He shell. There is thus a much greater power absorbed in those Si- and Fe-rich shells in model he3p30 than in models he6p00 and he12p00 in which the O-rich material was absorbing most of the power. Only 3.2\,\% of the decay power is absorbed in the ejecta, of which 96.8\,\% comes from the local deposition of positrons.

\begin{figure*}
\centering
\includegraphics[width=\hsize]{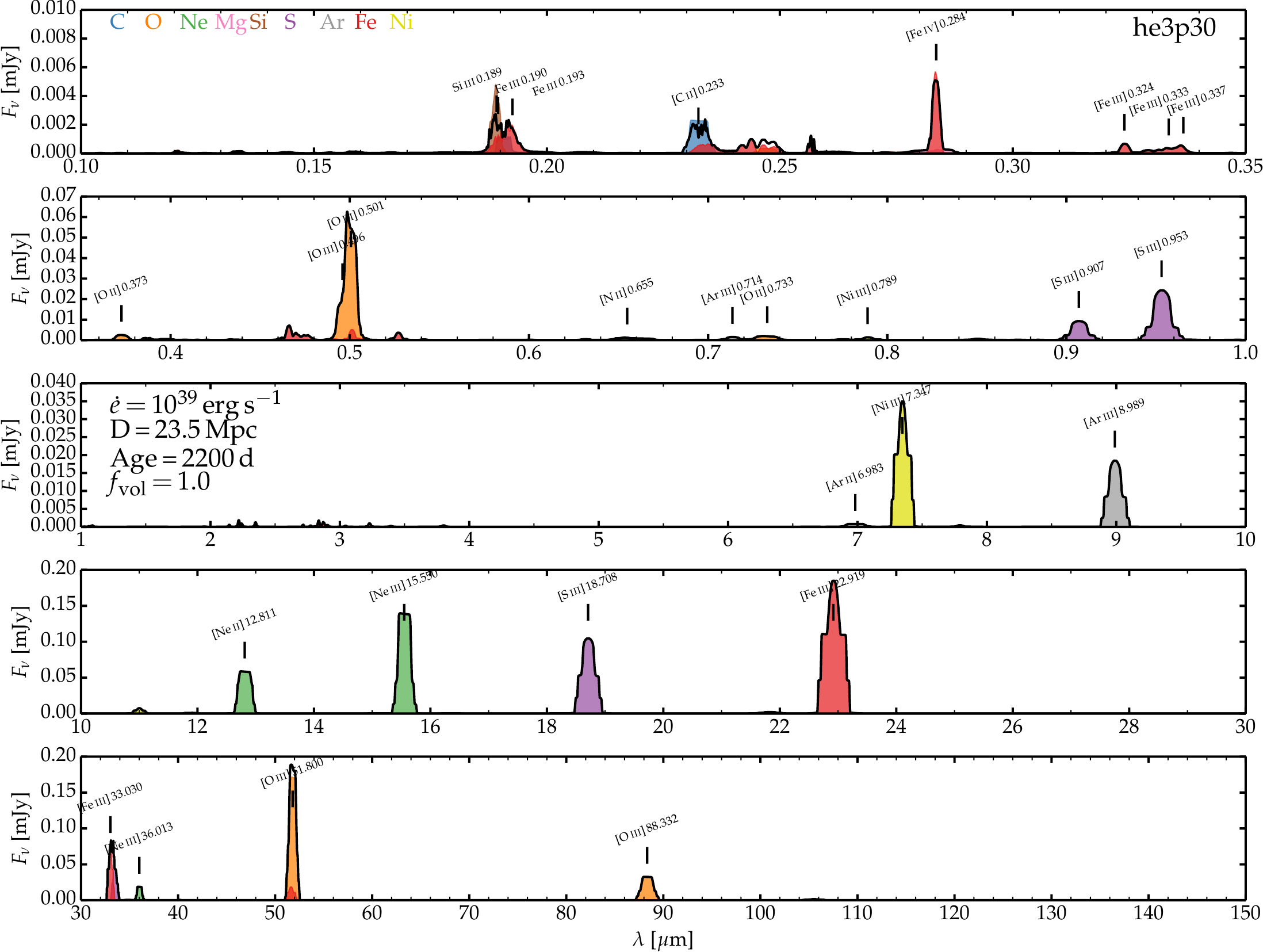}
\caption{Same as Fig.~\ref{fig_spec_uv_to_fir_he12p00_2200d} but now for model he3p30 at 2200\,d.
\label{fig_spec_uv_to_fir_he3p30_2200d}
}
\end{figure*}

The main coolants for each of these shells are shown in the bottom-row panel of Fig.~\ref{fig_cool_he3p30_2200d}. The outer He/N shell cools primarily through N\two\ collisional excitation and He\one\ nonthermal excitation but there is essentially no power absorbed in that outer region so no strong line will emerge from that He/N shell. The He/C/O shell cools primarily through C\two, O\two, and O\three\ collisional excitation. Because of the greater ionization in the O/Ne/Mg shell, O\three\ is the primary coolant with a smaller contribution from Ne\three\ and an even smaller one from Ne\two. Collisional excitation of O\three\ is a strong coolant for the O/Si and Si/S shells, but in the latter, collisional excitation of S\three\ dominates. In the Fe/He shell, Fe\three\ is the main coolant, assisted by Ni\three\ and Fe\four. Numerous other processes and coolants are shown in Fig.~\ref{fig_cool_he3p30_2200d} and they carry some small fraction of the total power absorbed. Table~\ref{tab_cooling_heating_he3p30opt_fvol1p0_pwr1e39_2200d_dv2e8} gives the integrated cooling power for that he3p30 model at 2200\,d. We find that 38.7\,\% of the magnetar power is radiated by O\three\ (nearly exclusively by \oiiidoub), followed with 14.2\,\% by Fe\three\ (numerous lines spread over the ultraviolet), with 11.3\,\%  by S\three. Numerous other coolants account for a few percent of the total and give rise to numerous metal lines mostly in the infrared with [Ni\three]\,7.347\mic, [Ar\three]\,8.989\mic, \neiifs, \neiiifs, [S\three]\,18.708\mic\ or [Fe\three]\,22.919\mic. A full spectrum covering from the ultraviolet to the far-infrared is shown in Fig.~\ref{fig_spec_uv_to_fir_he3p30_2200d}.

\begin{figure}
\centering
\includegraphics[width=\hsize]{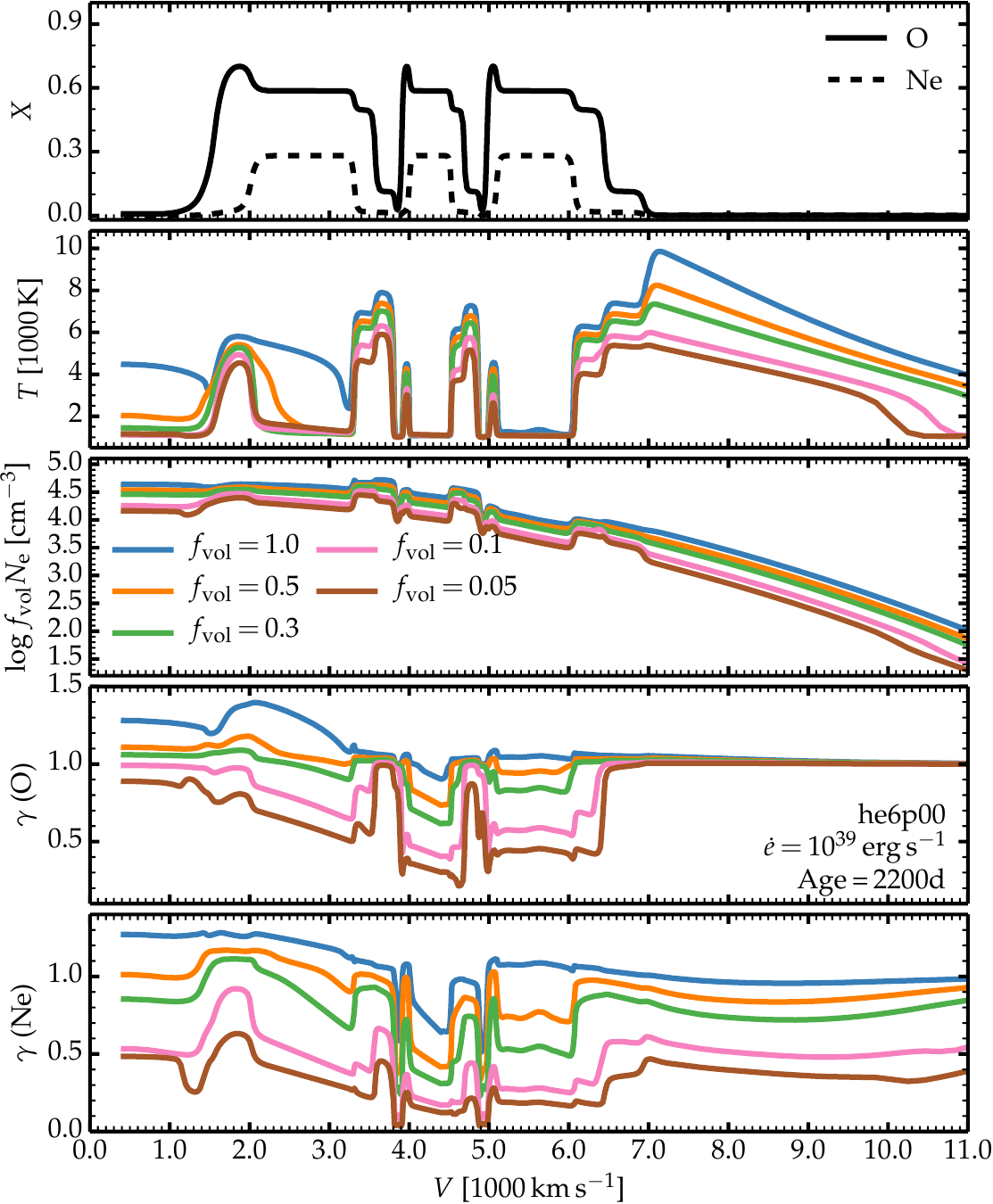}
\caption{Influence of clumping on ejecta properties for model he6p00 at 2200\,d and influenced by a magnetar power of 10$^{39}$\,\ergs. The adopted clumping is uniform and corresponds to a volume filling factor from 100\,\% (smooth ejecta) down to 5\,\%.  
\label{fig_he6p00_2200d_var_fvol_gas}
}
\end{figure}

\section{Dependency on ejecta clumping}
\label{sect_dep_fvol}

In core-collapse SNe, ejecta clumping should be driven by the various instabilities taking place during and after shock passage through the envelope (thus on a timescale of minutes in stripped-envelope SNe) as well as by the longer-term heating (on week timescale) from the radioactive decay of \nifs\ and \cofs\ \citep{fryxell_mueller_arnett_91,muller_87A_91,wongwathanarat_15_3d,gabler_3dsn_21}. This process may be stronger in Type II SNe than in stripped-envelope SNe because the latter have a greater kinetic energy per unit mass and become optically thin on a shorter timescale. All core-collapse SNe, even standard ones, should be inhomogeneous and clumped in their inner ejecta. In addition, this structure may be further affected by a prolonged and intense energy deposition from a compact remnant. Numerical simulations of this phenomenon show how chemical inhomogeneity and clumping of the ejecta is exacerbated \citep{chen_pm_2d_16,chen_3d_pm_20,suzuki_pm_2d_17,suzuki_pm_2d_21}. In this section, we explore the influence of varying levels of clumping on the gas and radiation properties of a magnetar-powered model (for a similar exploration but at times less than a year, see \citealt{d19_slsn_ic}). For simplicity, we adopt a uniform clumping at all depths and consider in this section only one ejecta model, one epoch, and one magnetar power (in practice, about half of the whole model grid has been computed with multiple levels of clumping). The treatment of clumping in \cmfgen\ is the same as in \citet{d18_fcl}.

Figure~\ref{fig_he6p00_2200d_var_fvol_gas} shows the impact of different levels of clumping (i.e., volume filling factor of 5, 10, 30, 50, and 100\,\%) on various gas quantities for model he6p00 at 2200\,d influenced by a magnetar power of 10$^{39}$\,\ergs\ and characterized by a deposition profile with $dV=$\,2800\,\kms. As is well known, increased clumping (i.e., smaller volume filling factor) leads to a reduction of the gas ionization, as evidenced by the profile for the density of free electrons (we show that quantity corrected for clumping so the offset in $f_{\rm vol} N{\rm e}$ reflects exclusively the change in ionization and not the change in mass density), the ionization of oxygen or that of neon. With increased clumping, the O mean ionization state in O-rich regions drops from slightly above one to about 0.3 as $f_{\rm vol}$ drops from 100 to 5\,\%. Neon exhibits a similar reduction for the same clumping variations, in part because the ionization potentials of their first ionization states are comparable. The variation of the temperature with clumping seems to go by steps, with a strong reduction when $f_{\rm vol}$ drops from 100 to 50\,\%, and little change for further clumping enhancements.

Because we adopt the same magnetar-power deposition profile, all five models have the same distribution of power amongst shells: 1\,\% goes to the outer He-rich shell, 12.5\,\% to the He/C/O shell, 9.4\,\% to the O/C shell, 52.9\,\% to the O/Ne/Mg shell, 11.2\,\% to the O/Si shell, 5.6\,\% to the Si/S shell, and 7.2\,\% to the Fe/He shell. With clumping, it is thus the variation in ionization that drives the change in coolants radiating the absorbed magnetar power and the spectral appearance. The ionization differences affect primarily the relative importance of doubly-ionized and single-ionized species, whereas Ne\two\ remains the dominant coolant in all five models (the fractional power radiated away is 45.22\,\% in the model with $f_{\rm vol}=$\,5\,\% compared to 35.2\,\% in the model with $f_{\rm vol}=$\,100\,\%). Hence, Ni\two, Fe\two, O\two\ are the most important secondary coolants for $f_{\rm vol}$ of 5 to 30\,\%, whereas O\three, Ne\three, S\three\, and Fe\three\ take over for $f_{\rm vol}$ of 50 to 100\,\% (i.e., the less clumped, more ionized ejecta models).

\begin{figure*}
\centering
\includegraphics[width=\hsize]{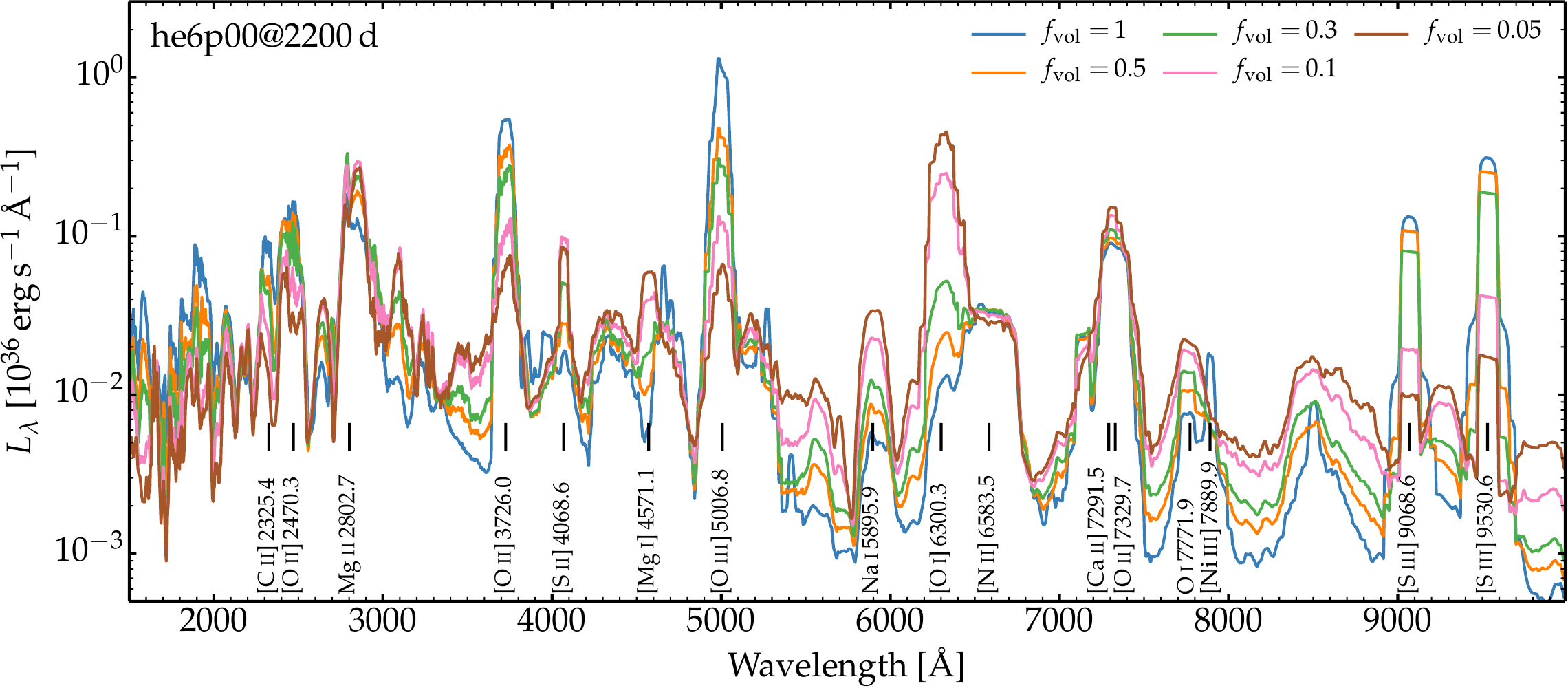}
\caption{Influence of clumping on the optical luminosity. We show the he6p00 model at 2200\,d influenced by a magnetar power of 10$^{39}$\,\ergs. The adopted clumping is uniform and corresponds to a volume filling factor covering from 100\,\% (smooth ejecta) down to 5\,\%. 
\label{fig_he6p00_2200d_var_fvol_spec}
}
\end{figure*}

These properties are readily visible in the optical spectra (Fig.~\ref{fig_he6p00_2200d_var_fvol_spec}) where about 20\,\% of the luminosity emerges (whereas about 75\,\% comes out in the infrared). Progressing redward from the blue edge of the optical, we see the strong variation of \oiiuv, \oiiidoub, \nad, \oidoub, \caiidoub, \oiidoub, \siiinz, and \siiinf. These modulations caused by clumping alter the relative strength of the three strongest optical lines of oxygen (i.e., \oiiidoub\ strong and \oidoub\ weak for a smooth ejecta, and \oiiidoub\ weak and \oidoub\ strong for the models with large clumping) but leave the total optical luminosity 
essentially unchanged. This occurs while the ejecta composition, mass, and kinetic energy are the same. This implies some uncertainty in interpreting observations (e.g., the relative strength the various oxygen lines in the optical range) since we do not have a robust knowledge of the clumping in magnetar-powered SNe.

\section{Influence of the treatment of injected power: high-energy electrons or X-ray irradiation}
\label{sect_pwr_vs_xray}

So far in this work, the magnetar power was injected in the ejecta in the form of high-energy electrons in order to mimic in a simple manner the radiation from the compact remnant. The advantage is that these particles are analogous to Compton-scattered thermal electrons in the presence of $\gamma$ rays from radioactive decay. This context is routinely treated in \cmfgen\ to handle nonthermal effects associated with radioactive decay. A first limitation of this approach is that not all the radiation from the magnetar may come from high-energy particles. There may also be high energy radiation such as X-rays. Secondly, we do not model the propagation of these high-energy electrons through the ejecta but instead specify where in the ejecta they are absorbed (i.e., we adopt a gaussian profile with a characteristic width set by $dV$ -- see Section~\ref{sect_setup}). In this section, we use an alternative approach for the magnetar power and introduce it instead in the form of an X-ray emitting source. 

We recently studied the impact of X-ray irradiation on the radiative properties of young SNe (typically less than 50\,d old; Dessart \& Hillier in prep.), but in that work the X-ray irradiation was used to mimic ejecta interaction with the progenitor wind and thus introduced in the outer parts of the ejecta. Here, the principles and the method are the same but the X-rays are injected in the inner ejecta. We assume the presence of an X-ray emitting plasma with a temperature of about five million degrees. For the X-ray emissivity, we adopt a profile that goes as $\exp(-(V/V_X)^3)$ (other profiles could be used), with $V_X$ set to 4000\,\kms\ in order to match the deposition profile in a model counterpart that treats magnetar power in the form of high energy electrons. The volume integral of the X-ray emissivity is set to a prescribed value, here of 10$^{39}$\,\ergs. 

\begin{figure*}
\centering
\includegraphics[width=\hsize]{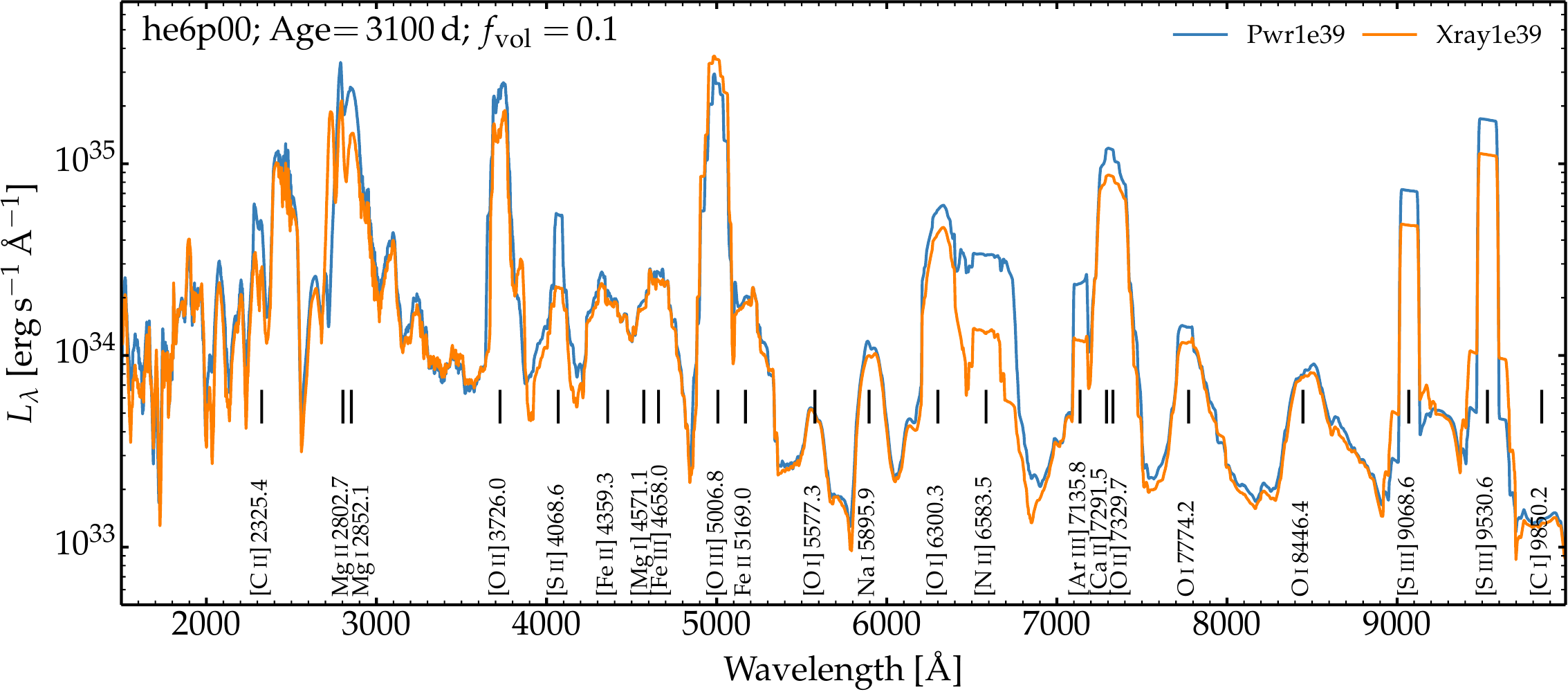}
\caption{Impact on the optical luminosity of the treatment for the power injection. We show the spectrum for the he6p00 model at 3100\,d, characterized by a uniform volume filling factor of 10\,\% and influenced by a magnetar power of 10$^{39}$\,\ergs\ treated either in the form of high-energy electrons (model ``Pwr1e39'') or X-ray irradiation (model ``Xray1e39''). The models have roughly the same power absorbed both in magnitude and deposition profile so only the nature of the power injected differs. [See Section~\ref{sect_pwr_vs_xray} for discussion.]
\label{fig_pwr_vs_xray}
}
\end{figure*}

Figure~\ref{fig_pwr_vs_xray} shows the luminosity over the optical range for the he6p00 model at 3100\,d, characterized by a uniform volume filling factor of 10\,\% and influenced by a magnetar power of 10$^{39}$\,\ergs\ treated either in the form of high-energy electrons (model ``Pwr1e39'') or X-ray irradiation (model ``Xray1e39''). The models have roughly the same power absorbed both in magnitude and location so only the nature of the power injected differs. We find that the two spectra are comparable with only a few differences (e.g., \niidoub) attributed to the fact that the model Pwr1e39 includes the outer ejecta layers all the way to  14,000\,\kms\ whereas model Xray1e39 extends out to 6000\,\kms\ (the reason for reducing the extent is that nearly 100\,\% of the injected power is absorbed below 6000\,\kms\ in both models -- the extra work of including the outer ejecta layers was superfluous). 

This insensitivity to the nature of the power source can be explained. In the simulation treating magnetar power in the form of high energy electrons, we find that the injected energy is channeled nearly entirely (i.e., $\sim$\,90\,\%) into heat, and little into nonthermal excitation and ionization (i.e., roughly 5\,\% for each). This heat channel dominates because the gas is ionized, whereas it is well known that nonthermal effects thrive under partial ionization or even neutral conditions \citep{KF92,d12_snibc}. The same effect was found in simulations of magnetar-powered superluminous SNe \citep{d19_slsn_ic}. With X-ray irradiation, the heating process is photoabsorption (followed by photoionization) in atoms and ions leading to ejection of K-shell electrons (mostly for O\one, O\two, C\two, Fe\two, Fe\three) or valence electrons (He\one, O\one, O\two, C\two, Mg\two). This process thus injects heat into the free-electron gas, just like in the alternate treatment of magnetar power in the form of high-energy electrons. Thus, the adopted treatment of the injected power does not impact the radiative properties of magnetar-powered SNe -- the deposition profile is much more critical (see next section).

In nature, the compact remnant will probably release its power both in the form of high-energy radiation and particles, whose mean free path may be very different. In our simulations with X-ray irradiation from the remnant, we find that 0.2\,\% of the X-rays injected on the grid escape the ejecta. Our he6p00 model at 3100\,d still has a high optical depth of about 10$^4$ at 1\,keV, which progressively decreases at longer wavelength, being about 100 in the Lyman continuum, and about one in the Balmer continuum (the total electron scattering optical depth is only about 0.001). So, the bulk of the optical and infrared radiation is optically thin (neglecting any presence of dust) but any high energy radiation below the Lyman edge is fully absorbed in the ejecta even at nearly ten years post explosion. Such X-ray attenuation by the ejecta compromises the detection of X-rays from magnetar-powered SNe at late times (see for example the results from the systematic survey of \citealt{margutti_slsn_xray_18}). 

The late-time magnetar power goes as $1/B_{\rm pm}^2$, where $B_{\rm pm}$ is the magnetic field of the magnetar, so events with a boost at early times from a large magnetic field are too faint later on. Large magnetar powers at late times would require modest values of $B_{\rm pm}$ which have no impact on the SN radiation at early times. With a magnetar field of order 10$^{14}$\,G, a magnetar power of 10$^{41}$\,\ergs\ results at 2000-3000\,d, of which less than 10$^{39}$\,\ergs\ would escape as X-rays according to our simulations.

\begin{figure*}
\centering
\includegraphics[width=\hsize]{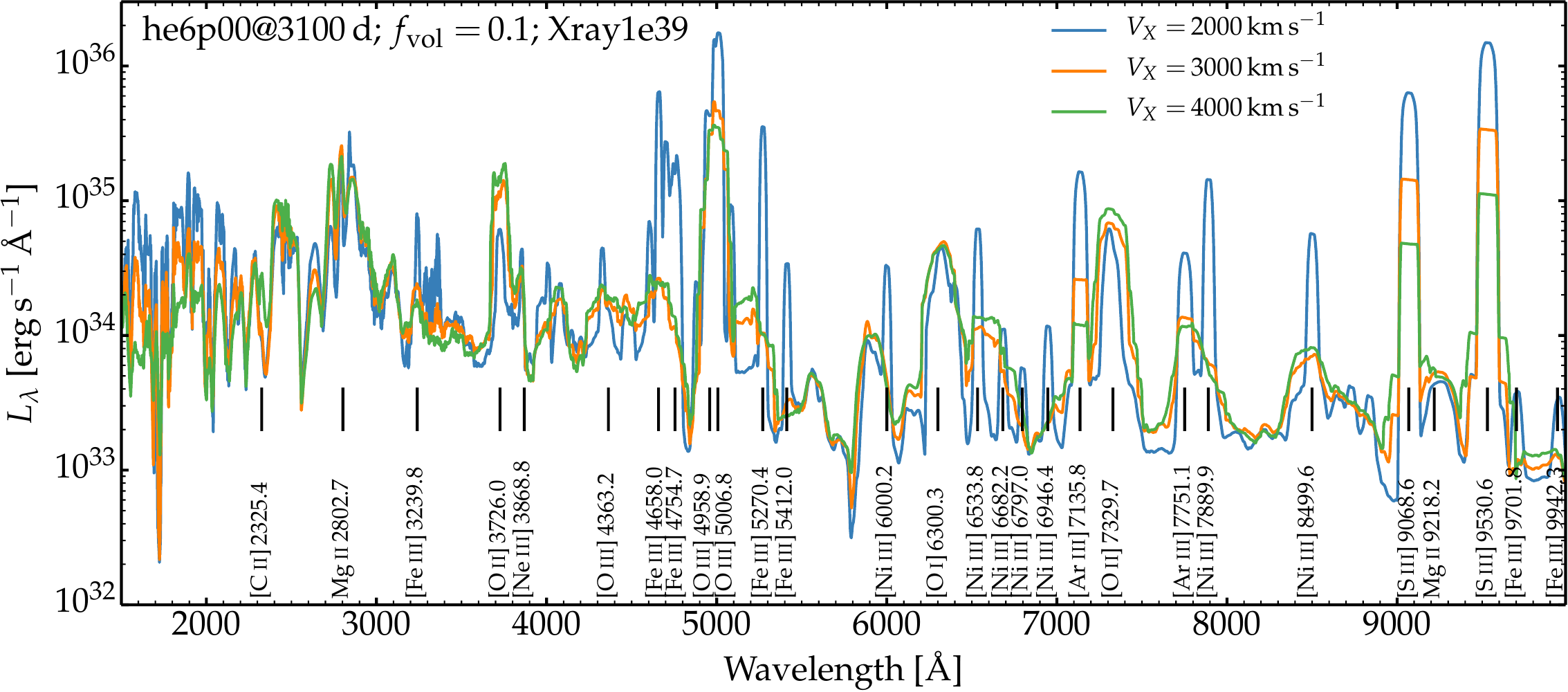}
\caption{Influence of the adopted magnetar-power deposition profile on the optical luminosity. We show the he6p00 model at 3100\,d influenced by a magnetar power of 10$^{39}$\,\ergs. The adopted clumping is uniform and corresponds to a volume filling factor of 10\,\%. Here, the power is injected in the form of X-ray irradiation from the central remnant with various choices of $V_X$. [See Section~\ref{sect_dep_edep} for discussion.] 
\label{fig_he6p00_2200d_var_dep_prof_spec}
}
\end{figure*}

\section{Dependency on deposition profile}
\label{sect_dep_edep}

In this section, we discuss the influence of the adopted deposition profile for the magnetar power. Simulations have proven difficult to converge when we make the deposition profile confined to the inner ejecta layers when we treat that power in the form of high-energy electrons. The temperature surges in those layers and the code struggles. So, we used instead X-ray irradiation from the remnant, adopting different velocity extent for the region containing the X-ray emission, with choices set to $V_X$ of 2000, 3000, and 4000\,\kms. Such variations lead to complicated and not necessarily obvious consequences. For example, injecting power over a restricted volume tends to boost the energy per unit volume or unit mass and can induce a rise in temperature and ionization. However, this power is then injected in slower-velocity regions, where the density is higher and thus where the ionization would tend to be lower. Conversely, energy deposition in the outer ejecta layers, at lower density, can lead to a higher ionization.

A second aspect is the range of ejecta shells where the power is absorbed. Only layers that absorb power can radiate. If the deposition is limited to the inner ejecta layer, a large part of the ejecta may not be energized and may remain invisible. For example, this might make an ejecta from a high mass He-star progenitor (with a massive O/Ne/Mg shell) appear similar to a modest-mass He-star progenitor (with a low mass O/Ne/Mg shell) because only the innermost layers rich in oxygen absorb the remnant power. Consequently, this can bias the absorption of magnetar power in favor of more metal-rich regions like the material from the Si/S and O/Si shells since there are located deep in the progenitor star. The magnitude of chemical mixing in the ejecta plays a critical role in this context since it determines what elements or shells absorb the magnetar power.

Figure~\ref{fig_he6p00_2200d_var_dep_prof_spec} shows the emergent optical luminosity of the he6p00 model at 3100\,d, assuming a total power for the X-ray irradiation of 10$^{39}$\,\ergs\ and a uniform volume filling factor of 10\,\%, but a velocity $V_X$ of 2000, 3000, or 4000\,\kms\ -- the X-ray irradiation goes as $\exp(-(V/V_X)^3)$. The model with $V_X=$\,4000\,\kms\ was already described in Section~\ref{sect_pwr_vs_xray}. Here, as $V_X$ is reduced, the spectrum changes initially little but for $V_X=$\,2000\,\kms\ its characteristics are vastly different, in particular with the presence of numerous lines of Fe\three\ and Ni\three, and much stronger \oiiidoub, \siiinz, and \siiinf. Many of these lines are also significantly narrower, in particular the Ni\three\ lines, because they form in the innermost ejecta layers. With this shift of the power absorbed to the inner ejecta layers, the \niidoub\ line is no longer present -- N\two\ is still a strong coolant for the He/N shell but that shell captures no power to radiate.

These properties are well understood when comparing the ejecta coolants in the models with $V_X$ of 2000 and 4000\,\kms. For the model with an extended deposition, the cooling is performed through collisional excitation of Ne\two\ for 57.3\,\%, by Ne\three\ for 8.1\,\%, Fe\three\ for 6.3\,\%, O\three\ for 5.0\,\%, and numerous other contributors at the few percent level. In the model with a more confined deposition, the primary coolants are Ne\two\ (27.7\,\%), S\three\ (23.4\,\%), O\three\ (11.6\,\%), Fe\three\ (10.5\,\%), Ne\three\ (8.9\,\%), Ni\three\ (6.9\,\%), and numerous other contributors at the few percent level. Here, the share of the cooling carried out by doubly-ionized ions is much greater. Interestingly, the change in strength of the Ne\two\ cooling accounts for half the change in infrared luminosity between the models with $V_X$ of 2000 and 4000\,\kms, and that part of the cooling lost by Ne\two\ goes essentially to O\three, S\three\, Fe\three, and Ni\three.

In all three cases, the optical spectra reveal lines having a range of ionization. For example, all three models show \oidoub, \oiidoub, and \oiiidoub, although the latter is typically ten times stronger than the other two lines. But these spectra are drastically different from those obtained at 1\,yr after explosion for the same ejecta (top row of Figs.~\ref{fig_spec_he12p00}--\ref{fig_spec_he3p30}) as well as those of standard Type Ib and Ic SNe (see, e.g., \citealt{modjaz_ibc_neb_08}; \citealt{milisavljevic_etal_10}; \citealt{fang_neb_22}). Magnetar power leads to a huge change in ionization and spectral characteristics. 

\begin{figure*}
\centering
\includegraphics[width=\hsize]{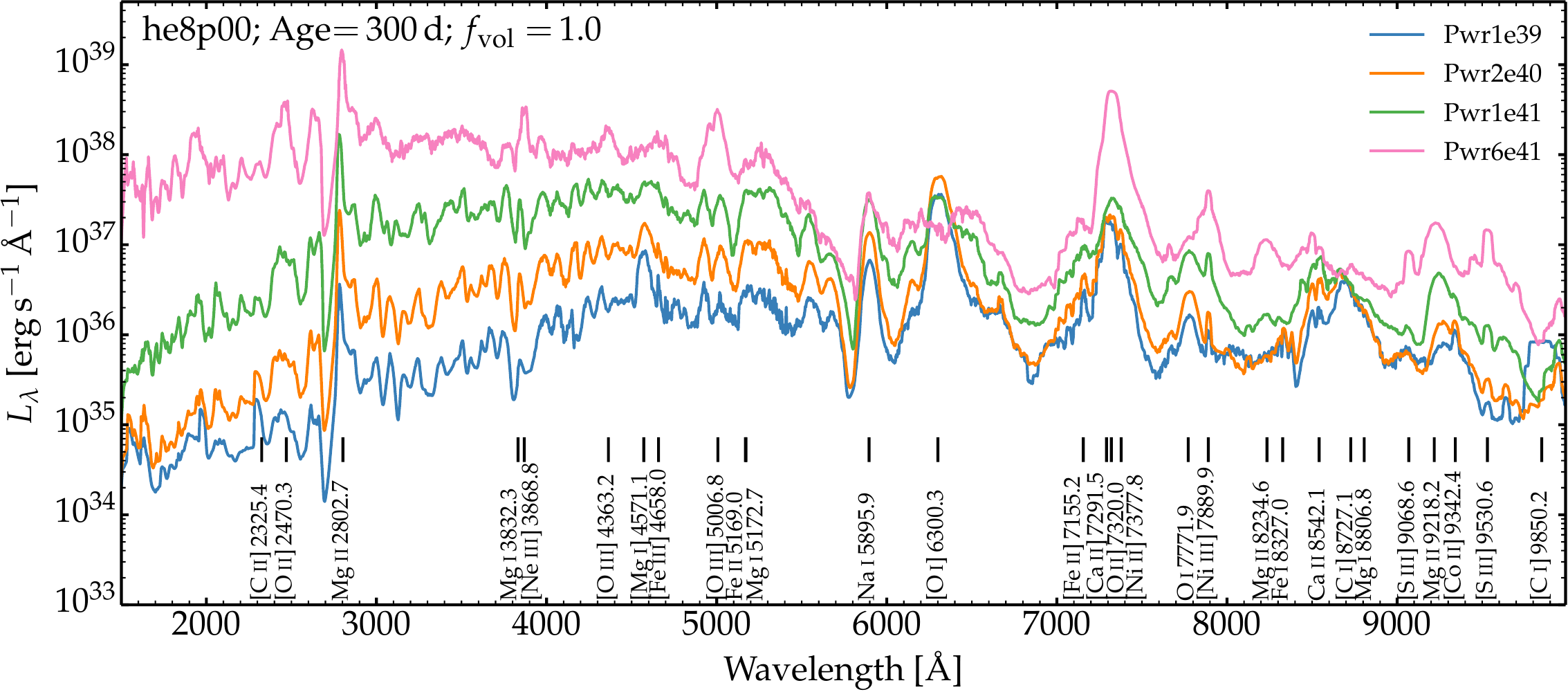}
\caption{Impact of variations in magnetar power on the optical luminosity of the he8p00 model at 300\,d after explosion. The magnetar power covers from 10$^{39}$ up to $6.0 \times 10^{41}$\,\ergs. Line identifications are provided at the bottom. [See Section~\ref{sect_pwr} for discussion.]
\label{fig_varpwr_he8_300d}
}
\end{figure*}

\section{Dependency on the magnitude of the injected magnetar power}
\label{sect_pwr}

In this section, we compare a set of models with different magnetar powers using the he8p00 model at 300\,d, assuming a smooth ejecta (i.e., volume filling factor of 100\,\%), and a deposition profile with $dV$ of 2000\,\kms. This leads to a deposition of the magnetar power within the inner 5000\,\kms\ of the ejecta. At that time, the radioactive decay power absorbed in the ejecta is $1.7 \times 10^{40}$\,\ergs. We run a set of simulations increasing the magnetar power from 10$^{39}$ to $6.0 \times 10^{41}$\,\ergs, thus going from a configuration where the magnetar power is 0.06 up to 35 times the decay power. 

Figure~\ref{fig_varpwr_he8_300d} shows the optical luminosity for the he8p00 model with a magnetar power of 10$^{39}$, $2.0 \times 10^{40}$, 10$^{41}$, and $6.0 \times 10^{41}$\,\ergs\ and for an age of 300\,d. For the lowest magnetar power, the optical spectrum is essentially unchanged from that in which  radioactive decay is the only power source. The ejecta ionization is low with He neutral throughout, O and Ne partially ionized, and C, Si, and Fe once ionized throughout the ejecta. The main coolants in this cool and essentially radioactively-powered model is dominated by collisional excitation of O\one, Mg\two, and Ca\two, followed by weaker contributions from collisional excitation of Fe\two, Ni\two\ and Ne\two, and nonthermal excitation of He\one.

With increasing magnetar power, the ejecta ionization rises and the coolants vary accordingly. For example, in the model with the magnetar power of $6.0 \times 10^{41}$\,\ergs, He is essentially once ionized throughout, O and Ne are once ionized, C is between once and twice ionized, and Si and Fe are twice ionized. These more ionized ejecta now cool through collisional excitation of Mg\two, C\two, O\two, Si\three, O\three, Fe\three, and Ne\three\ (in decreasing order of importance). This switch is visible in the optical luminosity shown in Fig.~\ref{fig_varpwr_he8_300d}, with a much bluer spectrum exhibiting strong flux in the ultraviolet, strong \mgiiuv, the absence of \oidoub\ but the presence of \oiidoub\ (which replaces the \caiidoub\ present in the models with a low magnetar power). Numerous high ionization lines are present, such as [Ne\three]\,3868.8\,\AA, [O\three]\,4363.3\,\AA, [Fe\three]\,4658.0\,\AA, [Ni\three]\,7889.9\,\AA, or \siiinf. The strong emission in the range 4000 to 5500\,\AA\ that appears as a pseudo-continuum is also primarily due to a forest of Fe\two\ lines.

Variations in injected power not only change the model luminosity but also strongly affect its colors and the lines through which the ejecta cool. This shift in ionization is reminiscent of the effect of clumping although clumping (treated as a radial compression only) leaves the power absorbed in the ejecta, and hence the bolometric luminosity, unchanged.


\section{Comparison to SN\,2012\lowercase{au}}
\label{sect_comp_obs}

In this section, we compare our models to the observations of the magnetar-powered SN candidate 2012au at about one and six years after explosion \citep{milisavljevic_12au_13,milisavljevic_12au_18,pandey_12au_21}. For SN\,2012au, we adopt from these references a distance of 23.5\,Mpc, a redshift of 0.0045, and a reddening $E(B-V)=$\,0.063\,mag. Assuming a rise time to $V$-band maximum of 16\,d \citep{milisavljevic_12au_13}, we present comparisons to SN\,2012au at a post-explosion age of about 289\,d (MMT spectrum taken on 19th of December 2012), 335\,d (data that combine optical MMT spectrum taken on 5th of February 2013 with near-infrared FIRE spectrum taken on 2nd of February 2013), and finally at 2269\,d (optical IMACS spectrum taken on 9th of June 2018, combined with another spectrum that extends to beyond one micron and thus covers the S\three\ lines). All spectra are from \citet{milisavljevic_12au_13,milisavljevic_12au_18}. We adjust the absolute flux level of the spectra at 289 and 335\,d so that they match the published photometry of \citet{milisavljevic_12au_13,milisavljevic_12au_18} and \cite{pandey_12au_21}. The spectrum at 2269\,d was not scaled since it was calibrated in absolute flux with a 30\,\% accuracy by \citet{milisavljevic_12au_18}. Although the presence of dust has been inferred from the observations of SN\,2012au at 6\,yr \citep{niculescu_duvaz_dust_22}, we currently ignore its influence on the spectra and defer its treatment to a forthcoming study.

Having a limited set of He-star masses in our grid, we tend to use the he8p00 model because it strikes a good balance between models he6p00 (high kinetic energy, light ejecta, large residual He mass) and he12p00 model (low kinetic energy, massive ejecta, low He content). This model he8p00 seems well suited to capture the salient properties of SN\,2012au. Being He-rich in its outer layers (the former $\sim$\,1\,\msun\ He/N shell occupies the ejecta layers above 6000\,\kms), it is compatible with the Type Ib classification. It is also rich in oxygen (total yield of 1.71\,\msun), which may explain the strong oxygen-dominated spectra of SN\,2012au at late times. But the explosion model is 1D and may be in tension with the observed light curve and spectral evolution at early times \citep{milisavljevic_12au_13,takaki_12au_13,pandey_12au_21}. As discussed in Section~\ref{sect_setup} and \citet{dessart_snibc_21}, our models have a fixed set of preSN mass, ejecta mass, explosion energy, and explosive yields and we make no attempt to try different combination of values for these characteristics. Our comparisons should thus be evaluated in a global sense to gauge whether they are broadly compatible with the observations and not whether any specific line is matched to within some given margin.

\begin{figure*}
\includegraphics[width=\hsize]{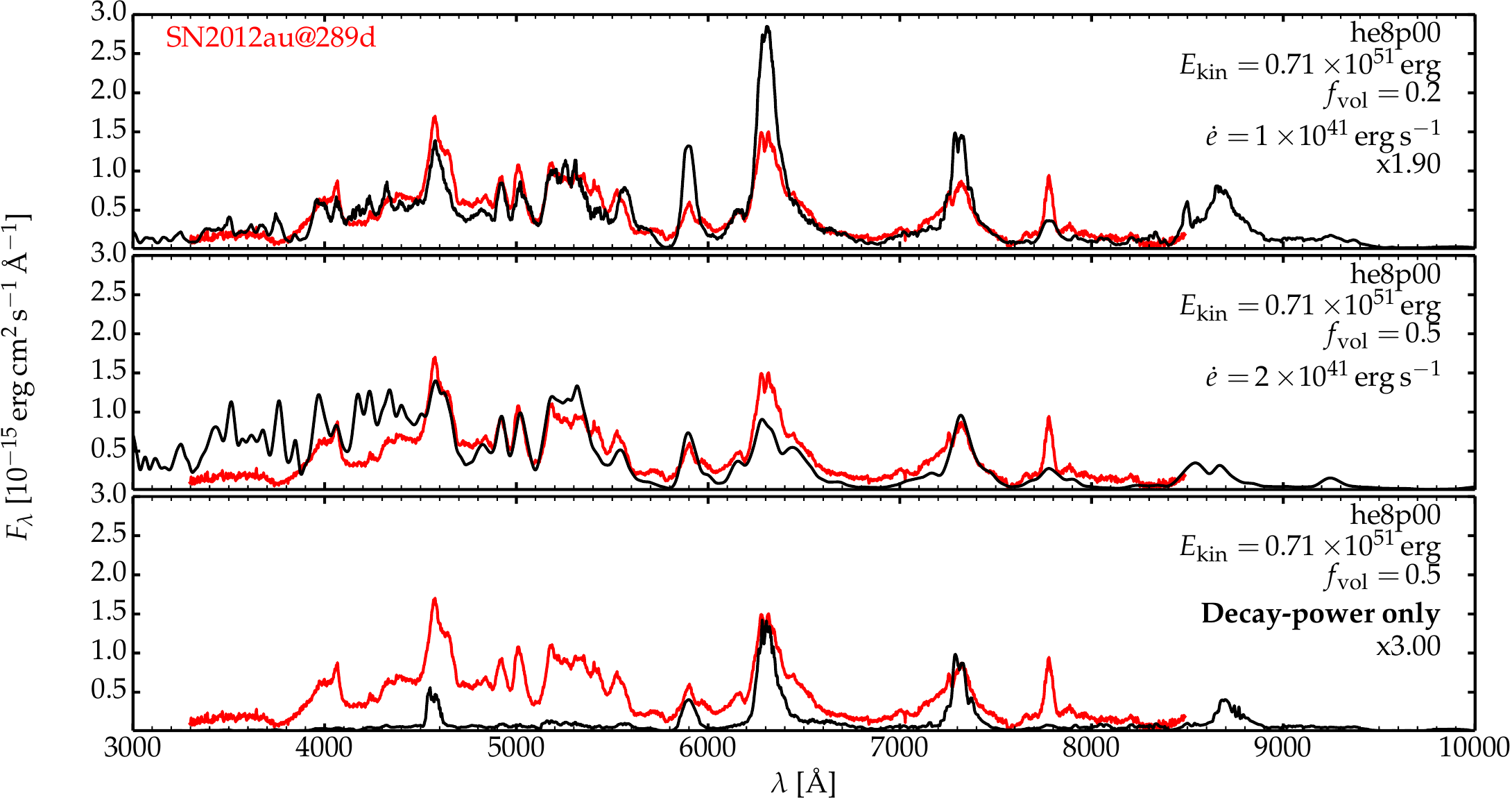}
\caption{Comparison between the observations of SN\,2012au at $\sim$\,289\,d after explosion and a set of contemporaneous He-star explosion models with magnetar power. For comparison, a model with decay power only is also shown (bottom-row panel). The observations have been corrected for redshift and extinction and the model has been scaled to the distance of SN\,2012au. Any additional flux scaling is indicated by a label at right. [See Section~\ref{sect_12au_289d} for discussion.]
\label{fig_12au_289d}
}
\end{figure*}

\subsection{The 289\,d spectrum} 
\label{sect_12au_289d}

Figure~\ref{fig_12au_289d} compares some of the magnetar-powered models computed in our grid with the observations of the Type Ib SN\,2012au at $\sim$\,289\,d after explosion. Because a significant amount of flux emerges in the infrared and because the SN age is only about a year, we had to raise the power well above the nominal value of 10$^{39}$\,\ergs\ adopted in the previous sections to $1-2\,\times\,10^{41}$\,\ergs. This magnetar power is more than ten times greater than the radioactive decay power at that time and causes a strong alteration of the spectrum, boosting its ionization and the optical color blueward (see discussion in Section~\ref{sect_pwr}). Most of this extra flux in the blue (i.e., below 5500\,\AA) is caused by Fe\two\ emission. To tame this color offset, we introduced clumping with a uniform volume filling factor of 20\,\% and 50\,\%. Striking the right balance is difficult: with greater clumping the optical color is well matched but the \oidoub\ (\nad\ is also a good analog) is too strong, whereas with weaker clumping there is too much flux in the blue and \oidoub\ is too weak. In both models, the 7300\,\AA\ emission is primarily due to \caiidoub\ and \nkiifs, with no contribution from \oiidoub. The big hump redward of \oidoub\ is predicted in the model to be due to Fe\one\ and Fe\two\ emission rather than \niidoub.

Overall, these two magnetar-powered models capture the essence of the unique properties of SN\,2012au, which is much bluer and luminous than typical Type Ib or Ic SNe at that phase. To make this distinction more compelling, we add in the bottom panel of Fig.~\ref{fig_12au_289d} the same he8p00 model with a 50\,\% uniform volume filling factor but without magnetar power. The model has been scaled by a factor of three to match the peak of \oidoub, which then also matches the peak flux of \caiidoub. However, this model powered by decay power only is much too faint. It entirely lacks the strong emission from Fe\two\ below 5500\,\AA. 

We also find more subtle features present in SN\,2012au at 289\,d and better matched by the magnetar-powered model. For example, there is a well defined and relatively narrow emission around 4070\,\AA\ that is also predicted in our models as being a combination of Ni\two\,4067.0\,\AA\ and [S\two]\,4068.6\,\AA\ (which overlaps with numerous lines of Fe\one\ and Fe\two). The O\one\,7774\,\AA\ is both narrow and strong and indicates the presence not only of oxygen at low velocity but also at relatively high density. This line is very weak in the model with only decay power because it forms through transitions between levels of relatively high excitation energy, but the two magnetar-powered models predict half the observed flux in that transition. Another temperature sensitive line (i.e., its lower level is above the ground state) that is observed here is [O\one]\,\,5577.3\,\AA\ and the magnetar-powered models reproduce its strength closely. A weak emission redward is possibly due to [Ni\three]\,7889.9\,\AA, although an alternate identification could be overlapping emission lines from iron. These transitions give some support to the idea that power is being injected in the inner ejecta of SN\,2012au, leading to emission not just from oxygen, but also from sulfur or nickel in the innermost layers of the ejecta. The associated narrow emission features also suggest that the late power injection is instrumental to make material shine in the inner ejecta while in general such material is invisible. This may indicate that SNe Ib and Ic may have more mass at low velocity than generally believed -- an indirect consequence of asymmetry. Here, in our he8p00 model from a 1D explosion, there is only 0.06\,\msun\ of material below 1000\,\kms. In contrast, an asymmetric explosion can naturally predict more mass both at higher and smaller velocities than in a 1D explosion of the same energy.

\begin{figure*}
\includegraphics[width=\hsize]{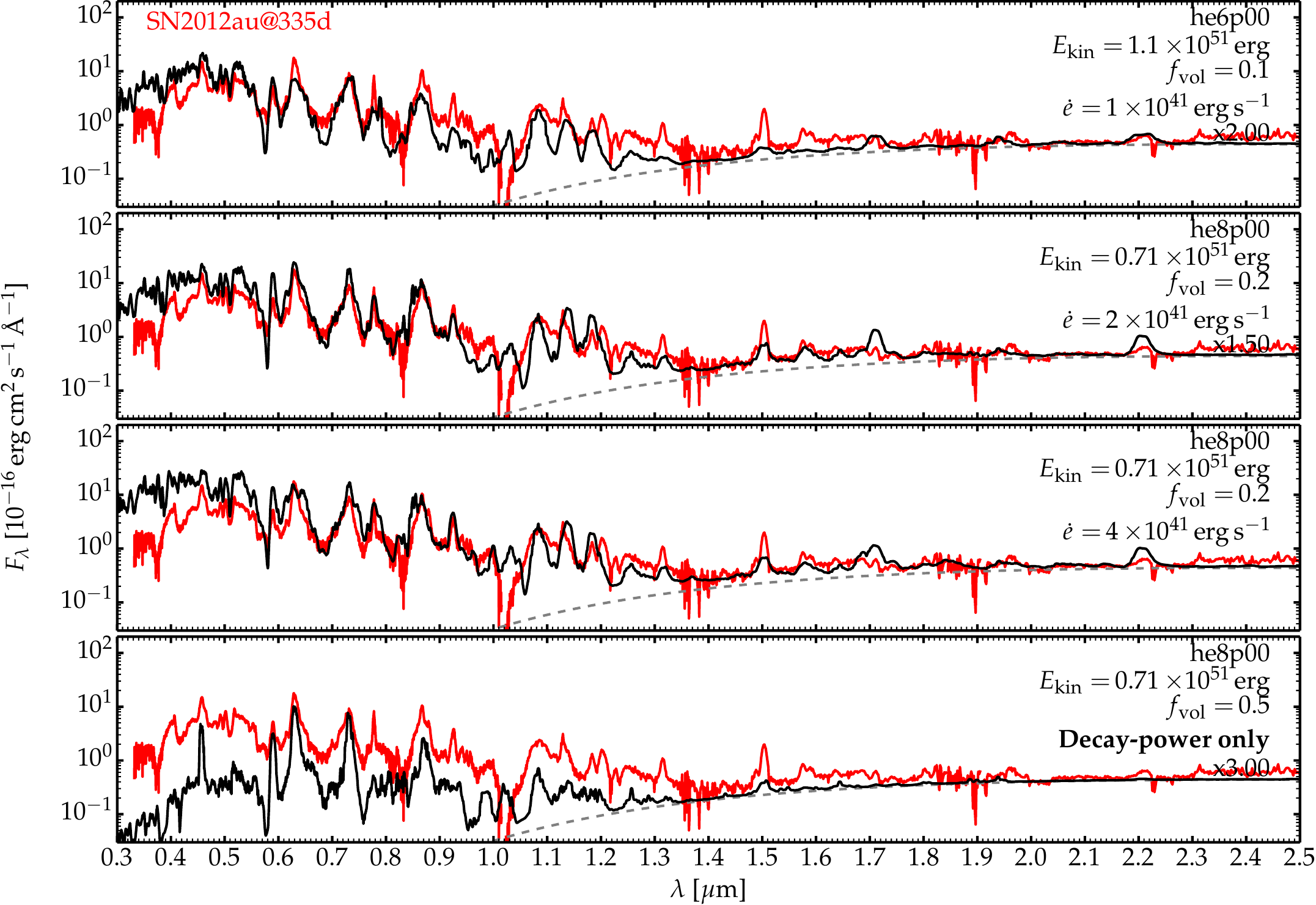}
\caption{Same as Fig.~\ref{fig_12au_289d} but now for the observations at 335\,d after explosion. [See Section~\ref{sect_12au_335d} for discussion.]
\label{fig_12au_335d}
}
\end{figure*}

\subsection{The 335\,d spectrum} 
\label{sect_12au_335d}

Figure~\ref{fig_12au_335d} compares some of the magnetar-powered models computed in our grid with the optical and near-infrared observations of the Type Ib SN\,2012au at $\sim$\,335\,d after explosion. Here, the models are similar to those used at 289\,d after explosion but the extended wavelength range gives access to additional lines and diagnostics. We defer any discussion of the origin of the near-infrared excess flux to future work. For now, we artificially add to our model flux a contribution from a blackbody with a temperature of 1200\,K and a radius of $8 \times 10^{15}$\,cm. 

The he8p00 model with a magnetar power of $2 \times 10^{41}$\,\ergs\ yields a good match to the overall distribution. We use a logarithmic scale for the flux to focus on the overall match rather than details. The optical properties were discussed in the preceding section but we can now comment on the near-infrared range. The magnetar-powered models predict the various lines observed, and specifically He\one\,1.083\mic, Na\one\,1.140\mic, Mg\one\,1.183\mic, [Fe\two]\,1.257\mic, O\one\,1.316\mic, Mg\one\,1.502\mic, Mg\one\,1.575\mic, Mg\one\,1.711\mic, and [Ni\two]\,1.939\mic\ (the observed feature is redshifted relative to this line and the model predicts no line emission at about 1.97\mic), and Na\one\,2.206\mic. By comparison, the model with decay-power only is much too faint and exhibits much stronger line emission relative to the `background' flux made essentially from Fe\one\ and Fe\two\ overlapping emission lines.

The he8p00 model with magnetar power reproduces satisfactorily the observations (with at most a 50\,\% scaling in flux). The main discrepancies are seen in the blue part of the optical where the flux is overly enhanced by the introduction of magnetar power. It is not clear how reliable the flux in the 1\mic\ region is but observations seem to indicate a very broad He\one\,1.083\mic\ line, which the model predicts systematically as narrow. The same applies for the Ca\two\,H\&\,K lines, which is much broader in the observations. Both line mismatches are in support of an asymmetric explosion and compatible with the spectral properties around maximum light \citep{pandey_12au_21}. 

\begin{figure*}
\centering
\includegraphics[width=\hsize]{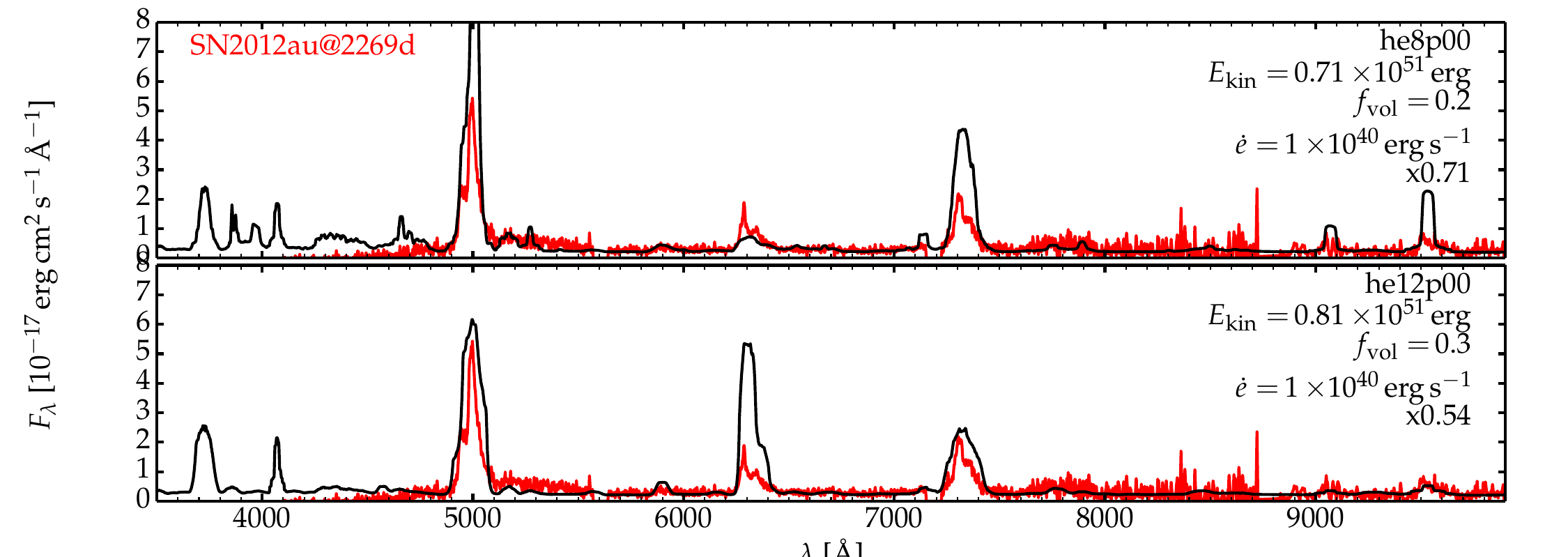}
\caption{Same as for Fig.~\ref{fig_12au_289d}, but now for the observations of SN\,2012au at 2269\,d and using the he8p00 or he12p00 models with a magnetar power of 10$^{40}$\,\ergs\ and different levels of clumping. In each case, the model is scaled to match the same optical luminosity as inferred from observations. We also shift both models up by $2 \times 10^{-18}$\ergs\,cm$^{-2}$\,\AA\ to match the observed flux present between the emission lines in SN\,2012au (the model predicts zero flux in those regions). [See Section~\ref{sect_12au_2269d} for discussion.]
\label{fig_12au_2269d}
}
\end{figure*}

\subsection{The 2269\,d spectrum} 
\label{sect_12au_2269d}

Figure~\ref{fig_12au_2269d} presents a comparison between the observations of SN\,2012au at 2269\,d after explosion with he8p00 and he12p00 models, both with a magnetar power of 10$^{40}$\,\ergs\ and thus a factor of ten below that used at 1\,yr -- these same models but without magnetar power would have a luminosity of about 10$^{36}$\,\ergs\ from the \iso{44}Ti decay chain. A volume filling factor of 20--30\,\% is introduced to tame the ionization and avoid having an optical spectrum dominated by the \oiiidoub\ line alone (see Sections~\ref{sect_he12p00}--\ref{sect_he3p30}). With the adopted magnetar power of 10$^{40}$\ergs, these models match the observations in the optical within a factor of about two. When considering the whole grid of models computed for this study, the simultaneous presence of \oidoub, \oiidoub, and \oiiidoub\ favors a higher mass progenitor such as he8p00 or he12p00 (ejecta masses of 3.95 and 5.32\,\msun), whereas he3p30 and he6p00 tend to exhibit a strong \oiiidoub. In both models, \siiinz\ and \siiinf\ are predicted but well matched in strength and shape only in the he12p00 model. With the shuffled-shell technique, our assumption of spherical symmetry, and our parametrized and simplistic treatment of magnetar-power deposition, our model spectra tend to produce line profiles in tension with observations: the model line profiles exhibit sharp jumps (reflecting the shells from which emission occurs) and may be too narrow or broad. In a 3D ejecta with chemical mixing, the various line emitting regions would spread more smoothly in velocity space and the line profiles would be more smooth (as coincidently obtained for \siiinz\ and \siiinf\ in the he12p00 model).

The off-the-shell he8p00 model (originally from \citealt{woosley_he_19}, \citealt{ertl_ibc_20}, and \citealt{dessart_snibc_21}), with some level of clumping ($f_{\rm vol}$ of $\sim$\,0.2) yields a good match to the evolution of SN\,2012au from 1 to 6\,yr post explosion (see, however, \citet{omand_pm_23} who find that none of their models fitted the whole 1--6\,yr evolution). This magnetar-powered model is broadly consistent over the optical and near-infrared when available, but also captures some unique features of SN\,2012au such as the presence of narrow and strong O\one\,7774\,\AA\ or the presence of forbidden sulfur lines at 4068\,\AA\ or 9530\,\AA. Because one can trade mass for clumping, our results may also be compatible with the he12p00 model although its kinetic energy is small and too helium poor (the he12p00 model would be in tension with observations during the photospheric phase and the Type Ib classification). 


\section{Magnetar versus circumstellar interaction}
\label{sect_pm_vs_csm}

Despite the relative success of the magnetar-powered models described above in explaining the late-time observations of SN\,2012au, interaction of the ejecta with CSM is, and has been proposed, as an alternative scenario. As we argue earlier, the inherent process is similar in both cases in the sense that power is ultimately injected in the plasma in the form of high-energy particles and photons, from the inside in the case of the magnetar and from the outside in the case of CSM interaction. Because this power may be absorbed over a range of ejecta velocities, it may be difficult to distinguish the two easily. We can consider various scenarios and associated implications for the observables.

If the SN\,2012au was in fact powered by interaction with a spherical CSM, the moderate line width observed in the 2269\,d spectrum would suggest near complete deceleration of the ejecta. This implies an extraction of over 10$^{51}$\,\ergs\ and would have made SN\,2012au one of the most luminous Type Ib SN ever observed. It was overluminous with peak luminosity of several 10$^{42}$\,\ergs\ but thus not at the level of interacting SNe. Furthermore, the observations would have revealed the broad-boxy emission features observed in SN\,1993J \citep{matheson_93j_00a} and the decrease of their widths in time -- this was not observed. So, this possibility seems excluded.

SN\,2012au may have resulted from interaction with an asymmetric CSM. SN\,2012au exhibits narrow lines at 1\,yr after explosion, but these are from O and S. This suggests the presence of metal-rich slow-velocity material. However, SN\,2012au is a Type Ib SN and thus any interaction with CSM should have revealed narrow He\one\ lines (or even possibly H\one\ lines) since the immediate CSM around SN\,2012au was helium rich (and possibly H rich). Such putative interaction yielding narrow emission lines in a late-time SN spectrum has been observed, both for H-rich CSM in Type Ib SNe (e.g., SN\,2014C; \citealt{milisavljevic_14C_15}; \citealt{margutti_14C_16}) and He-rich CSM in Type Ia SNe (e.g., SN\,2020eyj; \citealt{kool_iacsm_23}). Thus, interaction with an asymmetric CSM also seems excluded for SN\,2012au.

Numerous features in the time evolution of SN\,2012au support a central engine such as a magnetar. Its high peak luminosity was well above what may be expected from the representative \nifs\ mass produced in core-collapse SNe -- an extra power source is required. The presence of both very broad spectral lines at early times (and at late times with He\one\,1.983\mic), but also very narrow spectral lines at late times (i.e., O\one\,7774\,\AA) indicate an asymmetric explosion (i.e., low-velocity material is depleted as the explosion energy increases in a spherical blast). There is no broad nor narrow emission lines of He\one\ observed at late times as would be expected from the interaction of the SN\,2012au ejecta with a He-rich CSM. All these properties suggest that a magnetized, fast-rotating neutron star is what makes SN\,2012au stand apart from the majority of Type Ib SNe.


\begin{figure}
\centering
\includegraphics[width=\hsize]{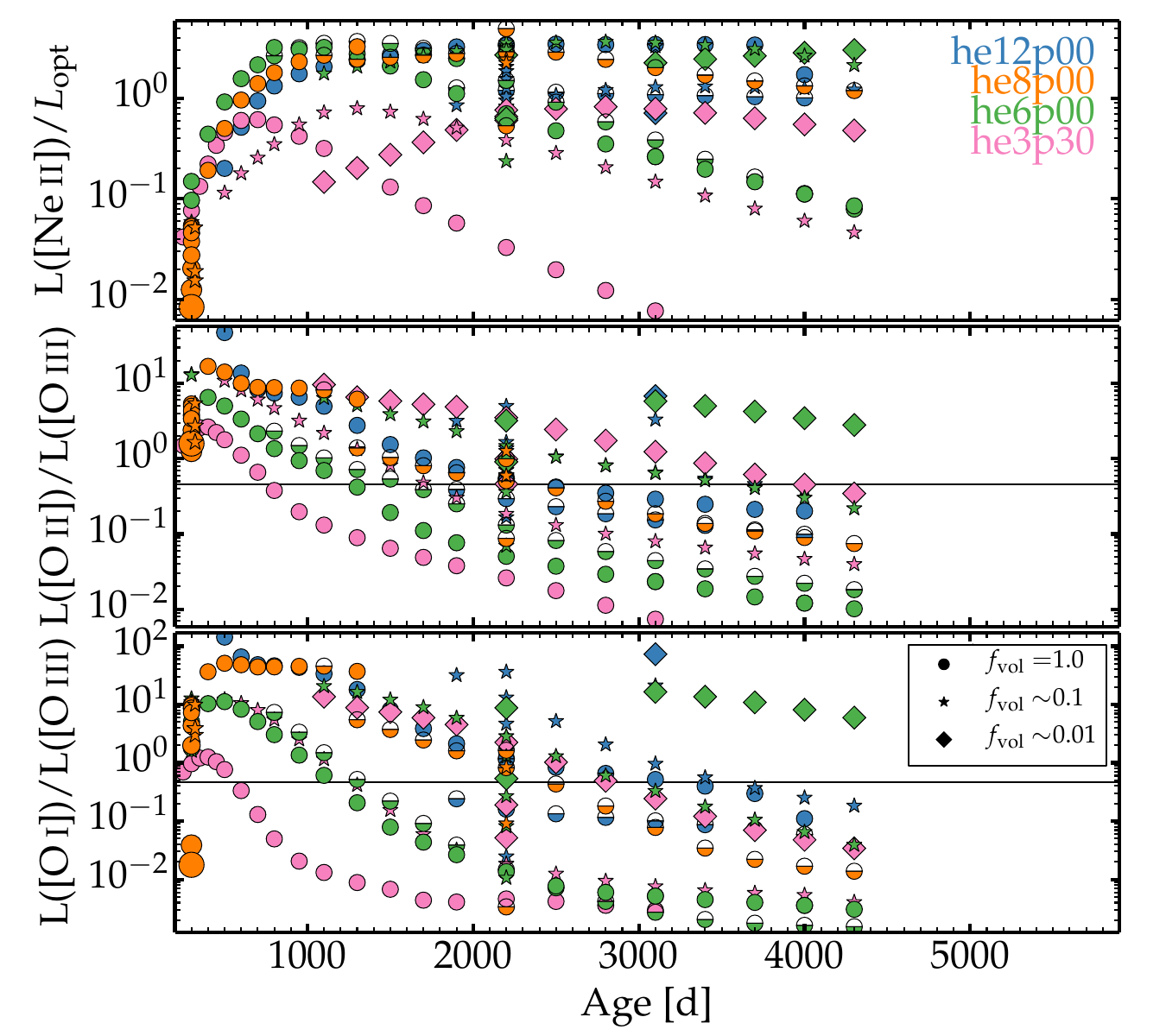}
\caption{Ratios of various line luminosities for the whole grid of magnetar-powered models as a function of time since explosion. We show the ratio of the \neiifs\ luminosity relative to the optical luminosity (top), the ratio of the \oiidoub\ and \oiiidoub\ luminosities (middle), and the ratio of the \oidoub\ and \oiiidoub\ luminosities (bottom). Different symbols are used to indicate the volume filling factor (i.e., clumping) and a color coding indicates the He-star mass. The horizontal black line corresponds to the observed counterparts for SN\,2012au at 2269\,d \citep{milisavljevic_12au_18}.
\label{fig_line_fluxes}
}
\end{figure}

\section{Conclusions}
\label{sect_conc}

We have presented a grid of magnetar-powered stripped-envelope SN models based on the He-star progenitors and explosions of \citet{woosley_he_19}, \citet{ertl_ibc_20}, and \citet{dessart_snibc_21}. Using the NLTE radiative transfer code \cmfgen\ \citep{hm98,HD12} and the treatment of \citet{d18_iptf14hls} and \citet{dessart_ibn_22} for the magnetar power, we have explored the influence of magnetar power in models he3p30, he6p00, he8p00, and he12p00 from about one to ten years after explosion and in particular how the resulting gas and radiative properties vary with ejecta clumping, the treatment of the power ejection (high-energy electrons versus X-ray irradiation, deposition profile, magnitude of the power). We also confronted our results to the observations of the magnetar-powered SN candidate 2012au. Our simulations are steady state and adopt a magnetar power of roughly 10$^{39}$ to 10$^{41}$\,\ergs\ to fall within a factor of ten of estimates for SN\,2012au \citep{milisavljevic_12au_13}.

Our simulations transition from decay powered to magnetar powered after 100-300\,d for a magnetar power of 10$^{41}$\,\ergs\ but that transition is delayed for weaker powers and occurs around 700\,d for a magnetar power of 10$^{39}$\,\ergs. After this transition, multiband light curves tend to show an inflection (e.g., $UVW1$ and $V$-band filters) but not necessarily since at such late times the ejecta cool exclusively through strong and isolated emission lines spread from the ultraviolet to the infrared.

The absorbed magnetar power is reradiated by means of collisional excitation and nonthermal excitation through a variety of lines. The strongest coolants vary with shell composition and ionization and their relative importance is dictated by the composition structure of the ejecta and the deposition profile of the magnetar power. These coolants may include species that are neutral, once, and twice ionized, that cover from carbon all the way to nickel, and that correspond to emission lines located from the ultraviolet to the infrared.

In higher mass models, the ejecta tend to be of lower ionization and being so oxygen-rich, they cool preferentially through O\one\ at early times, then through O\two, and finally through O\three. Ne\two\ is a critical coolant of the O-rich material under partially ionized conditions. The associated lines are \oidoub, \oiiuv, \oiidoub, \oiiidoub, and \neiifs, the later potentially channeling into the infrared 50\,\% or more of the total power absorbed by the ejecta. Being relatively small in comparison, the Fe- and Si-rich regions radiate little power and produce weak emission lines from S\two, S\three, Ar\two, Fe\two, Fe\three, Ni\two, or Ni\three. With a lower density, as in a model of lower mass or higher kinetic energy (i.e., he6p00), the ejecta conditions are typically more ionized and the cooling is done preferentially through higher ionization lines such as \oiiidoub. In a lighter model such as he3p30, the composition structure is a more balanced ensemble of shells of near-equal mass and distinct composition so the ejecta cool through a more diverse set of lines, in particular with a greater cooling from Si- and Fe-rich regions. Being of relatively low density, the ejecta are more ionized and cool at late times through O\three, Fe\three, or S\three.

These spectral properties are altered when ejecta clumping is introduced. With clumping, the ejecta density is increased, for the same ejecta mass, which boosts the recombination rates and facilitates recombination. The ejecta ionization is thus reduced with greater clumping, which modifies the coolants (e.g., O\three\ cooling might transition to O\two\ cooling etc.). Clumping introduces uncertainty when interpreting observations since it can considerably alter the spectrum even for the same ejecta composition, expansion rate, and mass. The change in ionization can also mitigate the amount of flux emitted in different spectral regions. For example, with a high ionization in the he12p00 model, most of the flux may emerge in \oiiidoub. But with a reduced ionization, most of the flux may emerge in the infrared through \neiifs\ while the optical spectrum contains a small fraction of the total power absorbed and radiates it through \oidoub, \oiidoub, and \oiiidoub. Multiwavelength (i.e., optical and infrared) observations of magnetar-powered SN candidates are essential to constrain the total power injected and the ejecta composition -- optical observations are not sufficient.

Treating the magnetar power in the form of X-ray irradiation yields similar results for the models and epochs tested here. This arises because at the microscopic level, the injection of high-energy electrons or X-rays photons is for the most part turned into heat in these ionized ejecta, and that heat is stored in thermal electrons. Hence, even though the two processes are distinct, their impact is similar. In nature, one may expect a combination of high-energy particles and high-energy photons from the magnetar, but in the end only the total power and its deposition profile within the ejecta matter. We also find that the ejecta are still optically thick to X-rays at ten years after explosion. The X-ray luminosity from magnetar-powered SNe should therefore be small and may escape detection \citep{margutti_slsn_xray_18}. Although our models evolve toward higher ionization, there is no dramatic ramp-up of the ionization and no X-ray breakout seems in the making \citep{metzger_pwn_14}. Our simplifying assumption of spherical symmetry has an adverse effect on X-ray escape but X-ray opacities should remain very large even at higher ionization since it is driven by K-shell photoabsorption, which will persist unless the material is fully ionized. High ionization is inhibited by the high cooling efficiency of these metal-rich, optically-thin ejecta (e.g., with \oiiidoub, \neiifs, or \neiiifs), whereby a single line may radiate the entire power absorbed.

A significant uncertainty in our work is the adopted deposition profile for the injected power. Ejecta asymmetry in both density and composition should influence the range of velocities over which the power is absorbed, what material absorbs this power, the coolants, and thus the spectral properties. Experimenting with the extent of the deposition profile, we find that more confined energy deposition (i.e., towards the inner ejecta layers) leads to a greater ionization, a greater preponderance of cooling from O\three, S\three, Fe\three, or Ni\three\ and narrower emission line profiles. This uncertainty could thus be reduced by means of spectroscopic observations. A similar shift towards higher ionization naturally follows from increasing the injected power.

We also explored whether some of the models in our grid were in rough agreement with the observations of SN\,2012au over the 1 to 6\,yr timespan. We find that the he8p00 model with clumping ($f_{\rm vol}$ of $\sim$\,0.2, corresponding to a density compression of five) yields a good match to the evolution of SN\,2012au from 1 to 6\,yr post explosion. This mitigates the result of \citet{omand_pm_23} who found that none of their models fitted the whole 1--6\,yr evolution. Our magnetar-powered he8p00 model is broadly consistent over the optical and near-infrared when available, but also captures some unique features of SN\,2012au such as the presence of narrow lines (e.g., O\one\,7774\,\AA\ or [S\two]\,4068\,\AA) and the strong emission blueward of 5500\,\AA\ mostly due to Fe\two\ lines.

Finally, we show in Fig.~\ref{fig_line_fluxes} various luminosity ratios for the models in our grid (not all are shown to limit the cluttering). This reemphasizes how diverse the strength of \oidoub, \oiidoub, \oiiidoub, and \neiifs\, taken relative to each other or to the total optical luminosity can be, both as a function of time, clumping, or progenitor model. In particular, we find that whenever O is partially ionized, the \neiifs\ luminosity is large and can even surpass the optical luminosity, in particular in models with a massive O/Ne/Mg since Ne\two\ becomes the primary coolant of that shell. This gives strong motivation for both optical and infrared observations of magnetar-powered SN candidates at late times.

\begin{acknowledgements}

We acknowledge discussions with John Hillier, Raffella Margutti, Ryan Chornock, and Danny Milisavljevic. LD thanks the Pittsburgh particle-physics, astrophysics, and cosmology center for financial support during a 2024 winter visit to the University of Pittsburgh. LD thanks the Max-Planck Institut f\"{u}r Astrophysik in Garching-bei-M\"{u}nchen  for their hospitality during the summer of 2024. LD thanks the Princeton Center for Theoretical Science for financial support to attend the workshop ``Forging a New Synthesis Between Supernova Theory and Observation'' held at Princeton University in December 2023. This work was supported by the ``Programme National Hautes Energies'' of CNRS/INSU co-funded by CEA and CNES. This work was granted access to the HPC resources of TGCC under the allocation 2023 -- A0150410554 on Irene-Rome made by GENCI, France. This research has made use of NASA's Astrophysics Data System Bibliographic Services.

\end{acknowledgements}


\end{document}